# Logic and Rational Languages of Scattered and Countable Series-Parallel Posets

## Amazigh Amrane
LITIS (EA 4108), Université de Rouen, Rouen, France
Amazigh.Amrane@etu.univ-rouen.fr

## Nicolas Bedon
LITIS (EA 4108), Université de Rouen, Rouen, France
Nicolas.Bedon@univ-rouen.fr

──── **Abstract** ────

Let $A$ be an alphabet and $SP^\diamond(A)$ denote the class of all countable N-free partially ordered sets labeled by $A$, in which chains are scattered linear orderings and antichains are finite. We characterize the rational languages of $SP^\diamond(A)$ by means of logic. We define an extension of monadic second-order logic by Presburger arithmetic, named P-MSO, such that a language $L$ of $SP^\diamond(A)$ is rational if and only if $L$ is the language of a sentence of P-MSO, with effective constructions from one formalism to the other. As a corollary, the P-MSO theory of $SP^\diamond(A)$ is decidable.

## 1 Introduction

Since they were independently established by Büchi [8], Elgot [13] and Trakhtenbrot [30], links between automata theory and formal logic have been widely developed. The fundamental result is the effective equivalence between Kleene automata [17] and sentences of Monadic Second-Order logic (MSO) for the description of languages of finite words. It provides, as examples, tools for languages classification, or decision algorithms on formal logic. This fundamental result have been later generalized to less restricted structures than finite words, relying on adapted notions of automata. Among them, let us cite words indexed by all the natural integers [9], ordinals [10], trees [26, 28, 11], linear orderings [7], etc. Among the first consequences of such generalizations are decision algorithms for second-order theories of countable ordinals [10], of two successors functions [26] and many others. Automata on a particular class $C$ of structures were often used to obtain decision procedures for a theory of $C$. However, it is to notice that automata on $C$ can also be used to obtain decision procedures for a theory of a seemingly unrelated class. As an example, the decidability of the second-order theory of countable linear orderings is reducible to the solvability of the decision problem of the second-order theory of two successors functions [26]. Apart from their interest in formal logic, generalization of Kleene automata are also used as models for processes. As an example, automata over $\omega$-words provide a basement for the model-checking theory of sequential processes. Automata over finite N-free (or equivalently, series-parallel [31]) partially ordered sets (posets for short) can be used as models for concurrent programs, where concurrency relies on fork-join rules. Recall that a set partially ordered by $<$ is *N-free* if it has no subset $X = \{a, b, c, d\}$ such that $< \cap X \times X = \{(a, b), (c, b), (c, d)\}$.

In this paper we focus on the class $SP^\diamond(A)$ of all countable N-free posets, labeled over an alphabet $A$, and whose chains are scattered linear orderings and antichains are finite. Recall that a linear ordering is *scattered* if it has no dense sub-ordering. A notion of automata, named *branching automata*, adapted to $SP^\diamond(A)$ has been introduced in [6] as well as equivalent rational expressions. These automata and rational expressions are a generalisation of those on finite N-free posets of Lodaya and Weil [20, 21, 22, 23] and those of Bruyère and Carton [7] on linear orderings. The logic P-MSO effectively equivalent to branching automata on finite N-free posets was introduced in [3]. Roughly speaking, P-MSO is a mix of MSO and Presburger arithmetic [25]. In [5] it was proved that the class of languages of countable linear orderings recognized by Bruyère and Carton automata is strictly included into MSO-definable languages, and that MSO and automata are effectively equivalent when linear orderings are restricted to be scattered. In [18, 19] Kuske proposed an extension of branching automata of Lodaya and Weil over finite N-free posets to N-free posets with finite antichains and $\omega$-chains, together with a connection with MSO in the particular case of languages of N-free posets with bounded-size antichains. This is extended in [2] to N-free posets with bounded-size antichains and scattered and countable chains.

Our main result is that a language of $SP^\diamond(A)$ is rational if and only if it is P-MSO definable (Theorem 36) with effective constructions from one formalism to the other. The decidability of the P-MSO theory of $SP^\diamond(A)$ follows as a corollary. P-MSO is defined in Section 4.

It is known from [4] that the class of rational languages of $SP^\diamond(A)$ is closed under boolean operations. Relying on this, the proof of the implication from right to left of Theorem 36 involves only well-known techniques: it is not developed in this paper. It is the purpose of the short Section 8. For the implication from left to right we had to develop new techniques. We were not able to generalize or extend those from the case of finite N-free posets [3], since they assume that every factor of a poset has a least upper or a greater lower bound: this is not true anymore for posets of $SP^\diamond(A)$. This paper is essentially devoted to the construction of a sentence $\varphi_e(X)$ of P-MSO from a rational expression $e$ such that $P$ satisfies $\varphi_e(P)$ if and only if $P \in L(e)$. It is developed using the following scheme. The first step is to transform the rational expression into an equivalent form, named $>1$-*expression*. In $>1$-expression, sequential products guarantee the sequential composition of at least two non-empty posets. Thus, the sequential composition of two languages necessarily contains only non-trivial sequential posets. The $>1$-expressions are introduced together with rational expressions in Section 3. The next step consists in computing a graph by induction on a $>1$-expression (Section 5). The nodes of the graph are labeled by letters, sequential operations involved in $>1$-expressions and Presburger formulæ. The operations labeling the nodes are all P-MSO expressible; in particular the sequential operations can be expressed with MSO using techniques from [5]. During the induction, we enforce properties on the resulting graph in order to calculate the Presburger formulæ that will appear in the final P-MSO formula. The next step is the transformation of the graph into a P-MSO formula. To each node $n$ we associate a P-MSO formula $\phi_n$, and the idea is to make $\phi_n$ dependent of $\phi_m$ if there is an edge $n \to m$. Unfortunately the graph is not always acyclic. Cycles can be broken by avoiding particular edges, named *special*, that are identified in each inductive step of the construction of the graph. In order to avoid circular dependencies between the $\phi_n$s we develop in Section 6 a technique named *s-coloring*, that permits to identify particular factors of posets by means of P-MSO. Using s-coloring, the effective transformation of the graph into a P-MSO formula is given in Section 7.

This paper is a long version, with proofs, of [1].

## 2  Notation, linear orderings and posets

We let $|E|$ denote the cardinality of a set $E$, $2^E$ its power-set, $[n]$ the set $\{1, \ldots, n\}$ (for any non-negative integer $n \in \mathbb{N}$), and $\pi_i(c)$ the $i^{\text{th}}$ component of a tuple $c$. We let $1_k$ denote the tuple of integers with 1 at the $k^{\text{th}}$ position and 0 at all other positions. Every time we use this notation the size $n$ of the tuple is clear from the context, and of course $k \in [n]$. Arithmetic operations on tuples of integers are componentwise. We also let $|s|$ denote the length $k$ of a finite sequence $s$ of $k$ elements.

### 2.1  Unlabeled posets and linear orderings

Recall that an ordering $<$ over a set $E$ is an irreflexive, transitive and asymmetric binary relation over the elements of $E$. It is *total*, or *linear*, when either $x < y$ or $y < x$ for all distinct $x, y \in E$. It is *dense* when for all $x, y$ such that $x < y$ there is some $z$ such that $x < z < y$. It is *scattered* if it has no dense non-trivial sub-ordering.

A partially ordered set, *poset* for short, $(P, <)$ consists of a set $P$ and an ordering relation $<$ over the elements of $P$. For simplicity we often denote $(P, <)$ by $P$. We let $0 = (\emptyset, \emptyset)$ denote the empty poset. A *chain* is a totally ordered set. An *antichain* $A$ is a set whose elements are pairwise incomparable: for all $x, y \in A$, neither $x < y$ nor $y < x$. An *interval* $I$ of $(P, <)$ is a subset $I \subseteq P$ such that for all $i_1, i_2 \in I$ and $p \in P$, if $i_1 < p < i_2$ then $p \in I$.

A poset $(P, <)$ is *N-free* if there is no $X = \{x_1, x_2, x_3, x_4\} \subseteq P$ such that $< \cap X^2 = \{(x_1, x_2), (x_3, x_2), (x_3, x_4)\}$. N-free posets play a particular role in computer science, since

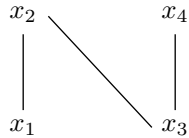

🟨 **Figure 1** The poset $N$. Exceptionally, in this Figure, the (increasing) ordering is bottom-up they are related to scheduling of processes, and model for example series and parallel electronic circuits, and concurrent processes relying on fork/join primitives. In this paper we particularly focus on N-free posets with finite antichains only.

We need the following operations on posets. The reversal $-(P, <)$ is $(P, <')$ defined by $x <' y$ if and only if $y < x$. Let $(P, <_P)$ and $(Q, <_Q)$ be two disjoint posets. The *union* (or *parallel composition*) $P \cup Q$ of $(P, <_P)$ and $(Q, <_Q)$ is the poset $(P \cup Q, <_P \cup <_Q)$. The *sum* (or *sequential composition*) $P + Q$ of $P$ and $Q$ is the poset $(P \cup Q, <_P \cup <_Q \cup P \times Q)$.

▶ **Example 1.** Figure 2 represents two posets $P_1 = (\{x_1, x_2, x_4, x_4\}, x_1 <_1 x_2, x_3, x_4)$ and $P_2 = (\{y_1, y_2, y_3, y_4\}, y_1, y_2 <_2 y_3, y_4)$, the parallel composition $P_1 \cup P_2$ and the sequential composition $P_1 + P_2$.

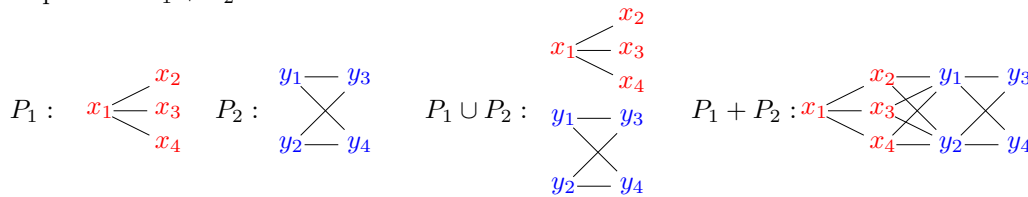

🟨 **Figure 2** The Hasse diagrams of two N-free posets $P_1$ and $P_2$, and their parallel and sequential compositions. In this Figure and all the others below, the ordering is from left to right

Before continuing let us focus on the particular cases of $(P, <)$ where the ordering $<$ is total. Linear orderings have a particular place in set theory. We refer the reader eg. to Rosenstein's book [27], entirely devoted to the subject. As it is conventional in the literature we let $\omega$ denote a representative of the class of linear orderings isomorphic to the linear ordering $(\mathbb{N}, <_{\mathbb{N}})$ of the natural integers. Ordinals are a particular case of scattered linear orderings. We let $\mathcal{O}$ and $\mathcal{S}$ denote respectively the class of countable ordinals and the class of countable scattered linear orderings (up to isomorphism). A *cut* $(K, L)$ of a linear ordering $(J, <)$ consists of a pair of two disjoint intervals $K$ and $L$ of $J$ such that $K \cup L = J$ and $k <_J l$ for all $(k, l) \in K \times L$. The set $\hat{J}$ of all cuts of $J$ is naturally equipped with the linear ordering $(K_1, L_1) <_J (K_2, L_2)$ if and only if $K_1 \subsetneq K_2$. By extension, we equip the set $J \cup \hat{J}$ with the linear ordering $<$ containing $<_J$ and $<_j$, and such that $j < (K, L)$ (resp. $(K, L) < j$) whenever $j \in K$ (resp. $j \in L$), for any $j \in J$ and $(K, L) \in \hat{J}$. We let $\hat{J}^*$ denote $\hat{J} \setminus \{(\emptyset, J), (J, \emptyset)\}$.

The sum of two posets can be generalized to any linearly ordered sequence $(P_j, <_j)_{j \in J}$ of pairwise disjoint posets by

$$\sum_{j \in J} P_j = (\bigcup_{j \in J} P_j, (\bigcup_{j \in J} <_j) \cup (\bigcup_{j, j' \in J, \ j < j'} P_j \times P_{j'}))$$

In this paper we consider posets of the following class:

▶ **Definition 2.** *The class $SP^\diamond$ of series-parallel scattered and countable posets is the smallest class of posets containing $0$, the singleton and being closed under disjoint finite parallel composition and disjoint sum indexed by countable scattered linear orderings. In this definition, posets are considered up to isomorphism.*

The following is an extension of a well-known result [31] on finite N-free posets.

▶ **Theorem 3** ([6]). *Let $NF^\diamond$ be the class of countable N-free posets with scattered chains only and without infinite antichains (up to isomorphism). Then $NF^\diamond = SP^\diamond$.*

▶ **Example 4.** For any $i \in \mathbb{N}$, let $A_i$ be an antichain of cardinality $|A_i| = i$. Set $P = \sum_{i \in \omega} A_i$. Then $P \in SP^\diamond$. Observe that there is no $n \in \mathbb{N}$ such that $|A| < n$ for all antichain $A$ of $P$.

We let $SP^{\diamond+}$ denote $SP^\diamond \setminus \{0\}$.

F. Hausdorff proposed in [16] an inductive definition of scattered linear orderings. In fact, each countable and scattered linear ordering is obtained using sums indexed by finite linear orderings, $\omega$ and $-\omega$. This has been adapted in [6] to $SP^\diamond$.

We denote by $\mathcal{C}_{\cup, +}(E)$ the closure of a set $E$ under finite disjoint union and finite disjoint sum. Recall that $0$ denotes the empty poset. We also let $1$ denote the singleton poset.

▶ **Definition 5.** *The classes of countable and scattered posets (equivalent up to isomorphism) $V_\alpha$ and $W_\alpha$ are defined inductively as follows:*

$$\begin{aligned}
V_0 &= \{0, 1\} \\
W_\alpha &= \mathcal{C}_{\cup, +}(V_\alpha) \\
V_\alpha &= \left\{ \sum_{i \in J} P_i : J \in \{\omega, -\omega\} \ and \ \forall i \in J, P_i \in \bigcup_{\beta < \alpha} W_\beta \right\} \cup \bigcup_{\beta < \alpha} W_\beta
\end{aligned}$$

*and the class $S_{sp}$ of countable and scattered posets by $S_{sp} = \bigcup_{\alpha \in \mathcal{O}} W_\alpha$.*

The following theorem extends a result of Hausdorff on linear orderings [16].

▶ **Theorem 6** ([6]). $S_{sp} = SP^{\diamond}$.

A unique ordinal can be associated to every poset of $SP^{\diamond}$, which can be used to prove properties inductively:

▶ **Definition 7.** *The* rank $r(P)$ *of* $P \in S_{sp}$ *is the smallest ordinal* $\alpha$ *such that* $P \in W_{\alpha}$.

The notion of a rank for posets of $SP^{\diamond}$ can be refined as follows. For every $\alpha \in \mathcal{O}$, the class $W_{\alpha}$ can be decomposed as the closure of $V_{\alpha}$ by finite disjoint union and finite disjoint sum.

▶ **Theorem 8** ([6]). *For all* $\alpha \in \mathcal{O}$, $i \in \mathbb{N}$, *let*

$$
\begin{aligned}
X_{\alpha,0} &= V_{\alpha} \\
Y_{\alpha,i} &= \left\{ P : \exists n \in \mathbb{N} \ P = \sum_{j \leq n} P_j \ such \ that \ P_j \in X_{\alpha,i} \ for \ all \ j \leq n \right\} \\
X_{\alpha,i+1} &= \left\{ P : \exists n \in \mathbb{N} \ P = \bigcup_{j \leq n} P_j \ such \ that \ P_j \in Y_{\alpha,i} \ for \ all \ j \leq n \right\}
\end{aligned}
$$

*Then* $W_{\alpha} = \bigcup_{i \in \mathbb{N}} X_{\alpha,i}$.

Define a well-ordering on $\mathcal{O} \times \mathbb{N}$ by $(\beta, j) < (\alpha, i)$ if and only if $\beta < \alpha$ or $\beta = \alpha$ and $j < i$. As a consequence of Theorems 6 and 8, for any $P \in SP^{\diamond}$ there exists a pair $(\alpha, i) \in \mathcal{O} \times \mathbb{N}$ as small as possible such that $P \in X_{\alpha,i}$.

▶ **Definition 9.** *The* X-rank $r_X(P)$ *of* $P \in SP^{\diamond}$ *is the smallest pair* $(\alpha, i) \in \mathcal{O} \times \mathbb{N}$ *such that* $P \in X_{\alpha,i}$.

We finish this section on unlabeled posets by some definitions that generalize to posets the usual notions of factors of words.

▶ **Definition 10.** *An interval* $I$ *of* $P \in SP^{\diamond}$ *is* good *if it is non empty and, for all* $p \in P$, *if there are* $x, y \in I$ *such that* $p < x$ *or* $x < p$ *and neither* $p < y$ *nor* $y < p$, *then* $p \in I$.

As emphasized by the following proposition, the notion of a good interval is deeply related with the usual notion of a factor. For that reason, in the sequel good intervals are called *factors*. Strictness of factors is relative to inclusion.

▶ **Proposition 11.** *Let* $P \in SP^{\diamond}$ *and* $I \subseteq P$. *Then* $I$ *is a good interval of* $P$ *if and only if there exist a non-empty* $J \in \mathcal{S}$ *and a sequence of non-empty posets* $(P_j)_{j \in J}$ *such that* $P = \sum_{j \in J} P_j$ *or* $J$ *is finite and* $P = \bigcup_{j \in J} P_j$, *and there exists* $j \in J$ *such that either* $I = P_j$ *or* $I \subsetneq P_j$ *and* $I$ *is a good interval of* $P_j$.

**Proof.** Observe that when $I$ is a good interval of $P$ and $X$ is such that $I \subseteq X \subseteq P$, then $I$ is also a good interval of $X$. As a consequence, the implication from left to right holds. For the converse, it suffices to note that when $I$ is a good interval of $I'$ which is itself a good interval of $P$, then $I$ is a good interval of $P$.                                                    ◀

▶ **Definition 12.** *Let* $P$ *be a poset and* $J \in \mathcal{S}$. *A* $J$-sequential-factorization *of* $P$, *also called* $J$-*factorisation* or sequential factorization *for short, is a sequence* $(P_j)_{j \in J}$ *of posets such that* $P = \sum_{j \in J} P_j$.

*A poset* $P$ *is* sequential *if it admits a* $J$-*factorization where* $J$ *contains at least two elements* $j \neq j'$ *with* $P_j, P_{j'} \neq 0$, *or* $P$ *is a singleton. It is* parallel *when* $P = P_1 \parallel P_2$ *for some* $P_1, P_2 \neq 0$. *A sequential factorization is* irreducible *when all the* $P_j$ *are either singletons or parallel posets. The notions of* parallel factorization *and* irreducible parallel factorization *are defined similarly.*

We let *Seq* denote the class of all sequential posets of $SP^{\diamond+}$. Note that every $P \in SP^{\diamond+}$ is either sequential or parallel, but not both. The empty poset is the only poset of $SP^{\diamond}$ which is neither sequential nor parallel. The poset $N$ of Figure 1 is neither sequential nor parallel, but $N \notin SP^{\diamond}$.

## 2.2 Labeled posets

An *alphabet* $A$ is a non-empty finite set whose elements are called *letters*. Recall also that a *language* of a set $S$ is a subset of $S$.

A poset $(P, <, l)$ *labeled* by $A$, also denoted by $P$ for short, consists of a poset $(P, <)$ and a *labeling* total map $l: P \to A$. Considered up to an isomorphism, labeled posets, also named *pomsets* in the literature, are a generalization of the usual notion of a word, since a word can be seen as a finite linear ordering labeled by $A$. The finite case was first investigated in [32, 15] from a systematic point of view. In order to be consistent with the usual notation on words, we let $\epsilon$ denote the unique empty labeled poset, and $a$ the singleton poset labeled by $a$. The class of posets of $SP^{\diamond}$ labeled by $A$ (or over $A$) is denoted by $SP^{\diamond}(A)$, and $SP^{\diamond+}(A) = SP^{\diamond}(A) \setminus \{\epsilon\}$. We also denote by $A^{\diamond}$ the restriction of $SP^{\diamond}(A)$ to posets with antichains of cardinality at most 1. Again, this notation is consistent with the words indexed by countable and scattered linear orderings, see eg. [7]. In order to match the words case, we also adapt the notation and nomenclature previously introduced for posets for the labeled case. The sequential and parallel compositions + and ∪ of posets are named *products*, and respectively denoted by · and ∥, when labeled posets are considered:

▶ **Definition 13.** *Let $(P, <_P, l_P)$ and $(Q, <_Q, l_Q)$ be two disjoint posets labeled by an alphabet $A$. The sequential product, or concatenation, $(P, <_P, l_P) \cdot (Q, <_Q, l_Q)$, or $(P, <_P, l_P)(Q, <_Q, l_Q)$ for short, is the poset $(P, <_P) + (Q, <_Q)$ labeled by $l_P \cup l_Q$. Similarly, the parallel product $(P, <_P, l_P) \parallel (Q, <_Q, l_Q)$ is the poset $(P, <_P) \cup (Q, <_Q)$ labeled by $l_P \cup l_Q$. The sequential product of a linearly ordered sequence of labeled posets is denoted by $\prod$.*

Let $A$ and $B$ be two alphabets, $P \in SP^{\diamond}(A)$, $L \subseteq SP^{\diamond}(B)$ and $\xi \in A$. The language of $SP^{\diamond+}(A \setminus \{\xi\} \cup B)$ consisting of the labeled poset $P$ in which each element labeled by the letter $\xi$ is non-uniformly replaced by a labeled poset of $L$ is denoted by $L \circ_\xi P$. By *non-uniformly* we mean that the elements labeled by $\xi$ may be replaced by different elements of $L$. This substitution $L \circ_\xi$ is the homomorphism from $(SP^{\diamond}(A), \parallel, \prod)$ into the power-set algebra $(2^{SP^{\diamond}(A \cup B)}, \parallel, \prod)$ with $\xi \mapsto L$ and $a \mapsto a$ for all $a \in A \setminus \{\xi\}$. In other words:

▶ **Definition 14.** *Let $A$ and $B$ be two alphabets, $P \in SP^{\diamond}(A)$, $L \subseteq SP^{\diamond}(B)$ and $\xi \in A$. Let $S_P$, $<_P$ and $l_P$ denote respectively the set of elements, the ordering relation, and the labeling map of $P$. Then $L \circ_\xi P$ consists of all $R \in SP^{\diamond}(A \setminus \{\xi\} \cup B)$ such that there exists $\nu_R: l_P^{-1}(\xi) \to L$ and*

$$S_R = (S_P \setminus l_P^{-1}(\xi)) \bigcup_{x \in l_P^{-1}(\xi)} S_{\nu_R(x)}$$

$$<_R = <_P|_{S_R} \bigcup_{x \in l_P^{-1}(\xi)} <_{\nu_R(x)} \bigcup_{(x,y) \in <_P \cap P \times l_P^{-1}(\xi)} \{x\} \times S_{\nu_R(y)} \bigcup_{(y,x) \in <_P \cap l_P^{-1}(\xi) \times P} S_{\nu_R(y)} \times \{x\}$$

$$l_R = l_P|_{S_R} \bigcup_{x \in l_P^{-1}(\xi)} l_{\nu_R(x)}$$

*where $f|_X$ is the restriction of the map or relation $f$ to $X$.*

▶ **Example 15.** Let $B = \{a, b\}$, $A = B \cup \{\xi\}$, $P = b \parallel (\xi \cdot \xi) \in SP^\diamond(A)$ and $L = \{a \parallel b, b \cdot a\} \subseteq SP^\diamond(B)$. Then $L \circ_\xi P = \{b \parallel ((a \parallel b) \cdot (a \parallel b)), b \parallel ((b \cdot a) \cdot (b \cdot a)), b \parallel ((a \parallel b) \cdot (b \cdot a)), b \parallel ((b \cdot a) \cdot (a \parallel b))\}$.

Sequential and parallel products are extended from labeled posets to languages of labeled posets in the usual way: when $L$ and $L'$ are languages of labeled posets and $op$ is either the sequential or the parallel product, then $L \; op \; L' = \{P \; op \; P' : P \in L, P' \in L'\}$.

## 3    Rational languages

In this section we define languages by means of expressions. Let $op$ be operations over a class of languages of structures labeled by an alphabet $A$. Recall that an *expression $e$* is a term of the free algebra over $\{\emptyset\} \cup A$ using the operations of $op$ as functions. The language $L(e)$ of $e$ is defined inductively using the definitions of the operations of $op$. Rational expressions describe rational languages. In a rational expression $e$, the union is usually denoted by $+$ instead of $\cup$.

In the first sub-section, we recall the definition of the class of rational languages of $SP^\diamond(A)$. By extension of a well-known result of Kleene [17] on languages of finite words, it is known from [6] that a language $L \subseteq SP^\diamond(A)$ is rational if and only if $L$ is the language of some automaton. In this paper, we need a slightly modified definition of rational languages of $SP^\diamond(A)$. In Sub-section 3.2, we define the class of >1-rational languages. A language $L \subseteq SP^\diamond(A)$ is rational if and only if it is >1-rational. Finally, in Sub-section 3.3, we recall the link between rational and semi-linear languages of finite commutative words.

### 3.1    Rational languages of $SP^\diamond(A)$

Let $A$ be an alphabet and $\xi \in A$. Define the following (disjoint) operations on the languages $L, L'$ of $SP^\diamond(A)$:

$$L \circ_\xi L' = \bigcup_{P \in L'} L \circ_\xi P \qquad\qquad L^* = \{\prod_{j \in [n]} P_j : n \in \mathbb{N}, P_j \in L\}$$

$$L^{*\xi} = \bigcup_{i \in \mathbb{N}} L^{i\xi} \text{ with } L^{0\xi} = \{\xi\} \text{ and } L^{(i+1)\xi} = (\bigcup_{j \le i} L^{j\xi}) \circ_\xi L$$

$$L^\omega = \{\prod_{j \in \omega} P_j : P_j \in L\} \qquad\qquad L^{-\omega} = \{\prod_{j \in -\omega} P_j : P_j \in L\}$$

$$L^\natural = \{\prod_{j \in \alpha} P_j : \alpha \in \mathcal{O}, P_j \in L\} \qquad L^{-\natural} = \{\prod_{j \in -\alpha} P_j : \alpha \in \mathcal{O}, P_j \in L\}$$

$$L \diamond L' = \{\prod_{j \in J \cup \hat{J}^*} P_j : J \in \mathcal{S} \setminus \{0\} \text{ and } P_j \in L \text{ if } j \in J \text{ and } P_j \in L' \text{ if } j \in \hat{J}^*\}$$

▶ **Definition 16.** *Let $A$ be an alphabet and $\xi \in A$. The class of* rational languages *[6] of $SP^\diamond(A)$ is the smallest class containing $\emptyset$, $\{\epsilon\}$, $\{a\}$ for all $a \in A$, and being closed under the operations of $op = \{\parallel, \circ_\xi, {}^{*\xi}, \cup, \cdot, *, \diamond, \omega, -\omega, \natural, -\natural\}$ under the following conditions:*

▬ $\epsilon \notin L$ in $L \circ_\xi L'$ and $L^{*\xi}$;
▬ *in $L^{*\xi}$, each element labeled by $\xi$ in a poset of $L$ must be incomparable with another element.*

The last condition excludes from the rational languages those of the form $(a\xi b)^{*\xi} = \{a^n \xi b^n : n \in \mathbb{N}\}$, for example, which is not rational in the usual language theory of finite words.

The definition above generalizes the notions of rational languages of several classes of structures. Actually:

- with $op = \{\cup, \cdot, *\}$ we get the rational languages of finite words of Kleene [17];
- with $op = \{\cup, \cdot, *, \|, \circ_\xi, *^\xi\}$ we get the rational languages of finite N-free posets of Lodaya and Weil [20, 21, 22, 23];
- with $op = \{\cup, \cdot, *, \diamond, \omega, -\omega, \natural, -\natural\}$ we get the rational languages of scattered and countable words of Bruyère and Carton [7].

For convenience we use the shortcut $L^\diamond$ for $L \diamond \epsilon + \epsilon$.

▶ **Example 17.** Let $A = \{a\}$ and $L = a \circ_\xi (a(\xi \| \xi))^{*\xi}$. Then $L$ is the smallest language containing $a$ and such that if $x, y \in L$ then $a(x \| y) \in L$. Thus $L = \{a, a(a \| a), a(a(a \| a) \| a), a(a(a \| a) \| a(a \| a)), \dots\}$. Furthermore, $L$ is a rational language of $SP^\diamond(A)$, since $L_1 = a(\xi \| \xi)$ and $L_2 = a$ are rational, $\epsilon \notin L_1, L_2$ and each $\xi$ is in parallel with the other $\xi$ in $L_1$.

The following result is fundamental in the sequel:

▶ **Theorem 18** ([4]). *Let $A$ be an alphabet. The class of rational languages of $SP^\diamond(A)$ is effectively closed under boolean operations.*

## 3.2   >1-**rational languages of** $SP^\diamond(A)$

In the remainder of the paper, we need sequentially operations to compose at least two non-empty posets. This is not the case for operations of the previous section, since for example $\{\epsilon\} \cdot \{a\} = \{a\}$. When $L, L' \subseteq SP^\diamond(A)$ define

$$L \cdot^{>1} L' = (L \setminus \{\epsilon\}) \cdot (L' \setminus \{\epsilon\}) \qquad\qquad L^{*^{>1}} = \{\prod_{i \in [n]} P_i : n > 1, P_i \in L \setminus \{\epsilon\}\}$$

$$L^{\omega^{>1}} = \{\prod_{i \in \omega} P_i : P_i \in L \text{ for all } i \in \omega \text{ and } P_i, P_j \neq \epsilon \text{ for some } i, j \text{ with } i \neq j\}$$

$$L \diamond^{>1} L' = \{\prod_{j \in J \cup \hat{J}^*} P_j : J \in \mathcal{S} \setminus \{0\}, P_j \in L \text{ if } j \in J, P_j \in L' \text{ if } j \in \hat{J}^*$$

$$\text{and } P_i, P_j \neq \epsilon \text{ for some } i, j \in J \cup \hat{J}^*, i \neq j\}$$

Define similarly $L^{-\omega^{>1}}$, $L^{\natural^{>1}}$ and $L^{-\natural^{>1}}$.

▶ **Definition 19.** *Let $A$ be an alphabet and $\xi \in A$. The class of* >1-*rational languages of $SP^\diamond(A)$ is the smallest class containing $\emptyset$, $\{\epsilon\}$, $\{a\}$ for all $a \in A$, and being closed under the operations of $op^{>1} = \{\|, \circ_\xi, *^\xi, \cup, \cdot^{>1}, *^{>1}, \diamond^{>1}, \omega^{>1}, -\omega^{>1}, \natural^{>1}, -\natural^{>1}\}$ under the following conditions:*

- $\epsilon \notin L$ in $L \circ_\xi L'$ and $L^{*\xi}$;
- in $L^{*\xi}$, each element labeled by $\xi$ in a poset of $L$ must be incomparable with another element.

We let $L +_c L'$ denote $L + L'$ when condition $c$ is verified, $L$ otherwise. Then

$$L \cdot L' = L \cdot^{>1} L' +_{\epsilon \in L} L' +_{\epsilon \in L'} L \qquad\qquad L^* = L^{*^{>1}} + L + \epsilon$$

$$L^\omega = L^{\omega^{>1}} +_{\epsilon \in L} L^* \qquad\qquad\qquad L \diamond L' = L \diamond^{>1} L' + L +_{\epsilon \in L} L'$$

Similar equalities hold for $-\omega$, $\natural$ and $-\natural$. Every rational expression can be transformed into a $>1$-expression. Considering the equalities above as rewriting rules this transformation is unique (up to associativity). As a consequence of Theorem 18, a language of $SP^\diamond(A)$ is rational if and only if it is $>1$-rational.

▶ **Example 20.** Consider the rational expressions $e_1 = a \circ_\xi (a(\xi \parallel \xi))^{*\xi}$ and $e_2 = e_1 \diamond b^\diamond$. Then $e_1' = a \circ_\xi (a \cdot^{>1}(\xi \parallel \xi))^{*\xi}$ is the $>1$-expression of $e_1$. The $>1$-expression of $e_2$ is $e_1' \diamond^{>1}(b \diamond^{>1}\epsilon + b + \epsilon) + e_1'$.

## 3.3 Rational languages of finite commutative words

Recall that in a monoid $(S, \cdot)$, the class of rational languages is the smallest containing the empty set, $\{s\}$ for all $s \in S$ and closed under the operations of $\{\cup, \cdot, ^*\}$. When the monoid is commutative we usually denote its product by $\parallel$ instead of $\cdot$ and its Kleene closure $^\circledast$ instead of $^*$. When $A$ is an alphabet, $A^\circledast$ is the class of all finite antichains over $A$, or equivalently, the class of all finite commutative words over $A$.

▶ **Definition 21.** *Let $(S, \parallel)$ be a commutative monoid. A subset $L$ of $S$ is linear if it has the form $L = a_1 \parallel \cdots \parallel a_k \parallel \left(\bigcup_{i \in I}(a_{i,1} \parallel \cdots \parallel a_{i,k_i})\right)^\circledast$ where $I$ is a finite set, $k, k_i \in \mathbb{N}$, $a_i, a_{i,j} \in S$ for all $i \in I$ and $j \in [k_i]$. It is semi-linear if it is a finite union of linear sets.*

▶ **Theorem 22** (see e.g. [12])**.** *In a commutative monoid, a language is rational if and only if it is semi-linear.*

▶ **Example 23.** Consider the language $L$ over the alphabet $A = \{a, b\}$ consisting of all commutative finite words with strictly more $a$s than $b$s. Then $L$ is the language of the rational expression $a \parallel a^\circledast \parallel (a \parallel a^\circledast \parallel b)^\circledast$ over the commutative monoid $A^\circledast$. It is linear, hence semi-linear, since it also has the form $a \parallel (a \parallel b \cup a)^\circledast$.

The following lemma is a direct consequence of a Parikh's Theorem [24, Theorem 2]:

▶ **Lemma 24.** *Let $A$ be an alphabet, $\xi \in A$, $L$ and $L'$ semi-linear languages of $A^\circledast$. Then $L \circ_\xi L'$ is a semi-linear language of $A^\circledast$, and if $\epsilon \notin L$ then so is $L^{*\xi}$.*

In the remainder of the paper, when an ordering is required over an alphabet $A = \{a_1, \ldots, a_n\}$ and is not specified explicitly, it is implicitly given by the indexes of the $a_i$s. Let $A = \{a_1, \ldots, a_n\}$ and $B = \{b_1, \ldots, b_m\}$ be two disjoint and totally ordered alphabets. For any $i \in [m]$ let us denote by $A \circ_{b_i} B$ the totally ordered alphabet $\{b_1, \ldots, b_{i-1}, a_1, \ldots, a_n, b_{i+1}, \ldots, b_m\}$. When $L$ and $L'$ are rational sets of respectively $A^\circledast$ and $B^\circledast$, $A$ and $B$ disjoint, then $L \circ_{b_i} L'$ is a rational set of $(A \circ_{b_i} B)^\circledast$. If $A = B$ then $L \circ_{b_i} L'$ is a rational set of $A^\circledast$.

Recall that every $u \in A^\circledast$ can be thought of as its Parikh's commutative image, that is a $n$-tuple $(|u|_{a_1}, \ldots, |u|_{a_n}) \in \mathbb{N}^n$, where $|u|_{a_i}$ denotes the number of occurrences of letter $a_i$ in $u$. Thus, $A^\circledast$ is isomorphic to $(\mathbb{N}^n, +)$.

## 4 Logic

Presburger arithmetic and Monadic Second-Order logic (MSO) are two classical logics in computer science. In this section we briefly recall those two logics and introduce P-MSO, which extends MSO with Presburger arithmetic.

### 4.1 Presburger arithmetic

Recall that Presburger arithmetic is the first-order logic of $(\mathbb{N}, +)$. The Presburger set $L(\rho)$ of a Presburger formula $\rho(x_1, \ldots, x_n)$ whose free variables are $x_1, \ldots, x_n$ consists of all interpretations of $(x_1, \ldots, x_n)$ which satisfy $\rho$. A language $L \subseteq \mathbb{N}^n$ is a *Presburger set* of $\mathbb{N}^n$ if it is the Presburger set of some Presburger formula. We let $\mathcal{P}_n$ denote the class of all Presburger formulæ with $n$ free variables and we set $\mathcal{P} = \bigcup_{i \in \mathbb{N}} \mathcal{P}_i$.

Presburger logic provides tools to manipulate semi-linear sets of commutative monoids with formulæ.

▶ **Theorem 25** ([14]). *Let $A = \{a_1, \ldots, a_n\}$ be a totally ordered alphabet. A language $L$ of $A^\circledast$ is semi-linear if and only if it is the Presburger set $L(\rho)$ of some Presburger formula $\rho(x_1, \ldots, x_n)$, i.e. $(|u|_{a_1}, \ldots, |u|_{a_n}) \in L(\rho)$ if and only if $u \in L$. Furthermore, the constructions from one formalism to the other are effective.*

Observe that the ordering of the free variables $x_1, \ldots, x_n$ of $\rho$ is related to the ordering of $A$. By convention $\{()\}$ is the Presburger set of any closed tautology $\rho$. When $\rho$ is used to define some language, we identify the empty tuple $()$ with $\epsilon$.

▶ **Example 26.** The Presburger set of the formula $\rho(x, y) \equiv \exists\, k, k'\ k = k' + 1 \land x = y + k$ of $\mathcal{P}_2$ is $\{(x, y) : y < x\}$. Let $A$ and $L$ be as in Example 23. Then $\rho(x, y)$ defines $L$ in $A^\circledast$.

With the help of Theorem 25 and Lemma 24 we get the following definition.

▶ **Definition 27.** *Let $\rho(x_1, \ldots, x_k)$ and $\rho'(x'_1, \ldots, x'_{k'})$ be Presburger formulæ and let $A = \{a_1, \ldots, a_k\}$ and $B = \{b_1, \ldots, b_{k'}\}$ be two totally ordered alphabets, either disjoint or equal. Consider the Presburger sets of $\rho$ and $\rho'$ as semi-linear languages $L$ and $L'$ of respectively $A^\circledast$ and $B^\circledast$. When $A$ and $B$ are disjoint, for all $i \in [k']$, $L \circ_{b_i} L'$ is a semi-linear language of $(A \circ_{b_i} B)^\circledast$ and also the Presburger set of some formula that we denote by $\rho \circ_{x'_i} \rho'$. When $A = B$, for all $i \in [k']$, $L \circ_{b_i} L'$ is a semi-linear language of $A^\circledast$ and the Presburger set of some formula that we denote by $\rho \bullet_{x'_i} \rho'$. If $(0, \ldots, 0) \notin L(\rho)$, for all $j \in \mathbb{N}$ and $i \in [k]$, $L^{*a_i}$ and $L^{j a_i}$ are semi-linear languages of $A^\circledast$ and also the Presburger sets of some formulæ that we denote respectively by $\rho^{*x_i}$ and $\rho^{j x_i}$.*

▶ **Example 28.** Let $\rho(x_1, x_2) = x_1 + x_2 = 1$ and $\rho'(x'_1, x'_2, x'_3) = x'_1 = 1 \land x'_2 + x'_3 \leq 1$. Then $L(\rho) = \{(1, 0), (0, 1)\}$ and $L(\rho') = \{(1, 0, 1), (1, 1, 0), (1, 0, 0)\}$. Let $A = \{a_1, a_2\}$ and $B = \{b_1, b_2, b_3\}$ disjoint. Consider the Presburger sets of $\rho$ and $\rho'$ as semi-linear languages $L$ and $L'$ of respectively $A^\circledast$ and $B^\circledast$. Then $L$, $L'$ and $L \circ_{b_3} L'$ are respectively the languages of the rational expressions $a_1 + a_2$, $b_1 \parallel (b_2 + b_3 + \epsilon)$ and $b_1 \parallel (b_2 + a_1 + a_2 + \epsilon)$ of $A^\circledast$, $B^\circledast$, and $(A \circ_{b_3} B)^\circledast$. The Parikh commutative image $\{(1, 0, 0, 0), (1, 1, 0, 0), (1, 0, 1, 0), (1, 0, 0, 1)\}$ of $L \circ_{b_3} L'$ is also the Presburger set of $\rho \circ_{x'_3} \rho' \equiv x_1 = 1 \land x_2 + x_3 + x_4 \leq 1$.

▶ **Lemma 29.** *Let $\rho(x_1, \ldots, x_n)$ be a Presburger formula and $i \in [n]$. For all $j \in \mathbb{N}$, $\rho^{(j+1)x_i} \equiv (\vee_{k \leq j} \rho^{k x_i}) \bullet_{x_i} \rho$. Furthermore, $L(\rho^{*x_i}) = \cup_{j \in \mathbb{N}} L(\rho^{j x_i})$.*

**Proof.** Consider the Presburger set of $\rho$ as a semi-linear language over the totally ordered alphabet $\{a_1, \ldots, a_n\}$. Let us start with the first part of the lemma. Following definitions,

$$L(\rho^{(j+1)x_i}) = L(\rho)^{(j+1)a_i} = (\cup_{k \leq j} L(\rho)^{k a_i}) \circ_{a_i} L(\rho)$$
$$= (\cup_{k \leq j} L(\rho^{k x_i}) \circ_{a_i} L(\rho) = L(\vee_{k \leq j} \rho^{k x_i}) \circ_{a_i} L(\rho) = L((\vee_{k \leq j} \rho^{k x_i}) \bullet_{x_i} \rho)$$

For the second part of the lemma, $L(\rho^{*x_i}) = L(\rho)^{*a_i} = \cup_{j \in \mathbb{N}} L(\rho)^{j a_i} = \cup_{j \in \mathbb{N}} L(\rho^{j x_i})$. ◀

▶ **Example 30.** Let $\rho(x_1, x_2) = (x_1 = 1 \wedge x_2 = 0) \vee (x_1 = 0 \wedge x_2 = 2)$. Let $A = \{a_1, a_2\}$ and consider the Presburger set of $\rho$ as a language $L$ of $A^{\circledast}$. Then $L = \{a_1, a_2 \parallel a_2\}$. Consider $L^{*a_2} = \bigcup_{i \in \mathbb{N}} L^{ia_2}$. By induction on $i$ it can be checked that

$$\bigcup_{j \leq i} L^{ja_2} = \{u \in A^{\circledast} : 1 \leq |u| \leq 2^i \text{ and } |u|_{a_1} \leq 2^{i-1}\}$$

It follows that $L^{*a_2} = A^{\circledast} \setminus \{\epsilon\}$, and as a consequence $\rho^{*x_2}(x_1, x_2) \equiv x_1 + x_2 \geq 1$. Now for any $i \in \mathbb{N}$ let $\rho_i(x_1, x_2) \equiv 1 \leq x_1 + x_2 \leq 2^i \wedge x_1 \leq 2^{i-1}$. Then $L(\rho_i)$ is the Parikh commutative image of $\bigcup_{j \leq i} L^{ja_2}$. Thus $\rho^{(i+1)x_2} \equiv \rho_i \bullet_{x_2} \rho$ and $\rho^{0x_2}(x_1, x_2) \equiv x_1 = 0 \wedge x_2 = 1$.

The following lemmas are direct consequences of the definitions above.

▶ **Lemma 31.** *Let $\rho(x_1, \ldots, x_k)$ and $\rho'(x'_1, \ldots, x'_{k'})$ be two Presburger formulæ where $k \in \mathbb{N}$, $k' \in \mathbb{N} \setminus \{0\}$ and $L(\rho), L(\rho') \neq \emptyset$, and let $i \in [k']$. Then $(v''_1, \ldots, v''_{k+k'-1}) \in L(\rho \circ_{x'_i} \rho')$ if and only if $(v''_1, \ldots, v''_{i-1}, v'_i, v''_{i+k}, \ldots, v''_{k+k'-1}) \in L(\rho')$, for some $v'_i \in \mathbb{N}$, and there exist $(v_{1,1}, \ldots, v_{1,k}), \ldots, (v_{v'_i,1}, \ldots, v_{v'_i,k}) \in L(\rho)$ such that $(v''_i, \ldots, v''_{i+k-1}) = (v_{1,1}, \ldots, v_{1,k}) + \cdots + (v_{v'_i,1}, \ldots, v_{v'_i,k})$.*

▶ **Lemma 32.** *Let $\rho(x_1, \ldots, x_k)$ and $\rho'(x'_1, \ldots, x'_k)$ be two Presburger formulæ where $k \in \mathbb{N} \setminus \{0\}$ and $L(\rho), L(\rho') \neq \emptyset$, and let $i \in [k]$. Then $(v''_1, \ldots, v''_k) \in L(\rho \bullet_{x'_i} \rho')$ if and only if there exist $(v'_1, \ldots, v'_k) \in L(\rho')$ and $(v_{j,1}, \ldots, v_{j,k}) \in L(\rho)$ for all $j \in [v'_i]$ such that for all $r \in [k]$*

$$v''_r = \begin{cases} \sum_{j \in [v'_i]} v_{j,r} & \text{when } r = i; \\ v'_r + \sum_{j \in [v'_i]} v_{j,r} & \text{otherwise.} \end{cases}$$

▶ **Lemma 33.** *Let $\rho(x_1, \ldots, x_k)$ be a Presburger formula for some $k \in \mathbb{N} \setminus \{0\}$ and such that $(0, \ldots, 0) \notin L(\rho)$. Let $i \in [k]$ and $t \in [k] \setminus \{i\}$. Then $1_t \in L(\rho)$ if and only if $1_t \in L(\rho^{*x_i})$.*

## 4.2 P-MSO

We refer the reader to [29] for a survey on MSO. Presburger Monadic Second-Order logic, P-MSO for short, is an extension of Monadic Second-Order logic (MSO) with Presburger arithmetic.

Let $A$ be an alphabet, $V_1 = \{x, y, \ldots\}$, $V_2 = \{X, Y, \ldots\}$ and $V_{\mathbb{N}}$ be three disjoint sets of variables, respectively named *first-order variables of P-MSO*, *second-order variables of P-MSO*, and *(first-order) variables of Presburger arithmetic*. Formulæ of P-MSO are inductively built according to the following grammar, where $a \in A$, $x, y \in V_1$, $X, Y, Z \in V_2$, $x_1, \ldots, x_n \in V_{\mathbb{N}}$, $\psi, \psi_1, \ldots, \psi_n$ are P-MSO formulæ, and $\rho$ a Presburger formula.

$$\psi ::= a(x) \mid x \in X \mid x < y \mid \psi_1 \vee \psi_2 \mid \psi_1 \wedge \psi_2 \mid \neg \psi$$
$$\mid \exists x \psi \mid \exists X \psi \mid \forall x \psi \mid \forall X \psi \mid \mathcal{Q}(Z, \psi_1, \ldots, \psi_n, \rho(x_1, \ldots, x_n))$$

This grammar restricted to the 10 first items produces exactly the formulæ of MSO. The last item extends MSO to P-MSO.

Let us turn to the semantics. Formulæ of P-MSO are interpreted over posets of $SP^{\diamond}(A)$. Note that we allow posets to be empty. The variables of $V_1$, $V_2$, $V_{\mathbb{N}}$ are respectively interpreted over the elements of the posets, sets of elements of the posets, and $\mathbb{N}$. For convenience we often do not make the distinction between a variable and its interpretation when the context is sufficiently explicit.

When $x$ and $y$ are interpreted over the elements of a poset $(P, <_P, l_P) \in SP^\diamond(A)$ and $X$ is interpreted as a subset of $P$, then the atomic formula $x < y$ is interpreted as $x <_P y$, and $a(x)$ as $l_P(x) = a$. The semantics of $x \in X$ is self-explanatory, and those of all the other forms of formulæ of the grammar, except the last one, are as usual. So let us turn to the semantics of $\psi(Z) \equiv \mathcal{Q}(Z, \psi_1, \ldots, \psi_n, \rho(x_1, \ldots, x_n))$. Let $P \in SP^{\diamond+}(A)$ and $Z \subseteq P$. Then $Z$ satisfies $\psi$ if it is a non-empty factor of $P$, and there exist $(v_1, \ldots, v_n) \in L(\rho)$ and sequential posets $Z_{1,1}, \ldots, Z_{1,v_1}, \ldots, Z_{n,1}, \ldots, Z_{n,v_n} \in SP^{\diamond+}(A)$ such that $Z = \|_{i \in [n]} \|_{j \in [v_i]} Z_{i,j}$ and $Z_{i,j}$ satisfies $\psi_i$ for all $i \in [n]$ and $j \in [v_i]$.

▶ **Example 34.** Let $A = \{a_1, a_2\}$, and let $\psi_i \equiv \exists x \, a_i(x), i \in [2]$. Let $\rho(x_1, x_2) \equiv \exists k_1, k_2 \, x_1 = 2k_1 \wedge x_2 = 2k_2 + 1 \in \mathcal{P}$. Let $P_1 = a_1(a_1 \parallel a_1)a_1$, $P_2 = a_2 a_2$ and $P_3 = (a_1 \parallel a_2)a_1$. Then $P = P_1 \parallel P_2 \parallel P_3$ satisfies $\mathcal{Q}(P, \psi_1, \psi_2, \rho(x_1, x_2))$ since there is $(K_1, K_2) = (\{1, 3\}, \{2\})$ with $(|K_1|, |K_2|) \in L(\rho)$, and for all $i \in K_j$, $j \in [2]$, $P_i$ satisfying $\psi_j$.

In this paper we use a lot of common shortcuts in formulæ, such as for example $\psi_1 \to \psi_2$ for $\neg \psi_1 \vee \psi_2$. As usual the logical equivalence of formulæ is denoted by $\equiv$.

A relation $R \subseteq P^m \times (2^P)^n$ is P-MSO (resp. MSO) *definable* in $P \in SP^\diamond(A)$ if there is a P-MSO (resp MSO) formula $\phi_R(x_1, \ldots, x_m, X_1, \ldots, X_n)$ which is true if and only if $(x_1, \ldots, x_m, X_1, \ldots, X_n) \in R$.

▶ **Example 35.** Let $P \in SP^\diamond(A)$, $X, Y, Z \subseteq P$. The following relations are MSO definable:

- $Y \subseteq Z$, $Y \cup Z = X$, $Y \cap Z = \emptyset$ and $Y, Z \neq \emptyset$. For example, $Y \subseteq Z \equiv \forall y \, y \in Y \to y \in Z$;
- $X < Y \equiv \forall x \forall y \, (x \in X \wedge y \in Y) \to x < y$
- "$(Y, Z)$ is a partition of $X$":
  $\texttt{Partition}(X, Y, Z) \equiv Y \cup Z = X \wedge Y \cap Z = \emptyset \wedge Y, Z \neq \emptyset$
- $|X| = n$, for any $n \in \mathbb{N}$:
  - $|X| = 0 \equiv \forall x \, x \notin X$
  - $|X| = n + 1 \equiv \exists Y, Z \; \texttt{Partition}(X, Y, Z) \wedge |Y| = 1 \wedge |Z| = n$
- $X = Y + Z \equiv |Y| \neq 0 \wedge |Z| \neq 0 \wedge \texttt{Partition}(X, Y, Z) \wedge Y < Z$
- $F(R, P)$ "$R$ is a factor of $P$", as a direct consequence of Proposition 11;
- $F_s(R, P)$ "$R$ is a sequential factor of $P$", since
  $F_s(R, P) \equiv F(R, P) \wedge (|R| = 1 \vee \exists R_1, R_2 \; \texttt{Partition}(R, R_1, R_2) \wedge R_1 < R_2)$

The main result of this paper is the following:

▶ **Theorem 36.** *Let $A$ be an alphabet. A language $L$ of $SP^\diamond(A)$ is rational if and only if it is P-MSO definable. Furthermore the constructions from one formalism to the other are effective.*

## 5 D-graphs

A *D-graph* is a particular case of rooted, directed and finite graph whose edges are partially ordered and nodes are labeled. More formally:

▶ **Definition 37.** *A D-graph $D = (V, E_S, E_N, r, out, A, \gamma)$ is a rooted, directed, finite and labeled graph whose edges are partially ordered:*

- *$V$ is a finite set whose elements are the* nodes *of $D$;*
- *the edges of $D$ consist of two disjoint sets $E_S \subseteq V \times V$ (special edges) and $E_N \subseteq V \times V$ (normal edges). An edge $(n, n')$ is also denoted $n \to n'$. Its source is $n$ and its destination is $n'$;*

- $r \in V$ *is the* root *of* $D$*;*
- out *is a total map that associates to every* $n \in V$ *a totally ordered and finite sequence* $e_1 \ldots e_k$ *containing all edges of* $D$ *of source* $n$ *(note that an edge may occur more than once in the sequence). We say that* $n$ *is* edged *by* $out(n) = e_1 \ldots e_k$*. A* leaf *is a node* $n$ *such that* $out(n)$ *has no* normal *edge;*
- $A$ *is the* alphabet *of* $D$*;*
- $\gamma: V \to A \cup \mathcal{P} \cup \{\cdot^{>1}, *^{>1}, \diamond^{>1}, \omega^{>1}, -\omega^{>1}, \natural^{>1}, -\natural^{>1}\}$ *is a total map labeling the nodes of* $D$ *verifying:*
  - *if* $\gamma(n) \in A$ *then* $|out(n)| = 0$*;*
  - *if* $\gamma(n) \in \mathcal{P}_k$ *then* $|out(n)| = k$*;*
  - *if* $\gamma(n) \in \{*^{>1}, \omega^{>1}, -\omega^{>1}, \natural^{>1}, -\natural^{>1}\}$ *then* $|out(n)| = 1$*;*
  - *if* $\gamma(n) \in \{\cdot^{>1}, \diamond^{>1}\}$ *then* $|out(n)| = 2$*.*

*The sets of special edges, normal edges, and the root of* $D$ *are respectively denoted by* $E_S(D)$*,* $E_N(D)$ *and* $r(D)$*.*

In a D-graph $D$, a sequence of nodes $n_1, \ldots, n_k$ is *consecutive* if there is an edge $n_i \to n_{i+1}$ for all $i \in [k-1]$. Similarly, a sequence of edges $e_1, \ldots, e_k$ is *consecutive* if the destination of $e_i$ is the source of $e_{i+1}$, for all $i \in [k-1]$.

Let $n$ and $p$ be two nodes of $D$. In $D$, we say that $p$ is an *ascendant* of $n$ if $D$ contains a sequence of consecutive nodes $m_1, \ldots, m_k$ for some $k \in \mathbb{N}$, such that $p, m_1, \ldots, m_k, n$ are consecutive too. When $k = 0$ we say that $p$ is a *direct ascendant* of $n$. The definitions of the notions of *descendant* and *direct descendant* are symmetrical. When $n$ is the destination of the $i^{\text{th}}$ edge of $out(p)$ then $n$ is the $i^{\text{th}}$ *direct descendant* of $p$. Note that $n$ may be simultaneously the $i^{\text{th}}$ and the $j^{\text{th}}$ direct descendant of $p$ with $i \neq j$ (see Example 54).

We often see sequences as words. For example, we let $e \circ_{e'} s$ denote the sequence of edges obtained by replacing in the sequence of edges $s$ each occurrence of the edge $e'$ by the edge $e$. When $i$ is the position of an element of $s$, then $s' \circ_i s$ is the sequence obtained by replacing the $i^{\text{th}}$ element of $s$ by $s'$. We also let $n' \circ_n^{\text{src}} s$ denote the sequence of edges obtained from $s$ by replacing every occurrence of $n$ in sources of edges by $n'$. When nothing is specified, replacing the source of a normal (resp. special) edge provides a normal (resp. special) edge.

▶ **Definition 38.** *Let* $A$ *be an alphabet and* $\xi \in A$*. A D-graph* $D = (V, E_S, E_N, r, out, A, \gamma)$ *is* $\xi$-normalized *when the following conditions are verified:*

1. *if a node labeled by* $\xi$ *is a direct descendant of the root then it cannot be direct descendant of another node;*
2. *for every* $n \in V$ *with* $\gamma(n) \in \mathcal{P}$*, if* $out(n)$ *is some* $out(n) = s_1 \to d_1, \ldots, s_n \to d_n$ *there is at most one* $i \in [n]$ *such that* $\gamma(d_i) = \xi$*.*

*The* $\xi$-normalization *of* $D$ *consists of the transformation given by Algorithm 1.*

In the algorithm above by $\epsilon \circ_{e_j} out(n)$ (resp. $out(n) \leftarrow (n \to y) \circ_f out(n)$) we remove all the occurrences of the edge $e_j$ (resp. $f$) from $out(n)$ and thus implicitly from the D-graph.

▶ **Definition 39.** *Let* $D = (V, E_S, E_N, r, out, A, \gamma)$ *be a D-graph. We say that* $D$ *has* Property PP *if it has no edge* $n \to m$ *such that* $\gamma(n), \gamma(m) \in \mathcal{P}$*. It has* Property SS *if it has no special edge* $n \to m$ *such that* $m$ *is labeled in* $\mathcal{P}$*. It has* Property DAG *if* $(V, \emptyset, E_N, r, out, A, \gamma)$ *is acyclic.*

Algorithm 2 transforms a D-graph with no consecutive edges $n \to n'$, $n' \to n''$ such that $\gamma(n), \gamma(n'), \gamma(n'') \in \mathcal{P}$ into a D-graph with Property PP.

Obviously, $\xi$-normalization and PP-suppression preserve Properties PP and SS. The $\xi$-normalization also preserves Property DAG, but not PP-suppression.

---

**Algorithm 1** $\xi$-normalization of a D-graph

---

// First step

**for** each edge $f : n \to x$ for some $n \neq r$, $x$ direct descendant of $r$ and with $\gamma(x) = \xi$ **do**

    Add a new node $y$ labeled by $\xi$ in $V$ and $n \to y$ in $E_N(D)$

    $\mathrm{out}(n) \leftarrow (n \to y) \circ_f \mathrm{out}(n)$

// Second step

**for all** $n \in V$ such that $\gamma(n)$ is some $\rho(x_1, \ldots, x_k)$

                             and $\mathrm{out}(n)$ is some $e_1 : n \to n_1, \ldots, e_k : n \to n_k$ **do**

    **if** $|\{i_1, \ldots, i_m \in [k] : \gamma(n_{i_j}) = \xi\}| > 1$ **then**

        Add a new node $n_0$ labeled by $\xi$ in $V$ and $n \to n_0$ in $E_N(D)$

        $\gamma(n) \leftarrow \exists x_{i_1}, \ldots, x_{i_m}(x_0 = \sum_{j \in [m]} x_{i_j}) \wedge \rho(x_1, \ldots, x_k)$

        $\mathrm{out}(n) \leftarrow (n \to n_0)(\epsilon \circ_{e_{i_1}} \ldots \epsilon \circ_{e_{i_m}} \mathrm{out}(n))$

        Remove all $n_{i_j}$, $j \in [m]$, that are not the destination of some remaining edge

---

**Algorithm 2** PP-suppression

---

**for all** node $n$ such that $\gamma(n)$ is some $\rho(x_1, \ldots, x_k)$ **do**

    **while** there is some $(p, i)$ such that $n$ is the $i^{\mathrm{th}}$ descendant of node $p$

                             and $\gamma(p)$ is some $\rho'(x'_1, \ldots, x'_{k'})$ **do**

        $\gamma(p) \leftarrow \rho \circ_{x'_i} \rho'$

        $\mathrm{out}(p) \leftarrow (p \circ_n^{\mathrm{src}} \mathrm{out}(n)) \circ_i \mathrm{out}(p)$

    Remove $n$ if it has no incoming edge

---

## 5.1 From $>1$-expressions to D-graphs

Let $e$ be a $>1$-expression. In this section we build a D-graph $D_e$ by induction on $e$. During the construction, we assume the D-graphs obtained by the induction to be $\xi$-normalized. Thus, at each inductive step we construct a D-graph, and if not $\xi$-normalized we implicitly transform it with Algorithm 1. The construction ensures that the D-graph constructed at each inductive step has Properties PP, SS and DAG.

Let us start the construction of $D_e = (V, E_S, E_N, r, \mathrm{out}, A, \gamma)$. Except when the contrary is specified, new edges added during the constructions of this section are normal.

- Case $e = \epsilon$ (resp. $e = a \in A$)

  $D_e$ is just a node labeled by any closed Presburger tautology (resp. labeled by $a$), without edges.

- Case $e = e_1 \ op \ e_2$ (resp. $e = e'^{op}$) with $op \in \{\cdot^{>1}, \diamond^{>1}\}$ (resp. $op \in \{*^{>1}, \omega^{>1}, -\omega^{>1}, \natural^{>1}, -\natural^{>1}\}$)

  Then $D_e$ is built from the union of $D_{e_1}$ and $D_{e_2}$ (resp. from $D_{e'}$), with one more node $n$ as a root, labeled by $op$, and edged by $n \to r(D_{e_1}), n \to r(D_{e_2})$ (resp. $n \to r(D_{e'})$).

- Case $e = e_1 \circ_\xi e_2$

  When $e_2 = \xi$ then $D_e$ is identical to $D_{e_1}$. Otherwise, $D_e$ is built from the union of $D_{e_1}$ and $D_{e_2}$ transformed by Algorithm 3. The root of the new D-graph is $r(D_{e_2})$.

- Case $e = e'^{*\xi}$

  This is the only case where special edges are added. This is also the only case likely to add to the edging $\mathrm{out}(n)$ of a node $n$ an edge which is already in $\mathrm{out}(n)$. The construction of $D_{e'^{*\xi}}$ from $D_{e'}$ relies on the same principle as the previous case: each node of $D_{e'}$ labeled by $\xi$ should be replaced by a copy of the root. Again, this may be not so simple in some cases because of the properties we want to ensure on the resulting D-graph. The

---

**Algorithm 3** Construction of $D_{e_1 \circ_\xi e_2}$

---

**for all** node $n$ of $D_{e_2}$ with $\gamma(n) = \xi$ **do**
  $\gamma(n) \leftarrow \gamma(r(D_{e_1}))$
  $\mathrm{out}(n) \leftarrow n \circ^{\mathrm{src}}_{r(D_{e_1})} \mathrm{out}(r(D_{e_1}))$
  Apply PP-suppression
Remove $r(D_{e_1})$ from the D-graph

---

construction follows Algorithm 4, starting from $D_{e'}$. We proceed in two steps. The first step ensures in particular that the root is labeled by some Presburger formula $\rho$ with a $i^{\mathrm{th}}$ direct descendant labeled by $\xi$ for exactly one $i$, and that $1_i$ is in the Presburger set of $\rho$. In the second step we proceed with the replacement of nodes labeled by $\xi$ by a copy of the root and we ensure Property PP. Note that after the first step, the D-graph fulfills

---

**Algorithm 4** Construction of $D_{e'^{*\xi}}$

---

// First step: root transformation
**if** $\gamma(r(D_{e'}))$ is some $\rho(x_1, \ldots, x_k)$ and $\mathrm{out}(r(D_{e'}))$ is some $e_1 \ldots e_k$ **then**
  **if** there exists $e_i \colon r(D_{e'}) \to n_i$, $i \in [k]$, such that $\gamma(n_i) = \xi$ **then**
                                                           ▷ Since $D_{e'}$ is $\xi$-normalized $i$ is unique
    $\gamma(r(D_{e'})) \leftarrow \rho^{*x_i}$
  **else**
    Add a new node $x$ labeled by $\xi$ and $r(D_{e'}) \to x$ in $E_N(D_{e'})$
    $\gamma(r(D_{e'})) \leftarrow (\rho(x_1, \ldots, x_k) \wedge x_{k+1} = 0) \vee (\wedge_{i \in [k]} x_i = 0 \wedge x_{k+1} = 1)$
    $\mathrm{out}(r(D_{e'})) \leftarrow \mathrm{out}(r(D_{e'}))(r(D_{e'}) \to x)$
**else**
  Consider $D_{e' + \xi}$ instead of $D_{e'}$ for the remainder of the construction
// Second step
**for** each node $n$ labeled by $\xi$ which is not a direct descendant of the root $r$ **do**
  $\gamma(n) \leftarrow \gamma(r)$
  $\mathrm{out}(n) \leftarrow n \circ^{\mathrm{src}}_r \mathrm{out}(r)$                                    ▷ All those new edges are special
  Apply PP-suppression

---

Property PP. In particular, none of the direct descendants $n_1, \ldots, n_k$ of the root $r$ has its label in $\mathcal{P}$. Since the new special edges have their destinations in $n_1, \ldots, n_k$ then the construction preserves Property SS.

∎ Case $e = e_1 + e_2$ (resp. $e = e_1 \parallel e_2$)
  $D_e$ is built from the union of $D_{e_1}$ and $D_{e_2}$, with a new node $n$ labeled by $\rho(x_1, x_2) \equiv \sum_{i \in [2]} x_i = 1$ (resp. $x_1 = x_2 = 1$), edged by $n \to r(D_{e_1})$, $n \to r(D_{e_2})$, which is the root of $D_e$. Apply PP-suppression if necessary.

▶ **Example 40.** Let $e = (b \cdot (a \parallel \xi^*))^{*\xi}$. The corresponding $>1$-expression is $f = (b \cdot {}^{>1}(a \parallel (\xi^{*>1} + \xi + \epsilon)))^{*\xi}$. The different steps of the construction of $D_f$ are detailed in Figure 3. The transformation of the Presburger formula during the PP-suppression step is detailed in Example 28.

In the remainder of the paper we let $D_e$ denote the D-graph of $e$ when $e$ is a $>1$-expression, or of the $>1$-expression of $e$ when $e$ is a rational expression.

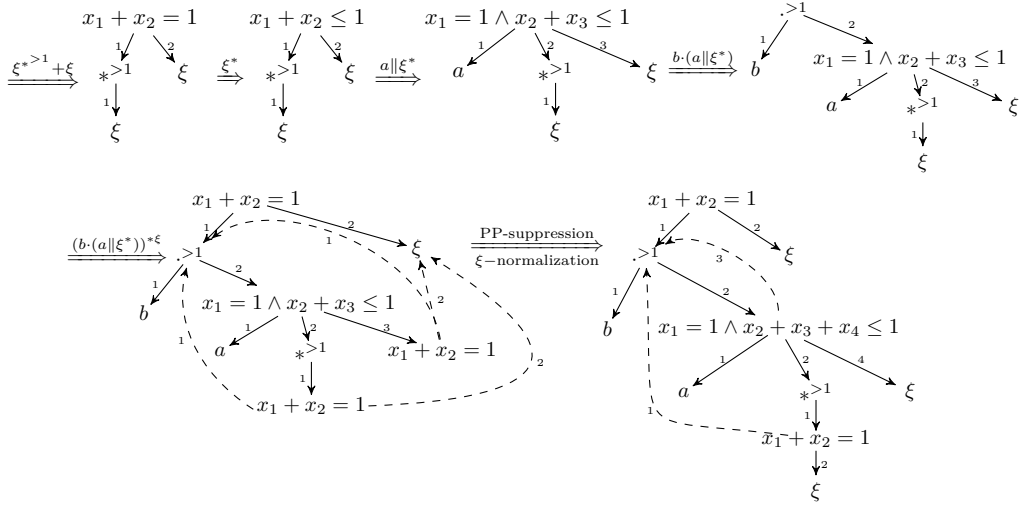

**Figure 3** The step-by-step construction of the D-graph of $(b \cdot^{>1}(a \parallel (\xi^{*>1} + \xi + \epsilon)))^{*\xi}$. PP-suppression is detailed only for the last step

## 5.2    Properties

This section is devoted to some structural properties of D-graphs of rational expressions. The proofs of these properties are essentially verification by induction on the rational expression from which the D-graphs are built.

The following lemma characterizes the D-graph $D_e$ of some $>1$-expression $e$ such that $\epsilon \in L(e)$.

▶ **Lemma 41.** *Let $e$ be a $>1$-expression. Then $\epsilon \in L(e)$ if and only if $r(D_e)$ is labeled by some Presburger formula $\rho(x_1, \ldots, x_k)$ and $(0, \ldots, 0) \in L(\rho)$, or $D_e$ has a unique node labeled by a closed Presburger tautology.*

**Proof.** By induction on $e$.

If $e = \epsilon$ then $D_e$ has a unique node labeled by a closed Presburger formula and the lemma immediately follows.

If $e = a$ then $D_e$ has a unique node labeled by $a$ and the lemma immediately follows.

If $e$ has the form $e = e_1 \; op \; e_2$ with $op \in \{\cdot^{>1}, \diamond^{>1}\}$ or $e$ has the form $e_1^{op}$ with $op \in \{*^{>1}, \omega^{>1}, -\omega^{>1}, \natural^{>1}, -\natural^{>1}\}$ then $\epsilon \notin L(e)$ and $r(D_e)$ is not labeled by a Presburger formula, so the lemma holds.

Assume now that $e$ has the form $e = e_1 \circ_\xi e_2$. Necessarily $\epsilon \notin L(e_1)$. If $e_2 = \xi$ then $D_e = D_{e_1}$ and the lemma follows from the induction hypothesis. Assume $e_2 \neq \xi$, so $D_e$ is built by modification of $D_{e_2}$, and the node root of $D_e$ is the (possibly modified) root of $D_{e_2}$. Assume first that $r(D_{e_2})$ is not modified during the construction. As $\epsilon \in L(e)$ if and only if $\epsilon \in L(e_2)$ the lemma directly follows from the induction hypothesis. Assume now that $r(D_{e_2})$ is modified during the construction. Then it is labeled by some Presburger formula $\rho(x_1, \ldots x_k)$, has a $i^{\text{th}}$ direct descendant $n$ labeled by $\xi$, $i \in [k]$, and $D_{e_1}$ has a root labeled by some Presburger formula $\rho'(x_1', \ldots, x_{k'}')$. By construction $r(D_e)$ is labeled by $\rho' \circ_{x_i} \rho$. Assume first $\epsilon \in L(e)$, and thus $\epsilon \in L(e_2)$. By induction hypothesis $(0, \ldots, 0) \in L(\rho)$ and by Lemma 31 $(0, \ldots, 0) \in L(\rho' \circ_{x_i} \rho)$. Reciprocally, assume $(0, \ldots, 0) \in L(\rho' \circ_{x_i} \rho)$. By Lemma 31 again, $(0, \ldots, 0) \in L(\rho)$. By induction hypothesis, $\epsilon \in L(e_2)$ and thus $\epsilon \in L(e)$.

Let us turn to the case $e = e'^{*\xi}$. For the implication from left to right notice that $\epsilon \in L(e)$ implies $\epsilon \in L(e')$, which is forbidden in the definition of $>1$-rational languages. So let us turn to the implication from right to left. Recall that the construction of $D_{e'^{*\xi}}$ from $D_{e'}$ is divided in two steps: the root of $D_{e'^{*\xi}}$ is built in the first step, and is not modified in the second step. After the first step, the root is labeled by a Presburger formula $\rho$. Step 1 is divided into three cases, and it is just verification to check that $\rho$ can not be closed and that $(0, \ldots, 0) \notin L(\rho)$.

The cases $e = e_1 + e_2$ and $e = e_1 \parallel e_2$ are also direct consequences of the construction and Lemma 31.

We have proved that the lemma holds at each step of the construction of $D_e$ from $e$ considering $D_e$ before Algorithm 1 is applied. To conclude, it suffices to observe that Algorithm 1 preserves the properties of the lemma.                                                                  ◀

Recall that by definition (Section 3.1), $L^{*\xi}$ is not rational when $L$ contains a poset of the form $P\xi P'$. The following lemma characterizes the D-graph $D_e$ of some rational expression $e$ such that $P\xi P' \in L(e)$ for some $P, P'$.

▶ **Lemma 42.** *Let $D_f$ be the D-graph of some rational expression $f$. Then either $D_f$ consists in a single node labeled by $\xi$, or there exists in $D_f$ a sequence of consecutive edges $e_1 \colon s_1 \to d_1, \ldots, e_n \colon s_n \to d_n$ such that*

- *$s_1 = r(D_f)$;*
- *$d_n$ is labeled by $\xi$;*
- *for all $i \in [n]$, if $s_i$ is labeled by some Presburger formula $\rho(x_1, \ldots, x_n)$, there is some $j \in [n]$ such that $e_i$ is the $j^{th}$ edge of $out(s_i)$ and $1_j \in L(\rho)$*

*if and only if $P_1 \xi P_2 \in L(f)$ for some $P_1, P_2 \in SP^\diamond(A)$.*

**Proof.** When $f$ is a rational expression the corresponding $>1$-expression is denoted by $e$. We proceed by induction on $f$. Before starting, recall that Algorithm 1 is used after each induction step of the construction of $D_e$. Observe that there is a sequence as in the statement of the lemma in $D_e$ after Algorithm 1 is finally applied if and only if such a sequence exists before Algorithm 1 is finally applied. Thus, we prove the lemma on the D-graph build by induction on $e$ before Algorithm 1 is finally applied.

If $f = a$ for some node $a \in A$ or $f = \epsilon$ then $D_e$ has no edges, and the lemma holds.

If $f$ has the form $f = f' \cdot f''$, then $e$ has the form $e = e'^{.>1}e'' +_{\epsilon \in L(e')} e'' +_{\epsilon \in L(e'')} e'$. Assume wlog. that $\epsilon \in L(e') \setminus L(e'')$, thus $e$ has the form $e = e'^{.>1}e'' + e''$. Assume first that $r(D_{e''})$ is not labeled by a Presburger formula. By construction, $D_e$ has a root labeled by the Presburger formula $\rho(x_1, x_2) \equiv x_1 + x_2 = 1$, and $out(r(D_e))$ has the form $out(r(D_e)) = r(D_e) \to n_1, r(D_e) \to n_2$, $n_2$ is the root of a copy of $D_{e''}$, $n_1$ is a node labeled by $.>1$ with two direct descendants $n_3$ and $n_4$, the roots of copies of respectively $D_{e'}$ and $D_{e''}$. Assume first that there exists in $D_e$ a sequence of consecutive edges $e_1 \colon s_1 \to d_1, \ldots, e_n \colon s_n \to d_n$ as in the statement of the lemma. There are five cases: $d_n$ is a descendant of $n_2$, $n_3$ or $n_4$, or $n_2$ or $n_4$ are nodes labeled by $\xi$ and $d_n$ is one of them. In the first case, it suffices to remove the first edge of the sequence and apply the induction hypothesis on the copy of $D_{e''}$ of root $n_2$. Similar arguments are used in the second case. The third case can be reduced to the first one, since $n_2$ and $n_4$ are both copies of $D_{e''}$. In the fourth case, $n = 1$, $(0, 1) \in L(\rho)$ and $\xi \in L(f)$. In the fifth case $n_2$ is also labeled by $\xi$. Assume now that $P_1 \xi P_2 \in L(f)$ for some $P_1, P_2 \in SP^\diamond(A)$. There are two cases. First case, there is some $P_{1,1}, P_{1,2} \in SP^\diamond(A)$ such that $P_1 = P_{1,1}P_{1,2}$ and $P_{1,2}\xi P_2 \in L(f'') = L(e'')$. It suffices to add $r(D_e) \to n_2$ at the beginning of the sequence of edges obtained by induction hypothesis applied on the copy of $D_{e''}$ of root $n_2$. Second case, there is some $P_{2,1}, P_{2,2} \in SP^\diamond(A)$ such that $P_2 = P_{2,1}P_{2,2}$ and $P_1\xi P_{2,1} \in L(f') = L(e')$. In this case it suffices to add $r(D_e) \to n_1, n_1 \to n_3$ at the beginning of the sequence of edges obtained by induction hypothesis applied on $D_{e'}$. When the label of $r(D_{e''})$ is a Presburger formula, Lemma 31 should be used in addition to the arguments above.

The cases of other sequential operators are similar.

If $f$ has the form $f = f' \circ_\xi f''$, then $e$ has the form $e = e' \circ_\xi e''$. If $e''$ is $\xi$ or $e'$ is $\xi$ then by construction $D_e$ is isomorphic respectively to $D_{e'}$ or $D_{e''}$ and it suffices to apply the induction hypothesis. So we may assume that $e'$ and $e''$ are not $\xi$, and that $D_e$ is build from $D_{e'}$ and $D_{e''}$. Assume first that there exists in $D_e$ a sequence of consecutive edges $e_1 \colon s_1 \to d_1, \ldots, e_n \colon s_n \to d_n$ as in the statement of the lemma. Since all the nodes labeled by $\xi$ have been replaced in $D_{e''}$, necessarily there exist $i \in [n]$ such that $e_i \colon s_i \to d_i$ is from a node of $D_{e''}$ to a node of $D_{e'}$. Then either the label of $s_i$ was $\xi$ and has been changed by the label of $r(D_{e'})$ during the construction, or $s_i$ was labeled by some Presburger formula $\rho(x_1, \ldots, x_k)$, $r(D_{e'})$ was labeled by a Presburger formula $\rho'(x'_1, \ldots, x'_{k'})$, $s_i$ had its $r^{\text{th}}$ direct descendant $s$ labeled by $\xi$ before the construction, $d_i$ is the $r'^{\text{th}}$ direct descendant of $r(D_{e'})$, and the label of $s_i$ has been changed to $\rho' \circ_{x_r} \rho$. Assume that the second case holds (the first case is simpler and uses similar arguments). By hypothesis $1_{r+r'-1} \in L(\rho' \circ_{x_r} \rho)$, and by Lemmas 41 and 31, $1_r \in L(\rho)$ and $1_{r'} \in L(\rho')$. The sequence $r(D_{e'}) \to d_i, s_{i+1} \to d_{i+1}, \ldots, s_n \to d_n$ of edges of $D_{e'}$ verifies the statement of the lemma, and by induction hypothesis there exist $P_1, P_2 \in SP^\diamond(A)$ such that $P_1\xi P_2 \in L(f')$. Consider now the nodes $s_j$, $j < i$ such that $s_j$ has been modified during the construction of $D_{e'\circ_\xi e''}$ and is labeled by some Presburger formula, necessarily of the form $\rho' \circ_{x_{r_j}} \rho_j$ where $\rho_j$ was the label of $s_j$ in $D_{e''}$ and $r_j$ is the index of $s_j \to x_j$ in $\text{out}(s_j)$ in $D_{e''}$, with $x_j$ labeled by $\xi$. By hypothesis, $1_{k_j} \in L(\rho' \circ_{x_{r_j}} \rho_j)$ for some $k_j$ such that $s_j \to d_j$ has index $k_j$ in $\text{out}(s_j)$ in $D_e$, and, since $s_j \to d_j$ is an edge of $D_{e''}$ and as a consequence of Lemma 31, $1_{k_{j'}} \in L(\rho_j)$ for some $k_{j'}$ such that $s_j \to d_j$ has index $k_{j'}$ in $\text{out}(s_j)$ in $D_{e''}$. The sequence of edges $s_1 \to d_1, \ldots, s_i \to s$ of edges of $D_{e''}$ verifies the conditions of the lemma, and by induction hypothesis $P'_1\xi P'_2 \in L(f'')$ for some $P'_1, P'_2 \in SP^\diamond(A)$. Thus $P'_1 P_1\xi P_2 P'_2 \in L(f' \circ_\xi f'')$. Let us turn now to the converse of the lemma, and assume that $P_1P_2 \in L(f' \circ_\xi f'')$ for some $P_1, P_2 \in SP^\diamond(A)$. Necessarily, there are some $P'_1, P''_1, P'_2, P''_2 \in SP^\diamond(A)$ such that $P'_1\xi P'_2 \in L(f'')$ and $P''_1\xi P''_2 \in L(f')$, $P_1 = P'_1 P''_1$ and $P_2 = P''_2 P'_2$. By induction hypothesis, there is a sequence $s_{1,1} \to d_{1,1}, \ldots, s_{n_1,1} \to d_{n_1,1}$ (resp. $s_{1,2} \to d_{1,2}, \ldots, s_{n_2,2} \to d_{n_2,2}$) of edges of $D_{e'}$ (resp. $D_{e''}$) that verifies the statement of the lemma. Assume first that $r(D_{e'})$ and $s_{n_2,2}$ are not both labeled by Presburger formulæ. The label of $d_{n_2,2}$ was $\xi$ in $D_{e''}$ and has been changed into the label of $r(D_{e'})$ in $D_{e'\circ_\xi e''}$. The sequence $s_{1,2} \to d_{1,2}, \ldots, s_{n_2,2} \to d_{n_2,2}, d_{n_2,2} \to d_{1,1}, \ldots, s_{n_1,1} \to d_{n_1,1}$ of edges of $D_{e'\circ_\xi e''}$ verifies the statement of the lemma. Indeed, by construction the sources $d_{n_2,2}, s_{2,1}, \ldots, s_{n_1,1}$ verify the statement of the lemma. If one of the sources $s_{i,2}$, $i \in [n_2]$, has been modified during the construction, necessarily $r(D_{e'})$ is labeled by some Presburger formula $\rho'$, this source can not be $s_{n_2,2}$, and its label is a Presburger formula $\rho$ in $D_{e''}$ which has been modified into $\rho' \circ_{x_j} \rho$ for some $j$ index of some $s_{i,2} \to s$ in $\text{out}(s_{i,2})$ in $D_{e''}$, with $s$ labeled by $\xi$. Since by induction hypothesis there is some $r$ such that $r$ is an index of $s_{i,2} \to d_{i,2}$ in $\text{out}(s_{i,2})$ in $D_{e''}$ and $1_r \in L(\rho)$, and thus $r \neq j$, then $1_k \in L(\rho' \circ_{x_j} \rho)$ by Lemma 31, for some $k$ index of $s_{i,2} \to d_{i,2}$ in $\text{out}(s_{i,2})$ in $D_e$. Now turn to the case where $r(D_{e'})$ and $s_{n_2,2}$ are both labeled by Presburger formulæ, respectively $\rho'$ and $\rho$. Then the sequence $s_{1,2} \to d_{1,2}, \ldots, s_{n_2,2} \to d_{1,1}, s_{2,1} \to d_{2,1}, \ldots, s_{n_1,1} \to d_{n_1,1}$ of edges of $D_{e'\circ_\xi e''}$ verifies the statement of the lemma. We argue as above, with an additional attention to the edge $s_{n_2,2} \to d_{1,1}$. By hypothesis, there exists $j$ such that $s_{n_2,2} \to d_{n_2,2}$ has index $j$ in $\mathrm{out}(s_{n_2,2})$ in $D_{e''}$, and $1_j \in L(\rho)$. Also by hypothesis, there exists $j'$ such that $r(D_{e'}) \to d_{1,1}$ has index $j'$ in $\mathrm{out}(r(D_{e'}))$ in $D_{e'}$, and $1_{j'} \in L(\rho')$. By Lemma 31 and by construction of $D_{e' \circ_\xi e''}$, $1_{j+j'-1} \in L(\rho' \circ_{x_j} \rho)$ and $j + j' - 1$ is an index of $s_{n_2,2} \to d_{1,1}$ in $\mathrm{out}(s_{n_2,2})$ in $D_{e' \circ_\xi e''}$.

If $f$ has the form $f = f' + f''$ similar arguments are used.

If $f$ has the form $f = f'^{*\xi}$, then $e$ has the form $e = e'^{*\xi}$. By definition, $\xi \in L(f)$ and by construction, there is a sequence (of length 1) of edges of $D_e$ satisfying the statement of the lemma.

If $f$ has the form $f = f' \parallel f''$ then $e$ has the form $e = e' \parallel e''$, and there are several cases. If $\epsilon \notin L(f') \cup L(f'')$ then none of the two members of the equivalence of the lemma is true, and the lemma holds. So assume wlog. $\epsilon \in L(f'')$. By Lemma 41, $r(D_{e''})$ is labeled by some Presburger formula $\rho_2$, such that either $\rho_2$ is a closed tautology or $(0, \ldots, 0) \in L(\rho_2)$. Assume that $P_1 \xi P_2 \in L(f')$ for some $P_1, P_2 \in SP^\diamond(A)$, and wlog. that $r(D_{e'})$ is not labeled by a Presburger formula. By construction $r(D_e)$ is labeled by $\rho \equiv \rho_2 \circ_{x_2} (x_1 = x_2 = 1)$, and $1_1 \in L(\rho)$ by Lemma 31. It suffices to add the edge $r(D_e) \to r(D_{e'})$ at the beginning of the sequence of edges obtained by induction hypothesis on $D_{e'}$ to build the desired sequences of edges of $D_e$. Let us turn to the other implication of the lemma, and assume the existence of the sequence of edges in $D_e$. Assume also wlog. that $r(D_{e'})$ and $r(D_{e''})$ are both labeled by Presburger formulæ, respectively $\rho_1$ and $\rho_2$, thus the label of $r(D_e)$ is $\rho \equiv \rho_1 \circ_{x_1} \rho_2 \circ_{x_2} (x_1 = x_2 = 1)$. By Lemma 31, one of $\rho_1, \rho_2$, say wlog. $\rho_1$ is a closed tautology or has $(0, \ldots, 0)$ in its language, thus $\epsilon \in L(f')$ by Lemma 41. Using Lemma 31 again and the induction hypothesis on $D_{e''}$, we have $P_1 \xi P_2 \in L(f'')$ for some $P_1, P_2 \in SP^\diamond(A)$ thus $P_1 \xi P_2 \in L(f)$. ◀

Note that in the statement of Lemma 42, the D-graph is obtained from the $>1$-expression of a rational expression, and not from an arbitrary $>1$-expression. This is a necessary condition (consider for example the $>1$-expression $\xi \cdot^{>1} \epsilon$, whose language is empty, and its D-graph).

In the construction of a D-graph from a $>1$-expression, the only place where special edges are added is Algorithm 4. In this step, if a new special edge $e \colon n \to x$, with $x$ labeled by $\xi$, is inserted, then necessarily $x$ is a direct descendant of the root. Algorithm 1 is applied after Algorithm 4. Its first step removes $e$. As a consequence:

▶ **Lemma 43.** *The D-graph $D_e$ of some $>1$-expression $e$ has no special edge whose destination is labeled by $\xi$.*

▶ **Proposition 44.** *Let $D_f$ be the D-graph of some rational expression $f$. For any sequence $\alpha = e_1 \colon s_1 \to d_1, \ldots, e_l \colon s_l \to d_l$ of consecutive edges of $D_f$ such that $s_1 = r(D_f)$ and $e_l \in E_S(D_f)$, there exists $i \in [l]$ such that $s_i$ is labeled by some Presburger formula $\rho(x_1, \ldots, x_m)$, and for all $r \in [m]$ such that $e_i$ has index $r$ in $\mathrm{out}(s_i)$ and for all $(y_1, \ldots, y_m) \in L(\rho)$, if $y_r > 0$ then $\sum_{i \in [m]} y_i > 1$.*

**Proof.** When $f$ is a rational expression the corresponding $>1$-expression is denoted by $e$. By induction on the rational expression $f$.

Recall that Algorithm 1 is applied after each step of the induction. If the proposition is true on the D-graph $D$ before Algorithm 1 is applied, then it is still true after (D-graph $D'$). Indeed, Algorithm 1 does not add special edges; it adds only normal edges whose destinations are nodes labeled by $\xi$. Thus, if there is a sequence $\alpha$ of consecutive edges that ends with a special edge in $D'$, then this sequence exists in $D$, but the label and edging of some sources of the edges may have been modified if they are Presburger formulæ. Assume that the label of the source $s_i$ of some edge $e_i$ of $\alpha$ has been changed; this has occurred in Step 2 of the algorithm and necessarily the label of $s_i$ is a Presburger formula. By Lemma 43 there is no $d_j$, $j \in [l]$, labeled by $\xi$. Thus there is a one-to-one correspondence $f$ between the set of positions of $e_i$ in $\mathrm{out}(s_i)$ in $D$ and the set of positions of $e_i$ in $\mathrm{out}(s_i)$ in $D'$. Let $\rho(x_1, \ldots, x_k)$ be the label of $s_i$ in $D$, $\rho'(x'_1, \ldots, x'_{k'})$ its label in $D'$, $r$ some position of $e_i$ in $\mathrm{out}(s_i)$ in $D$ and $r'$ some position of $e_i$ in $\mathrm{out}(s_i)$ in $D'$ such that $f(r) = r'$. By construction of $\rho'$ from $\rho$ we have $(y_1, \ldots, y_k) \in L(\rho)$, $y_r > 0$, $\sum_{i \in [k]} y_i > 0$ if and only if $(y'_1, \ldots, y'_{k'}) \in L(\rho')$, $y'_{r'} > 0$, $\sum_{i \in [k']} y'_i > 0$.

The arguments for the proof of the proposition are very similar to those used in the proof of Lemma 42 in all cases, except $\circ_\xi$ and $*^\xi$. We focus on those two cases.

Assume $f$ has the form $f = f' \circ_\xi f''$, and that we are not in the trivial cases $f'' = \xi$ or $f' = \xi$. Thus $e$ has the form $e = e' \circ_\xi e''$. Since the construction of $D_e$ from $D_{e'}$ and $D_{e''}$ does not add new special edges, all special edges are from either $D_{e'}$ or $D_{e''}$. Assume first that $e_l \in D_{e''}$. By construction, $e_1 \ldots e_l$ are all from $D_{e''}$. The nodes $s_i$, $i \in [l]$, may have been modified by the construction (labels, edgings, or both). The modified $s_i$'s were necessarily labeled by Presburger formulæ in $D_{e''}$, and are still labeled by Presburger formulæ in $D_e$. Let $K = \{k \in [l] : s_k$ has its label in $\mathcal{P}$ in $D_e\}$. If $|K| = 0$ we have a contradiction with the induction hypothesis. So $|K| > 0$. Assume by contradiction that for all $k \in K$, $s_k$ is labeled by some Presburger formula $\rho_k(x_{1,k}, \ldots, x_{m_k,k})$, and there is some $r_k \in [m_k]$ such that $e_k$ has index $r_k$ in $\mathrm{out}(s_k)$ in $D_e$ and $1_{r_k} \in L(\rho_k)$. If for all $k \in K$ the label of $s_k$ has not been modified during the construction of $D_e$ we get a contradiction with the induction hypothesis. Let $K'$ be the subset of $K$ such that $k' \in K'$ if and only if the label of $s_{k'}$ has been modified during the construction of $D_e$. By construction, $r(D_{e'})$ is labeled by a Presburger formula $\rho_0(x_{1,0}, \ldots, x_{k_0,0})$ and for each $k' \in K'$, $s_{k'}$ was labeled by some $\rho'_{k'}(x_{1,k'}, \ldots, x_{m'_{k'},k'})$ in $D_{e''}$ and $\rho_{k'} \equiv \rho_0 \circ_{x_{i_{k'},k'}} \rho'_{k'}$ for some $i_{k'} \in [m'_{k'}]$. As $\epsilon \notin L(e')$, by Lemma 41 then $(0, \ldots, 0) \notin L(\rho_0)$. So there are $(y_{1,0}, \ldots, y_{k_0,0}) \in L(\rho_0)$ and $(y_{1,k'}, \ldots, y_{m'_{k'},k'}) \in L(\rho'_{k'})$ such that either $\sum_{i \in [k_0]} y_{i,0} = 1$, $y_{i_{k'},k'} = 1$ and $\sum_{i \in [m'_{k'}]} y_{i,k'} = 1$, or $y_{i_{k'},k'} = 0$ and $\sum_{i \in [m'_{k'}]} y_{i,k'} = 1$. In both cases, we get a contradiction with the induction hypothesis on $D_{e''}$.

Assume now that $e_l \in D_{e'}$. Then by construction there is a $k \in [l]$ such that $s_k$ belongs to $D_{e''}$ and with outgoing edges to nodes of $D_{e'}$. If, in the construction of $D_e$, $s_k$ has been modified to be a copy of $r(D_{e'})$ then it suffices to apply the induction hypothesis on $D_{e'}$ to get the proposition. Otherwise, the label of $s_k$ in $D_e$ is $\rho(x_1, \ldots, x_k) \circ_{x'_i} \rho'(x'_1, \ldots, x'_{k'})$ with $\rho(x_1, \ldots, x_k)$ the label of $r(D_{e'})$ and $\rho'(x'_1, \ldots, x'_{k'})$ the label of $s_k$ in $D_{e''}$, for some $i \in [k']$. The proposition follows from Lemma 31 and from the induction hypothesis applied on $D_{e'}$.

Let us turn to the case $f = f'^{*\xi}$, so $e$ is of the form $e = e'^{*\xi}$. By contradiction assume that the proposition is false for some consecutive sequence of edges $e_1, \ldots, e_l$ that we may assume as small as possible. We may assume that it is false before Algorithm 1 is applied. Necessarily, the only special edge of the sequence is $e_l$, otherwise the sequence would not be as small as possible. Note that in the construction of $D_{e'^{*\xi}}$ from $D_{e'}$, all the new edges are special except the new edges of $\mathrm{out}(r(D_{e'^{*\xi}}))$. Thus, if there is a new edge in $e_1, \ldots, e_{l-1}$, it is necessarily $e_1$. Let $K \subseteq [l]$ be such that $k \in K$ if and only if $s_k$ is a node of $D_{e'}$ and $s_k$ has been modified during the construction of $D_{e'^{*\xi}}$ from $D_{e'}$. If $K = \emptyset$ we get a contradiction with the induction hypothesis, either when $s_1$ is a new node or not. Thus $K \neq \emptyset$. Let $k \in K$, and assume that $s_k$ was labeled by $\xi$ in $D_{e'}$. Then the new label of $s_k$ is identical to the label of $r(D_{e'^{*\xi}})$ and $\mathrm{out}(s_k) = s_k \circ^{\mathrm{src}}_{r(D_{e'^{*\xi}})} \mathrm{out}(r(D_{e'^{*\xi}}))$, with all edges of source $s_k$ in $D_{e'^{*\xi}}$ special. As the sequence $e_1, \ldots, e_l$ is as small as possible then necessarily $k = l > 1$.

Assume first that all edges $e_1, \ldots, e_{l-1}$ exist in $D_{e'}$. Assume also that $s_l$ was labeled by $\xi$ in $D_{e'}$. For all $k \in K \setminus \{1, l\}$, $s_k$ is labeled in $D_{e'*\xi}$ by some Presburger formula $\rho_k(x_1, \ldots, x_{n_k})$ of the form $\rho_k \equiv \rho \circ_{x'_{t_k}} \rho'_k$ for some $t_k \in [n'_k]$, with $\rho'_k(x'_1, \ldots, x'_{n'_k})$ its label in $D_{e'}$, and $\rho(x''_1, \ldots, x''_{n''})$ the label of $r(D_{e'*\xi})$. Also, $\text{out}(s_k) = (s_k \circ^{\text{src}}_{r(D_{e'*\xi})} \text{out}(r(D_{e'*\xi}))) \circ_{t_k} w$ with $w = w_1, \ldots, w_{n'_k} = \text{out}(s_k)$ in $D_{e'}$. Assume that $e_k$ appears at some position $k_1 \in [n_k]$ in $\text{out}(s_k)$ in $D_{e'*\xi}$. If $1_{k_1} \in L(\rho_k)$ then necessarily, by definition of $\rho_k$ and because $(0, \ldots, 0) \notin L(\rho)$ since $\epsilon \notin L(e')$ (Lemma 41), there is some $k_2 \in [n'_k]$ such that $k_2$ is an index of $e_k$ in $\text{out}(s_k)$ in $D_{e'}$ and $1_{k_2} \in L(\rho'_k)$ as a consequence of Lemma 31. Now, consider $s_1$. As it is an element of $D_{e'}$ then necessarily it is labeled by a Presburger formula $\rho$ in $D_{e'}$, and $\rho(x_1, \ldots, x_n)$ can have been changed by the construction in two ways. If $s_1$ has its $i^{\text{th}}$ direct descendant labeled by $\xi$ in $D_{e'}$, then $\rho$ has been changed to $\rho^{*x_i}$ in $D_e$. According to Lemma 33, if $1_j \in L(\rho^{*x_i})$ for some $j \in [n] \setminus \{i\}$ then $1_j \in L(\rho)$. Otherwise, if $s_1$ has no direct descendant labeled by $\xi$ in $D_{e'}$ then $\rho$ has been changed to $\rho'(x_1, \ldots, x_{n+1}) \equiv (\rho(x_1, \ldots, x_n) \wedge x_{n+1} = 0) \vee (\wedge_{i \in [n]} x_i = 0 \wedge x_{n+1} = 1)$, and if $1_j \in L(\rho')$ for some $j \in [n]$ then $1_j \in L(\rho)$. It follows that the sequence of edges $e_1, \ldots, e_{l-1}$ in $D_{e'}$ satisfies the statement of Lemma 42, and thus there is some $P_1, P_2 \in SP^\diamond(A)$ such that $P_1 \xi P_2 \in L(e')$ which is in contradiction with the $^{*\xi}$ case of the definition of rational languages. When $s_l$ was labeled by some Presburger formula in $D_{e'}$ we argue as above.

Finally, when $e_1$ is not an edge of $D_{e'}$, then $r(D_{e'*\xi})$ is a new node labeled by $x_1 + x_2 = 1$ and $\text{out}(r(D_{e'*\xi})) = e_1, f$ for some edge $f$ to a new node labeled by $\xi$, and the result also follows from the arguments above. ◄

▶ **Lemma 45.** *Let $e$ be a >1-expression. Let $S$ be the set of all special edges of $D_{e*\xi}$ added in the construction of $D_{e*\xi}$ from $D_e$. Then for all nodes $s, d$ of $D_{e*\xi}$, if $s \to d \in S$ then $r(D_{e*\xi}) \to d$ is an edge of $D_{e*\xi}$.*

*Furthermore, for any $s \to d \in S$, denote the labels of $s$ and of $r(D_{e*\xi})$ in $D_{e*\xi}$ by respectively $\rho_s(x_1^s, \ldots, x_{n_s}^s)$ and $\rho_r(x_1^r, \ldots, x_{n_r}^r)$, and assume that $i$ is an index of $s \to d$ in $\text{out}(s)$ in $D_{e*\xi}$. If $1_i \in L(\rho_s)$ then there is some $j$ index of $r(D_{e*\xi}) \to d$ in $\text{out}(r(D_{e*\xi}))$ such that $1_j \in L(\rho_r)$.*

**Proof.** The first part of the lemma is immediate by construction of $D_{e*\xi}$. For the second part there are two cases: either $s$ was labeled by $\xi$ in $D_e$ or not. If it was, by construction $\rho_s \equiv \rho_r$ and the result holds. Otherwise, $s$ was the direct ascendant of a node $s'$ labeled by $\xi$ in $D_e$, and was labeled by some $\rho(x_1, \ldots, x_n)$ in $D_e$. Let $k$ be the index of $s \to s'$ in $\text{out}(s)$ in $D_e$. Then $\rho_s \equiv \rho_r \circ_{x_k} \rho$. Thus, if $1_i \in L(\rho_s)$, then $1_k \in L(\rho_r)$ and $1_{i-k+1} \in L(\rho_r)$, as a consequence of Lemma 31. By definition of $\text{out}(s)$ in the construction of $D_{e*\xi}$, $i - k + 1$ is an index of $r(D_{e*\xi}) \to d$ in $\text{out}(r(D_{e*\xi}))$. ◄

The proof of the following lemma uses arguments similar to those of proof of Lemma 45.

▶ **Lemma 46.** *Let $e$ and $e'$ be two >1-expressions and $f \colon s \to d$ an edge of $D_{e \circ_\xi e'}$ from a node $s$ of $D_{e'}$ to a node $d$ of $D_e$ added in the construction of $D_{e \circ_\xi e'}$ from $D_{e'}$ and $D_e$. Then $r(D_e) \to d$ is an edge of $D_e$.*

*Furthermore, the label of $s$ in $D_{e \circ_\xi e'}$ is a Presburger formula $\rho_s(x_1^s, \ldots, x_{n_s}^s)$ if and only if the label of $r(D_e)$ is a Presburger formula $\rho_r(x_1^r, \ldots, x_{n_r}^r)$. In this case, assume that $i$ is an index of $f$ in $\text{out}(s)$. If $1_i \in L(\rho_s)$ then there is some $j$ index of $r(D_e) \to d$ in $\text{out}(r(D_e))$ such that $1_j \in L(\rho_r)$.*

Now, starting the sequence of edges of Proposition 44 from any special edge, we have:

▶ **Proposition 47.** *Let $D_f$ be the D-graph of some rational expression $f$. For any sequence $\alpha = e_1 \ldots e_l$ of consecutive edges of $D_f$ with $e_1, e_l \in E_S(D_f)$, $l > 1$, there exists a node $n$* source of some $e_i$, $i \in [l]$, such that $n$ is labeled by some Presburger formula $\rho(x_1, \ldots, x_m)$, and for all $r \in [m]$ such that $e_i$ has index $r$ in $out(n)$ and for all $(y_1, \ldots, y_m) \in L(\rho)$, if $y_r > 0$ then $\sum_{i \in [m]} y_i > 1$.

**Proof.** When $f$ is a rational expression the corresponding $>1$-expression is denoted by $e$. We proceed by induction on $f$.

If $f = a$ for some $a \in A$ or $f = \epsilon$ then $e$ has the same form and $D_e$ has no edges, and the proposition holds. If $f$ is of the form $f = f'^{op}$ for some $op \in \{*, \omega, -\omega, \natural, -\natural\}$, or of the form $f = f'\ op\ f''$ for some $op \in \{\cdot, \diamond\}$ it suffices to apply the induction hypothesis.

If $f = f_1 \parallel f_2$ then $e$ has the form $e = e_1 \parallel e_2$. Observe that the sequence of edges of the statement is entirely composed of edges that all belong to $D_{e_1}$, or all belong to $D_{e_2}$. Since the destinations and sources of those edges are not modified by the construction, it suffices to use the induction hypothesis.

The case $f = f_1 + f_2$ is exactly the same as $f = f_1 \parallel f_2$.

Assume that $f$ has the form $f = f' \circ_\xi f''$. When all the edges $e_1, \ldots, e_l$ are from $D_{f'}$ it suffices to apply the induction hypothesis. When all the edges $e_1, \ldots, e_l$ are from $D_{f''}$ the argumentation is identical to the one of the proof of Proposition 44. Otherwise, there exists $i \in [l]$ such that $e_i$ goes from $D_{f''}$ to $D_{f'}$, all edges $e_1, \ldots, e_{i-1}$ are from $D_{f''}$ and all edges $e_{i+1}, \ldots, e_l$ from $D_{f'}$. Then the proposition is a direct consequence of Lemma 46 and Proposition 44.

Assume finally that $f$ has the form $f = f'^{*\xi}$. When every $e_i$, $i \in [l]$, exists in $D_{f'}$ the proposition follows from the induction hypothesis and arguments similar to those used in the proof of Proposition 44. So assume there is some $i \in [l]$ such that $e_i$ is an edge added in the construction of $D_{f'^{*\xi}}$ from $D_{f'}$. The edge $e_i$ has been added in Step 1 or 2 of the construction. If it is from Step 1, then $e_i$ is not a special edge and its source is the root of $D_{f'^{*\xi}}$, thus $i = 1$, which provides a contradiction. Thus $e_i$ comes from Step 2. The proposition follows from Lemma 45 and Proposition 44. ◀

## 5.3 Languages

In this section we define the language $L(D_e)$ of a D-graph $D_e$ of a rational expression $e$, such that $L(D_e) = L(e)$.

▶ **Definition 48.** *Let $e$ be the $>1$-expression of some rational expression. Let $n$ be a node of $D_e$ and $P \in SP^\diamond(A)$. A path from $n$ in $D_e$ and labeled by $P$ is an ordered labeled tree $T_P$ such that:*

1. *when $P$ is a singleton labeled by $a \in A$ and $n$ is a leaf labeled by $a$ then $T_P$ is a singleton labeled by $(n, a)$;*

2. *when $P$ is a sequential poset and $n$ is labeled by $\cdot^{>1}$ and edged by $out(n) = n \to n_1, n \to n_2$ then $T_P$ has the form $T_P = (m, T_{P_1}, T_{P_2})$ with $m$ labeled by $(n, 2)$ and $T_{P_i}$ is a path in $D_e$ from $n_i$ labeled by some non-empty $P_i$, $i \in [2]$, such that $P = P_1 P_2$;*

3. *when $P$ is a sequential poset and $n$ is labeled by $*^{>1}$ and edged by $out(n) = n \to n_1$ then $T_P$ has the form $T_P = (m, T_{P_1}, \ldots, T_{P_k})$ with $m$ labeled by $(n, k)$ and $T_{P_i}$ is a path in $D_e$ from $n_1$ labeled by some $P_i$, $i \in [k]$, such that $P = P_1 \ldots P_k$. There must exist $i, i' \in [k]$, $i \neq i'$, such that $P_i, P_{i'} \neq \epsilon$;*

4. *when $P$ is a sequential poset and $n$ is labeled by $\omega^{>1}$ and edged by $out(n) = n \to n_1$ then $T_P$ has the form $T_P = (m, (T_{P_i})_{i \in \omega})$ with $m$ labeled by $(n, \omega)$ and $T_{P_i}$ is a path in $D_e$ from $n_1$ labeled by some $P_i$, $i \in \omega$, such that $P = \prod_{i \in \omega} P_i$. There must exist $i, i' \in \omega$ such that $i \neq i'$ and $P_i, P_{i'} \neq \epsilon$;*

5. *the construction when $P$ is a sequential poset and $n$ is labeled by $-\omega^{>1}$ is symmetrical;*

6. *when $P$ is a sequential poset and $n$ is labeled by $\natural^{>1}$ and edged by $out(n) = n \to n_1$ then $T_P$ has the form $T_P = (m, (T_{P_i})_{i \in \alpha})$ for some $\alpha \in \mathcal{O} \setminus \{0, 1\}$ with $m$ labeled by $(n, \alpha)$ and $T_{P_i}$ is a path in $D_e$ from $n_1$ labeled by some $P_i$, $i \in \alpha$, such that $P = \prod_{i \in \alpha} P_i$. There must exist $i, i' \in \alpha$ such that $i \neq i'$ and $P_i, P_{i'} \neq \epsilon$;*

7. *the construction when $P$ is a sequential poset and $n$ is labeled by $-\natural^{>1}$ is symmetrical;*

8. *when $P$ is a sequential poset and $n$ is labeled by $\diamond^{>1}$ and edged by $out(n) = n \to n_1, n \to n_2$ then $T_P$ has the form $T_P = (m, (T_{P_j})_{j \in J \cup \hat{J}^*})$ for some $J \in \mathcal{S} \setminus \{0, 1\}$ with $m$ labeled by $(n, J \cup \hat{J}^*)$ and $T_{P_j}$ is a path in $D_e$ from $n_1$ when $j \in J$, from $n_2$ when $j \in \hat{J}^*$, and labeled by some $P_j$, $j \in J \cup \hat{J}^*$, such that $P = \prod_{j \in J \cup \hat{J}^*} P_j$. There must exist $j, j' \in J \cup \hat{J}^*$ such that $j \neq j'$ and $P_j, P_{j'} \neq \epsilon$;*

9. *when $n$ is labeled by a Presburger formula $\rho(x_1, \ldots, x_k)$ and edged by $out(n) = n \to n_1, \ldots, n \to n_k$ then $T_P$ has the form $T_P = (m, (T_{P_{i,j}})_{i \in [k], j \in [y_i]})$ with $m$ labeled by some $(n, (y_1, \ldots, y_k))$ with $(y_1, \ldots, y_k) \in L(\rho)$, and $T_{P_{i,j}}$ is a path in $D_e$ from $n_i$ labeled by some non-empty $P_{i,j}$, $i \in [k]$, $j \in [y_i]$, such that $P = \|_{i \in [k]} \|_{j \in [y_i]} P_{i,j}$. Note that since the parallel product of posets commutes, the ordering of the sequence $(T_{P_{i,j}})_{i \in [k], j \in [y_i]}$ has no consequence.*

In the last case, when $T_P$ has the form $T_P = (m, (T_{P_{i,j}})_{i \in [k], j \in [y_i]})$ and $n \to n_t \in E_S(D_e)$ for some $t \in [k]$, we say that $P_{t,j}$ is *marked* by $n \to n_t$ in $T_P$, and that $n \to n_t$ *starts* $T_{P_{t,j}}$, for all $j \in [y_t]$. Marking is an hereditary notion: every factor of $P_i$ marked by some special edge $e$ in $T_{P_i}$ is also considered marked by $e$ in $T_P$. The class of paths of $D_e$ from $n$ labeled by $P$ is denoted $\mathcal{R}_P(D_e, n)$.

Note that $\epsilon$ can be the label of a path only by Case 9 when $(0, \ldots, 0) \in L(\rho)$ or when $\rho$ is a closed tautology. Note also that Property PP ensures that all $P_{i,j}$ are non-empty in Case 9. Finally observe also that a path may have infinite height (with the usual definition of height on trees). Proposition 50 below shows that Definition 48 is well-founded.

▶ **Definition 49.** *Let $T$ and $T'$ be two paths in the D-graph of some $>1$-expression. Then $T'$ is a (resp. strict) sub-path of $T$ if $T' \subseteq T$ (resp. $T' \subsetneq T$). A strict sub-path $T'$ of $T$ is* direct *when it is not a strict sub-path of strict sub-path of $T$.*

▶ **Proposition 50.** *Definition 48 is well-founded.*

**Proof.** Consider the class of all $(T_P, P)$ formed of a path $T_P$ and its label $P$ partially ordered by the relation $(T_P, P) < (T_{P'}, P')$ if and only if $T_P$ is a strict sub-path of $T_{P'}$. We claim that $<$ is a well-ordering. Indeed, assume by contradiction that there exists an infinite decreasing sequence $\cdots < (T_{P_i}, P_i) < \cdots < (T_{P_1}, P_1) < (T_{P_0}, P_0)$. We may assume that $T_{P_{i+1}}$ is a direct sub-path of $T_{P_i}$, for all $i$. Each $T_{P_i}$ is a path from a node $n_i$. Thus, there is an edge $e_i : n_i \to n_{i+1}$ in $D_e$ for each $i$. The sequence $e_0, e_1, \ldots$ is infinite and consecutive. Since a loop of consecutive edges in a D-graph must contain a special edge, there is among the $e_i$'s a special edge $e$ that occurs infinitely often in the sequence. Let $i_0, i_1, \ldots$ be the sequence of all indexes $i_j$ such that $e_{i_j} = e$. The construction of $P_{i_j}$ from $P_{i_j+1}$ involves at least one parallel product by a non-empty labeled poset, as a consequence of Proposition 47. Since the destination of a special edge is not labeled by a Presburger formula, the construction of $P_{i_j}$ from $P_{i_j+1}$ also involves a sequential product by a non-empty sequential labeled poset. Thus necessarily $r_X(P_{i_j}) > r_X(P_{i_j+2})$. The sequence $r_X(P_{i_0}) > r_X(P_{i_2}) > \ldots$ is an infinitely decreasing sequence of X-ranks, which can not exist since the ordering of X-ranks is well-founded. ◀

▶ **Definition 51.** *Let $D$ be the D-graph of some rational expression. Let $n$ be a node of $D$. The* language $L(n)$ *of $n$ consists of all labels of paths from $n$ in $D$. The* language *of $D$ is $L(D) = L(r(D))$.*

Note that when $D$ is the D-graph of some rational expression and $n$ is a node of $D$, if $a \in L(n)$ for some $a \in A$, then necessarily $n$ is labeled by $a$, or $n$ is labeled in $\mathcal{P}$ and has a direct descendant labeled by $a$.

It is just verification to check that when they are applied during the construction of the D-graph of a rational expression, Algorithms 1 and 2 preserve the languages of the D-graphs as well as Properties PP, SS and DAG.

Property PP is used in particular in order to compute, during the construction of $D_e$ from $e$, the Presburger formulæ that will appear later in the P-MSO formula built from $D_e$. Property SS ensures that $L(n)$ do not contain parallel posets when $n$ is the destination of a special edge.

▶ **Remark 52.** As a consequence of Proposition 47, when there is a path from $r(D_e)$ labeled by some $P$, if it contains two different sub-paths labeled by $F_1$ and $F_2$ both marked by the same special edge, then $F_1$ and $F_2$ are necessarily sequential posets (Property SS), and either

$(C_1)$ $F_1 \cap F_2 = \emptyset$. Possibly, $F_1 F_2$ is a sequential factor of $P$;

$(C_2)$ one is strictly included into the other, wlog. $F_1 \subsetneq F_2$. In this case, there is some $x \in F_2 \setminus F_1$ such that $x$ is incomparable to all the elements of $F_1$.

▶ **Example 53.** Let $L_1$ and $L_2$ be the languages of respectively $e_1 = a \circ_\xi (a(\xi \parallel \xi))^{*\xi}$ (see also Example 20) and of $e_2 = b^\diamond$. Consider the D-graphs $D_e, D_{e_1}$ and $D_{e_2}$ of respectively $e = e_1 \diamond e_2$, $e_1$ and $e_2$ pictured on the left side of Figure 4. We let $\zeta$ denote the linear ordering of all integers (negative, 0 and positive) and by $\mathbf{3}$ the finite linear ordering $\{x_1, x_2, x_3\}$ such that $x_1 < x_2 < x_3$. Then $\hat{\mathbf{3}}^* = \{(\{x_1\}, \{x_2, x_3\}), (\{x_1, x_2\}, \{x_3\})\}$. Let $x_4 = (\{x_1\}, \{x_2, x_3\})$ and $x_5 = (\{x_1, x_2\}, \{x_3\})$. Then $\mathbf{3} \cup \hat{\mathbf{3}}^* = \{x_1, \ldots, x_5\}$ ordered by $x_1 < x_4 < x_2 < x_5 < x_3$. The same reasoning can be applied to infer the infinite set of elements and the ordering of $\zeta \cup \hat{\zeta}^*$. On the right side of the figure is pictured a path $T$ of $D_e$ from $r(D_e)$ labeled by $P = \prod_{j \in \mathbf{3} \cup \hat{\mathbf{3}}^*} F_j$, where $F_{x_1} = a(a(a \parallel a) \parallel a), F_{x_4} = \epsilon, F_{x_2} = a, F_{x_5} = b^\zeta$ and $F_{x_3} = a(a \parallel a)$. Note that $F_{x_1}, F_{x_2}, F_{x_3} \in L(a \circ_\xi (a(\xi \parallel \xi))^{*\xi})$ and $F_{x_4}, F_{x_5} \in L(b^\diamond)$. Note also that $F_{x_5} = \prod_{j \in \zeta \cup \hat{\zeta}^*} F_{5,j}$ where $F_{5,j} = b$ when $j \in \zeta$ and $F_{5,j} = \epsilon$ when $j \in \hat{\zeta}^*$. Observe that the path $T$ has a unique direct sub-path $T'$ also labeled by $P$. Let $T'_{x_1}, T'_{x_4}, T'_{x_2}, T'_{x_5}, T'_{x_3}$ be the direct sub-paths of $T'$ taken in order from the left to the right. Then observe that each $T'_{x_i}$ is labeled by $F_{x_i}$.

▶ **Example 54.** Let $e = ((a \parallel \xi^*)^* + b(c \parallel d))^{*\xi}$ be a rational expression. The D-graph $D_e$ of the >1-expression of $e$ and the poset $P = a \parallel (\xi(a \parallel (b(c \parallel d)b(c \parallel d))))$ of $L(D_e)$ are represented on the left side of Figure 5. Let $n_1, \ldots, n_{11}$ be the pre-order traversal of $D_e$ without its special edges. We have $E_S(D_e) = \{n_4 \to n_2, n_4 \to n_3, n_4 \to n_7\}$. The leaves of $D_e$ are $n_2, n_5, n_6, n_8, n_{10}, n_{11}$. Note that in $\text{out}(n_4) = \{n_4 \to n_2, n_4 \to n_3, n_4 \to n_2, n_4 \to n_3, n_4 \to n_5, n_4 \to n_7\}$ each of $n_4 \to n_2$ and $n_4 \to n_3$ occurs twice. On the right side of Figure 5 is pictured a path $T_P$ in $D_e$ from $n_1$ labeled by $P$. Its root is labeled by $(n_1, (1, 1, 0, 0))$ with $(1, 1, 0, 0) \in L(\rho_1)$. The path $T_P$ has two direct sub-paths $T_{P_1} = (n_2, a)$ and $T_{P_2} = ((n_3, 2), T_{P_{2,1}}, T_{P_{2,2}})$ labeled respectively by $P_1 = a$ and $P_2 = P_{2,1} P_{2,2}$, with $P = P_1 \parallel P_2$, $T_{P_{2,1}}$ and $T_{P_{2,2}}$ direct sub-paths of $T_{P_2}$ from $n_4$ and respectively labeled by $P_{2,1} = \xi$ and $P_{2,2} = a \parallel (b(c \parallel d)b(c \parallel d))$. In continuation, $T_{P_{2,1}}$ and $T_{P_{2,2}}$ have respectively the form $((n_4, (0, \ldots, 0, 1, 0)), (n_5, \xi))$ and $((n_4, (1, 1, 0, \ldots, 0)), T_{F_4}, T_{F_3})$ with $(0, \ldots, 0, 1, 0), (1, 1, 0, \ldots, 0) \in L(\rho_2)$ and $(n_5, \xi)$, $T_{F_4}$ and $T_{F_3}$ are from respectively $n_5$, $n_2$ and $n_3$ labeled by respectively $\xi$, $F_4 = a$ and $F_3 = b(c \parallel d)b(c \parallel d)$. Note that there is a path from $n_4$, different from $T_{P_{2,2}}$ and with the same label. It differs only by its root which is labeled by $(n_4, (0, 0, 1, 1, 0, 0))$. Observe that $T_{F_3}$ and $T_{F_4}$ are started by respectively $n_4 \to n_3$ and $n_4 \to n_2$. Similarly, $T_{F_1}$ and $T_{F_2}$ are sub-paths of $T_{F_3}$ from $n_7$ labeled by $F_1 = F_2 = b(c \parallel d)$ and started by $n_4 \to n_7$. Each $F_i$ is marked by the edge that starts $T_{F_i}$, $i \in [4]$. As a consequence of Remark 52, the edge marking $F_3$ is necessarily different from those which mark $F_1$ and $F_2$.

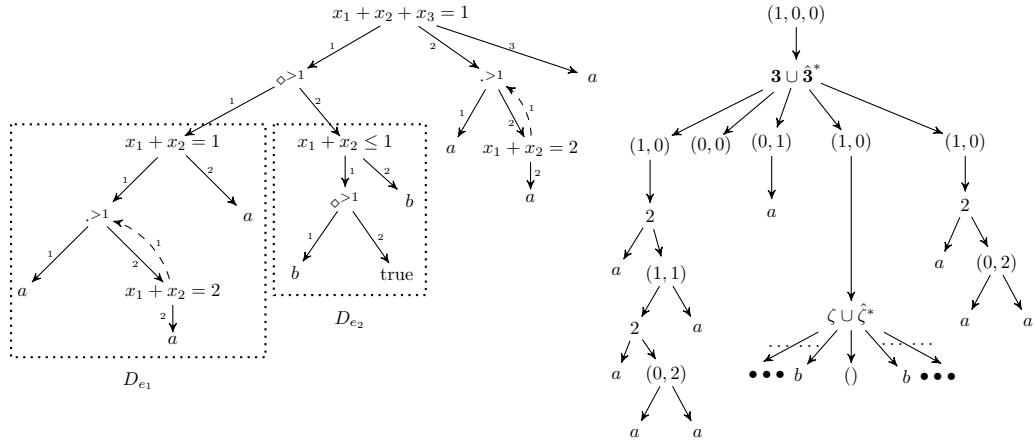

**Figure 4** The D-graph $D_e$ of $e = a \circ_\xi (a(\xi \parallel \xi))^{*\xi} \diamond b^\diamond$ and a path in $D_e$ from $r(D_e)$. In order to lighten figures, nodes of paths are labeled by $s$ instead of $(n, s)$ where $n$ is a node of $D_e$

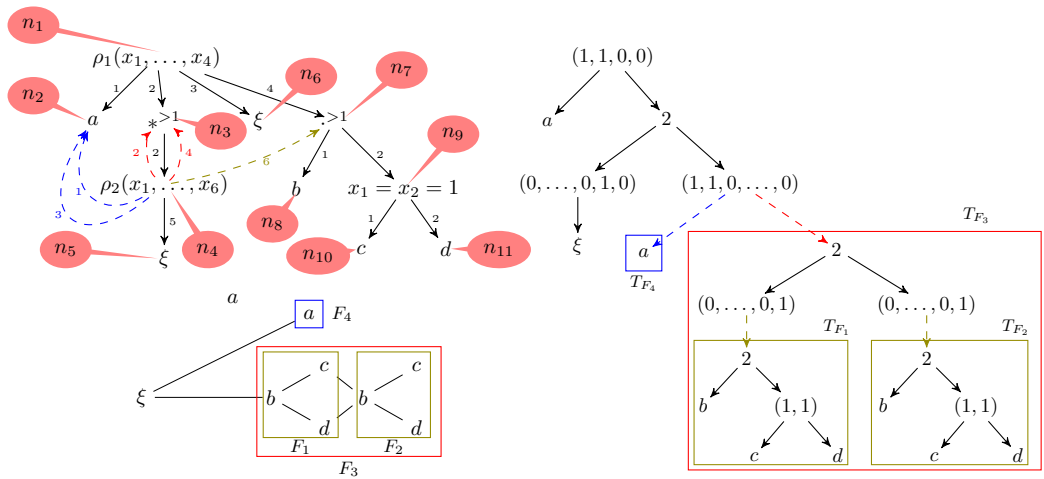

**Figure 5** The D-graph $D_e$ of the >1-expression of $e = ((a \parallel \xi^*)^{*\xi} + (b(c \parallel d)))^{*\xi}$, a poset $P = a \parallel (\xi(a \parallel (b(c \parallel d)b(c \parallel d))))$ of $L(e)$ and a path in $D_e$ from $r(D_e)$ labeled by $P$. Here $\rho_1(x_1, \ldots, x_4) \equiv (x_1 = x_2 = 0 \land x_3 + x_4 = 1) \lor (x_1 \geq 1 \land x_2 + x_3 + x_4 \leq 1)$ and $\rho_2(x_1, \ldots, x_6) \equiv (x_1 \geq 1 \land x_2 \leq 1 \land_{3 \leq i \leq 6} x_i = 0) \lor (x_1 \geq 0 \land x_2 = 0 \land ((x_3 = x_4 = 0 \land x_5 + x_6 = 1) \lor (x_3 \geq 1 \land x_4 + x_5 + x_6 \leq 1)))$

### 5.4 Correctness of the construction of a D-graph from a rational expression

This section is devoted to the proof of the following proposition.

▶ **Proposition 55.** *Let $e$ be the $>1$-expression of some rational expression. Then $L(D_e) = L(e)$.*

Note that the nodes of $D_e$ may be the nodes of other D-graphs obtained by induction hypothesis, with potential modifications on their labelings or edgings. When $n$ is such a node, the notation $L(n)$ may be confusing and we use instead $L(G, n)$ ($L(n)$ in D-graph $G$) in order to avoid ambiguities.

#### 5.4.1 Case $e = e' \circ_\xi e''$

In the following we prove the correctness of the construction of $D_{e' \circ_\xi e''}$ for some rational expressions $e'$ and $e''$. Recall that $D_{e' \circ_\xi e''}$ is built from the union of $D_{e'}$ and $D_{e''}$, which are obtained by induction hypothesis, with several modifications detailed in Algorithm 3. Always by induction hypothesis, we know that $L(D_{e'}) = L(e')$ and $L(D_{e''}) = L(e'')$. Thus, since the rationality of $e' \circ_\xi e''$ implies that $\epsilon \notin L(D_{e'})$ then as a consequence of Lemma 41, when $r(D_{e'})$ is labeled by some $\rho(y_1, \ldots, y_m) \in \mathcal{P}_m$ for some $m \in \mathbb{N}$ neither $\rho$ is a closed tautology nor $(0, \ldots, 0) \in L(\rho)$.

▶ **Lemma 56.** *Let $D_{e'}$ and $D_{e''}$ be the D-graphs of some rational expressions $e'$ and $e''$ and let $e = e' \circ_\xi e''$. Then for any common node $n$ of $D_{e''}$ and $D_e$, $L(r(D_{e'})) \circ_\xi L(D_{e''}, n) = L(D_e, n)$.*

**Proof.** We start by proving that $L(r(D_{e'})) \circ_\xi L(D_{e''}, n) \subseteq L(D_e, n)$. Let $P \in L(r(D_{e'})) \circ_\xi L(D_{e''}, n)$. By definition of substitution operation, there exists $P' \in L(D_{e''}, n)$ such that $P \in L(r(D_{e'})) \circ_\xi P'$. We proceed by induction on paths to prove that $P \in L(D_e, n)$. Let $\gamma, \gamma_1, \gamma_2$ be the maps labeling respectively $D_e$, $D_{e'}$, $D_{e''}$. Let us begin by the cases where $\text{out}(n)$ is empty in $D_{e''}$.

Assume that $\gamma_2(n) = \xi$. By construction, $\gamma(n) = \gamma_1(r(D_{e'}))$ and $\text{out}(n) = n \circ_{r(D_{e'})}^{\text{src}} \text{out}(r(D_{e'}))$ in $D_e$. In this case, $\mathcal{R}_P(D_e, n)$ is isomorphic to $\mathcal{R}_P(D_{e'}, r(D_{e'}))$ and $L(D_e, n) = L(r(D_{e'}))$. As $L(D_{e''}, n) = \{\xi\}$, by definition of substitution operation, $P \in L(D_e, n)$.

Otherwise, if $\gamma_2(n) = a \neq \xi \in A$ then, by construction, $\gamma(n) = \gamma_2(n)$. We have $L(D_e, n) = L(r(D_{e'})) \circ_\xi L(D_{e''}, n) = \{a\}$.

Assume now that $\text{out}(n) = e_{2,1} : n \to n_1, \ldots, e_{2,k} : n \to n_k$ in $D_{e''}$, for some $k \in \mathbb{N}$. Let $T \in \mathcal{R}_{P'}(D_{e''}, n)$. If $\gamma_2(n) \in \{\cdot^{>1}, *^{>1}, \diamond^{>1}, \omega^{>1}, -\omega^{>1}, \natural^{>1}, -\natural^{>1}\}$ then $P'$ admits a $J$-factorization of the form $\prod_{j \in J} P'_j$ for some $J \in \mathcal{S} \setminus \{0, 1\}$ such that there exist at least $j, j' \in J$ with $j \neq j'$, such that $P'_j, P'_{j'} \neq \epsilon$ and for all $j \in J$, $P'_j \in L(D_{e''}, n_t)$ is the label of some direct sub-path of $T$ in $D_{e''}$ from $n_t$, for some $t \in [k]$. Therefore by definition of substitution operation, $P = \prod_{j \in J} P_j$ where each $P_j \in L(r(D_{e'})) \circ_\xi P'_j$. In this case, it suffices to apply the induction hypothesis to prove that $P \in L(D_e, n)$.

Finally, assume that $\gamma_2(n) = \rho_2(x_1, \ldots, x_k)$ for some $\rho_2(x_1, \ldots, x_k) \in \mathcal{P}_k$. In this case, the root of $T$ is labeled by some $(n, (x_1, \ldots, x_k))$ with $(x_1, \ldots, x_k) \in L(\rho_2)$. In addition, by construction $\gamma(n) \in \mathcal{P}$. Assume that $\gamma(n) = \rho(z_1, \ldots, z_l)$ for some $l \in \mathbb{N}$ and $\rho(z_1, \ldots, z_l) \in \mathcal{P}_l$. Assume also that $\text{out}(r(D_{e'})) = e_{1,1} : r(D_{e'}) \to r_1, \ldots, e_{1,m} : r(D_{e'}) \to r_m$ for some $m \in \mathbb{N}$. There are two cases.

1. If there exists $i \in [k]$ such that $\gamma_2(n_i) = \xi$ and $\gamma_1(r(D_{e'})) \in \mathcal{P}_m$ then $l = k + m - 1$.
   Note that $i$ is unique since $D_{e''}$ is $\xi$-normalized by hypothesis. Assume that $\gamma_1(r(D_{e'})) =$

$\rho_1(y_1, \ldots, y_m)$ for some $\rho_1(y_1, \ldots, y_m) \in \mathcal{P}_m$. Then $\rho \equiv \rho_1 \circ_{x_i} \rho_2$ and $\mathrm{out}(n) = e_1, \ldots, e_l$ in $D_e$ where by construction, $e_1, \ldots, e_{i-1}, e_{i+m}, \ldots, e_l = e_{2,1}, \ldots, e_{2,i-1}, e_{2,i+1}, \ldots, e_{2,k}$ and $e_i, \ldots, e_{i+m-1} = n \circ^{src}_{r(D_{e'})} \mathrm{out}(r(D_{e'}))$. There are three cases.

**a.** If $(x_1, \ldots, x_k) = (0, \ldots, 0)$ then $T$ is a leaf and $P' = P = \epsilon$. By Lemma 31, $(0, \ldots, 0) \in L(\rho)$. Thus $P \in L(D_e, n)$;

**b.** otherwise, if there exists $t \in [k]$ such that $(x_1, \ldots, x_k) = 1_t$ then there are two cases according to $t$: if $t \neq i$ then as a consequence of Lemma 31 and the construction, there exists $t' \in [l]$ such that $1_{t'} \in L(\rho)$ and $e_{t'} = e_{2,t}$. We deduce that $L(D_e, n_{t'}) \subseteq L(D_e, n)$. Furthermore, $T = (d, T_{P'})$ for some $d$ labeled by $(n, 1_t)$ and $T_{P'} \in \mathcal{R}_{P'}(D_{e''}, n_t)$. Thus, as $T_{P'}$ is a sub-path of $T$ from a common node of $D_e$ and $D_{e''}$, then by induction hypothesis $L(r(D_{e'})) \circ_\xi P' \subseteq L(D_e, n_t)$. We have $P \in L(D_e, n)$. Otherwise, $P' = \xi$ and by definition of substitution operation $P \in L(r(D_{e'}))$. Then $P$ is the label of a tree $T_P \in \mathcal{R}_P(D_{e'}, r(D_{e'}))$ whose root is labeled by some $(r(D_{e'}), (y_1, \ldots, y_m))$ with $(y_1, \ldots, y_m) \in L(\rho_1)$. By Lemma 31, there exists $(x_1, \ldots, x_{i-1}, y_1, \ldots, y_m, x_{i+1}, \ldots, x_k) \in L(\rho)$. In addition recall that by construction $e_i, \ldots, e_{i+m-1} = n \to r_1, \ldots, n \to r_m$. We have $P \in L(D_e, n)$;

**c.** otherwise, $P' = P'_1 \parallel P'_2$ where $P'_1 = \parallel_{j \in [k]} \parallel_{s \in [x_j]} P'_{1,j,s}$, $P'_{1,j,s} \in L(D_{e''}, n_j)$ and $_{j \neq i}$
$P'_2 = \xi^{\parallel x_i}$. Thus, by definition of substitution operation, $P = P_1 \parallel P_2$ where $P_1 = \parallel_{j \in [k]} \parallel_{s \in [x_j]} P_{1,j,s}$, each $P_{1,j,s} \in L(r(D_{e'})) \circ_\xi P'_{1,j,s}$, $P_2 = \parallel_{s \in [x_i]} P_{2,s}$ and for $_{j \neq i}$
all $s \in [x_i]$, $P_{2,s} \in L(r(D_{e'}))$ is the label of some $T_{2,s} \in \mathcal{R}_{P_{2,s}}(D_{e'}, r(D_{e'}))$ of root labeled by some $(r(D_{e'}), (y_{s,1}, \ldots, y_{s,m}))$ for some $(y_{s,1}, \ldots, y_{s,m}) \in L(\rho_1)$. Furthermore each $P'_{1,j,s}$ is the label of some direct sub-path of $T$ in $D_{e''}$ from some common node of $D_{e''}$ and $D_e$. Then by induction hypothesis, each $P_{1,j,s} \in L(D_e, n_j)$. In addition, as a consequence of Lemma 31, there exists $(z_1, \ldots, z_l) \in L(\rho)$ such that $(z_1, \ldots, z_{i-1}, z_{i+m}, \ldots, z_l) = (x_1, \ldots, x_{i-1}, x_{i+1}, \ldots, x_k)$ and $(z_i, \ldots, z_{i+m-1}) = (y_{1,1}, \ldots, y_{1,m}) + \cdots + (y_{x_i,1}, \ldots, y_{x_i,m})$. We have $P_1 \parallel P_2 \in L(D_e, n)$;

**2.** otherwise, by construction, $\gamma(n) = \gamma_2(n)$ and the edging of $n$ in $D_{e''}$ is the same in $D_e$. There are three cases.

**a.** If $(x_1, \ldots, x_k) = (0, \ldots, 0)$ or $\rho$ is a closed tautology then $T$ is a leaf and $P = P' = \epsilon$. As $L(\rho_2) = L(\rho)$ then $\epsilon \in L(D_e, n)$;

**b.** otherwise, if there exists $t \in [k]$ such that $1_t \in L(\rho)$ then $T = (d, T_{P'})$ for some $d$ labeled by $(n, 1_t)$ and $T_{P'} \in \mathcal{R}_{P'}(D_{e''}, n_t)$. In addition, as $T_{P'}$ is a sub-path of $T$ from a common node of $D_e$ and $D_{e''}$, then by induction hypothesis, whether $\gamma_2(n_t) = \xi$ or not, $L(r(D_{e'})) \circ_\xi L(D_{e''}, n_t) \subseteq L(D_e, n_t)$. We have $P \in L(D_e, n)$;

**c.** otherwise, $P' = \parallel_{j \in [k]} \parallel_{s \in [x_j]} P'_{j,s}$ where each $P'_{j,s} \in L(D_{e''}, n_j)$. Then by definition of substitution operation, $P = \parallel_{j \in [k]} \parallel_{s \in [x_j]} P_{j,s}$ where each $P_{j,s} \in L(r(D_{e'})) \circ_\xi P'_{j,s}$. The lemma follows from the induction hypothesis.

The proof that $L(D_e, n) \subseteq L(r(D_{e'})) \circ_\xi L(D_{e''}, n)$ uses similar arguments. ◄

The construction in Algorithm 3 states that $r(D_{e' \circ_\xi e''}) = r(D_{e''})$ for some rational expressions $e'$ and $e''$. Then thanks to Lemma 56, we have:

▶ **Corollary 57.** *Let $D_{e'}$ and $D_{e''}$ be the D-graphs of some rational expressions $e'$ and $e''$. Then $L(r(D_{e' \circ_\xi e''})) = L(r(D_{e'})) \circ_\xi L(r(D_{e''}))$.*

### 5.4.2 Case $e = e'^{*\xi}$

Let us turn to prove the correctness of the construction of $D_{e'^{*\xi}}$ for some rational expression $e'$. The D-graph $D_{e'^{*\xi}}$ is built from $D_{e'}$ which is obtained by induction hypothesis. Always by induction hypothesis, we know that $L(D_{e'}) = L(e')$. Thus, since the rationality of $e'^{*\xi}$ implies that $e' \neq \xi$ and $\epsilon \notin L(e')$ then $D_{e'}$ is neither a single node labeled by $\xi$ nor a D-graph rooted by a node labeled by some $\rho(y_1, \ldots, y_m) \in \mathcal{P}_m$ for some $m \in \mathbb{N}$ such that $\rho$ is a closed tautology or $(0, \ldots, 0) \in L(\rho)$ (Lemma 41). Furthermore, when $e' = a \in A \setminus \{\xi\}$, $D_{e'}$ is a single node labeled by $a$ without edges, and by construction, $D_{e'^{*\xi}}$ is a D-graph containing three nodes. A node $n_1$ labeled by $a$, a node $n_2$ labeled by $\xi$ and a node labeled by $x_1 + x_2 = 1$ which is the root of $D_{e'^{*\xi}}$ edged by $r(D_{e'^{*\xi}}) \to n_1, r(D_{e'^{*\xi}}) \to n_2$. In this case, observe that $L(e'^{*\xi}) = L(r(D_{e'}))^{*\xi} = L(D_{e'^{*\xi}}) = \{a, \xi\}$. So, in the followings lemmas, this case is not treated.

Recall that the first step of Algorithm 4 ensures in particular that the root of $D_{e'^{*\xi}}$ is labeled by some Presburger formula $\rho$ with a $i^{\text{th}}$ direct descendant labeled by $\xi$ for exactly one $i$, and that $1_i$ is in the Presburger set of $\rho$. We have:

▶ **Lemma 58.** *Let $D_e$ be the D-graph of some rational expression $e$. Then $\xi \in L(r(D_{e^{*\xi}}))$.*

**Proof.** Let $\gamma$ and $\gamma'$ be the maps labeling respectively $D_e$ and $D_{e^{*\xi}}$. Assume that $\text{out}(r(D_e)) = r(D_e) \to r_1, \ldots, r(D_e) \to r_k$ and $\text{out}(r(D_{e^{*\xi}})) = r(D_{e^{*\xi}}) \to r'_1, \ldots, r(D_{e^{*\xi}}) \to r'_m$ for some $k, m \in \mathbb{N} \setminus \{0\}$. There are two cases according to $\gamma(r(D_e))$.

1. If $\gamma(r(D_e)) \notin \mathcal{P}$ then by construction, $m = 2$, $\gamma'(r(D_{e^{*\xi}})) = \rho'(x_1, x_2) \equiv x_1 + x_2 = 1$ and $\gamma'(r'_2) = \xi$. Observe that $1_2 \in L(\rho')$ and $\mathcal{R}_\xi(D_{e^{*\xi}}, r'_2) \neq \emptyset$. Thus $\xi \in L(r(D_{e^{*\xi}}))$;

2. otherwise, assume that $\gamma(r(D_e)) = \rho(x_1, \ldots, x_k)$ then, by construction, $r(D_e) = r(D_{e^{*\xi}})$. There are two cases according to $\gamma(r_j)$, for all $j \in [k]$.

   a. If there exists $i \in [k]$ such that $\gamma(r_i) = \xi$ then $i$ is unique since $D_e$ is $\xi$-normalized by hypothesis and, by construction, $m = k$, $\gamma'(r(D_{e^{*\xi}})) = \rho'(x_1, \ldots, x_k) \equiv \rho^{*x_i}$ and $\gamma(r_i) = \gamma'(r'_i)$. Since $L(\rho^{0x_i}) \subseteq L(\rho')$, $1_i \in L(\rho')$. In addition, observe that $\mathcal{R}_\xi(D_{e^{*\xi}}, r'_i) \neq \emptyset$. Thus $\xi \in L(r(D_{e^{*\xi}}))$;

   b. otherwise, by construction, $m = k+1$, $\gamma'(r(D_{e^{*\xi}})) = \rho'(x_1, \ldots, x_m) \equiv (\rho(x_1, \ldots, x_k) \wedge x_m = 0) \vee (\wedge_{j \in [k]} x_j = 0 \wedge x_m = 1)$ and $\gamma'(r'_m) = \xi$. Observe that $1_m \in L(\rho')$ and $\mathcal{R}_\xi(D_{e^{*\xi}}, r'_m) \neq \emptyset$. Thus $\xi \in L(r(D_{e^{*\xi}}))$.

◀

▶ **Lemma 59.** *Let $D_e$ be the D-graph of some rational expression $e$ and let $d$ be the direct descendant of $r(D_{e^{*\xi}})$ labeled by $\xi$. For any common node $n$ of $D_e$ and $D_{e^{*\xi}}$, if the following conditions are true:*

- $n \neq r(D_{e^{*\xi}})$;
- $n \neq d$ *when $r(D_e)$ is labeled in $\mathcal{P}$;*

*then $L(r(D_{e^{*\xi}})) \circ_\xi L(D_e, n) \subseteq L(D_{e^{*\xi}}, n)$.*

**Proof.** As the second step of Algorithm 4 is similar to Algorithm 3, the proof uses arguments similar to those of the proof of Lemma 56. In fact, the arguments used to prove the lemma when $n$ is labeled by $a \in A \setminus \{\xi\}$ or in $\{.^{>1}, *^{>1}, \diamond^{>1}, \omega^{>1}, -\omega^{>1}, \natural^{>1}, -\natural^{>1}\}$ are the same than those used in the proof of Lemma 56 when $n$ is labeled similarly. In addition, when $n$ is labeled by $\xi$ in $D_e$, then $n$ is labeled in $D_{e^{*\xi}}$ exactly as $r(D_{e^{*\xi}})$ is, and $\text{out}(n) = n \to y \circ_{n \to d} (n \circ^{\text{src}}_{r(D_{e^{*\xi}})} \text{out}(r(D_{e^{*\xi}})))$ in $D_{e^{*\xi}}$ with $y$ a new node labeled by $\xi$ too. Therefore, $\mathcal{R}_P(D_{e^{*\xi}}, n)$ is isomorphic to $\mathcal{R}_P(D_{e^{*\xi}}, r(D_{e^{*\xi}}))$ for all $P \in SP^\diamond(A)$ and $L(D_{e^{*\xi}}, n) = L(r(D_{e^{*\xi}}))$. As $L(D_e, n) = \{\xi\}$, the lemma holds by definition of substitution operation. Finally, the case when $n$ is labeled in $\mathcal{P}$ uses also arguments similar to those used in the proof of Lemma 56. Indeed, since $r(D_{e^{*\xi}})$ is always labeled in $\mathcal{P}$ by construction, the arguments used in the case where $n$ has a direct descendant labeled by $\xi$ (different from $d$ thanks to $\xi$-normalization) are similar to those used in Item 1 of the proof of Lemma 56. The opposite case uses similar arguments to those used in Item 2 of the proof of Lemma 56. ◄

▶ **Lemma 60.** *Let $D_e$ be the D-graph of some rational expression $e$ where $r(D_e)$ is labeled by some Presburger formula $\rho(x_1, \dots, x_k)$, $out(r(D_e)) = r(D_e) \rightarrow r_1, \dots, r(D_e) \rightarrow r_k$ for some $k \in \mathbb{N} \setminus \{0\}$, and $r_i$ is labeled by $\xi$, for some $i \in [k]$, and let $(x'_1, \dots, x'_k) \in L(\rho^{*x_i})$. Then*

$$\left( L(r(D_e))^{*\xi} \circ_\xi L(D_e, r_1)^{\|^{x'_1}} \| \cdots \| \xi^{\|^{x'_i}} \| \cdots \| L(r(D_e))^{*\xi} \circ_\xi L(D_e, r_k)^{\|^{x'_k}} \right) \subseteq L(r(D_e))^{*\xi}$$

**Proof.** Let $L = L(r(D_e))^{*\xi} \circ_\xi L(D_e, r_1)^{\|^{x'_1}} \| \cdots \| \xi^{\|^{x'_i}} \| \cdots \| L(r(D_e))^{*\xi} \circ_\xi L(D_e, r_k)^{\|^{x'_k}}$. By Lemma 29 there exists some $t \in \mathbb{N}$, that we may assume as small as possible, such that $(x'_1, \dots, x'_k) \in L(\rho^{tx_i})$. We proceed by induction on $t$.

If $t = 0$ then $(x'_1, \dots, x'_k) = 1_i$. Then $L = \{\xi\}$ and by definition of iterated substitution $L \subseteq L(r(D_e))^{0\xi}$.

If $t = 1$ then $(x'_1, \dots, x'_k) \in L(\rho)$. Let $P \in L$. By definition of iterated substitution, there exists $j \in \mathbb{N}$ such that $P$ is in

$$\left( \bigcup_{j' \leq j} L(r(D_e))^{j'\xi} \circ_\xi L(D_e, r_1)^{\|^{x'_1}} \right) \| \cdots \| \xi^{\|^{x'_i}} \| \cdots \| \left( \bigcup_{j' \leq j} L(r(D_e))^{j'\xi} \circ_\xi L(D_e, r_k)^{\|^{x'_k}} \right)$$

In addition, as $\xi \in \bigcup_{j' \leq j} L(r(D_e))^{j'\xi}$, then

$$P \in \left( \bigcup_{j' \leq j} L(r(D_e))^{j'\xi} \right) \circ_\xi \left( L(D_e, r_1)^{\|^{x'_1}} \| \cdots \| \xi^{\|^{x'_i}} \| \cdots \| L(D_e, r_k)^{\|^{x'_k}} \right)$$

We know that $L(r(D_e)) = \bigcup_{(x_1, \dots, x_k) \in L(\rho)} L(D_e, r_1)^{\|^{x_1}} \| \cdots \| \xi^{\|^{x_i}} \| \cdots \| L(D_e, r_k)^{\|^{x_k}}$. Thus, by definition of iterated substitution, $P \in L(r(D_e))^{(j+1)\xi}$.

If $t > 1$ then $(x'_1, \dots, x'_k) \in L((\bigvee_{t' < t} \rho^{t'x_i}) \bullet_{x_i} \rho)$ by Lemma 29. In this case by Lemma 32 there exist $(x_1, \dots, x_k) \in L(\rho)$ and $(x_{1,1}, \dots, x_{k,1}), \dots, (x_{1,x_i}, \dots, x_{k,x_i}) \in L(\bigvee_{t' < t} \rho^{t'x_i})$ such that for all $l \in [k]$, $l \neq i$, $x'_l = x_l + \sum_{s \in [x_i]} x_{l,s}$ and $x'_i = \sum_{s \in [x_i]} x_{i,s}$. Therefore,

$$L = \left( L(r(D_e))^{*\xi} \circ_\xi L(D_e, r_1)^{\|^{x_1}} \right) \| \cdots \| L_1 \| \cdots \| L_{x_i} \| \cdots \| \left( L(r(D_e))^{*\xi} \circ_\xi L(D_e, r_k)^{\|^{x_k}} \right)$$

where for all $s \in [x_i]$,

$$L_s = \left( L(r(D_e))^{*\xi} \circ_\xi L(D_e, r_1)^{\|^{x_{1,s}}} \right) \| \cdots \| \xi^{\|^{x_{i,s}}} \| \cdots \| \left( L(r(D_e))^{*\xi} \circ_\xi L(D_e, r_k)^{\|^{x_{k,s}}} \right)$$

By induction hypothesis, for all $s \in [x_i]$, $L_s \subseteq L(r(D_e))^{*\xi}$. Then

$$L \subseteq L(r(D_e))^{*\xi} \circ_\xi \left( L(D_e, r_1)^{\|^{x_1}} \| \cdots \| \xi^{\|^{x_i}} \| \cdots \| L(D_e, r_k)^{\|^{x_k}} \right)$$

Let $P \in L$. Then there exists $j \in \mathbb{N}$ such that

$$P \in \left( \bigcup_{j' \le j} L(r(D_e))^{j'\xi} \right) \circ_\xi \left( L(D_e, r_1)^{\|x_1} \| \cdots \| \xi^{\|x_i} \| \cdots \| L(D_e, r_k)^{\|x_k} \right)$$

Observe that $L(D_e, r_1)^{\|x_1} \| \cdots \| \xi^{\|x_i} \| \cdots \| L(D_e, r_k)^{\|x_k} \subseteq L(r(D_e))$. Then $P \in L(r(D_e))^{(j+1)\xi}$. ◀

▶ **Lemma 61.** *Let $D_e$ be the D-graph of some rational expression $e$ and let $d$ be the direct descendant of $r(D_{e^*\xi})$ labeled by $\xi$. For any common node $n$ of $D_e$ and $D_{e^*\xi}$, if the following conditions are true:*

- $n \ne r(D_{e^*\xi})$;
- $n \ne d$ when $r(D_e)$ is labeled in $\mathcal{P}$;

*then $L(D_{e^*\xi}, n) \subseteq L(r(D_e))^{*\xi} \circ_\xi L(D_e, n)$.*

**Proof.** Let $P \in L(D_{e^*\xi}, n)$. We prove by induction on paths that $P \in L(r(D_e))^{*\xi} \circ_\xi L(D_e, n)$. Let $\gamma, \gamma'$ be the maps labeling respectively $D_e$, $D_{e^*\xi}$. We prove only the case where $\gamma(n) = \xi$. The arguments to prove the other cases are similar to those of Lemma 56.

Assume that $\gamma(n) = \xi$ then $L(D_e, n) = \{\xi\}$. Thus by definition of substitution operation, we have to prove that $P \in L(r(D_e))^{*\xi}$ or equivalently that there exists $t \in \mathbb{N}$ such that $P \in L(r(D_e))^{t\xi}$. By Algorithm 4 $\gamma'(r(D_{e^*\xi})) \in \mathcal{P}$, $\gamma'(n) = \gamma'(r(D_{e^*\xi}))$ and $out(n) = n \circ_{r(D_{e^*\xi})}^{src} out(r(D_{e^*\xi}))$ in $D_{e^*\xi}$. As $d$ is a direct descendant of $r(D_{e^*\xi})$, then Algorithm 4 produces a non $\xi$-normalized D-graph. Thus Algorithm 1 transforms $out(n)$ to $n \to y \circ_{n \to d} (n \circ_{r(D_{e^*\xi})}^{src} out(r(D_{e^*\xi})))$ where $y$ is a new node labeled by $\xi$.

Assume that $out(n) = n \to n_1, \ldots, n \to n_k$ in $D_{e^*\xi}$ and $out(r(D_e)) = r(D_e) \to r_1, \ldots, r(D_e) \to r_m$ for some $m, k \in \mathbb{N} \setminus \{0\}$. Assume also that $\gamma'(n) = \rho'(x_1, \ldots, x_k)$. Observe that $\mathcal{R}_P(D_{e^*\xi}, n) \ne \emptyset$ and for any $T \in \mathcal{R}_P(D_{e^*\xi}, n)$, the root of $T$ is labeled by some $(n, (x'_1, \ldots, x'_k))$ with $(x'_1, \ldots, x'_k) \in L(\rho')$. There are two cases according to $\gamma(r(D_e))$:

1. If $\gamma(r(D_e)) \notin \mathcal{P}$ then by construction, $k = 2$, $\gamma'(n) = \rho'(x_1, x_2) \equiv x_1 + x_2 = 1$, $n_1 = r(D_e)$, $n_2 = y$, $\gamma'(n_1) = \gamma(r(D_e))$ and $\gamma'(n_2) = \xi$. Observe that $L(\rho') = \{1_1, 1_2\}$. If $(x'_1, x'_2) = 1_2$ then $P = \xi$ and, by definition of iterated substitution, $P \in L(r(D_e))^{0\xi}$. Otherwise, $T = (v, T_P)$ for some $v$ labeled by $(n, 1_1)$ and $T_P \in \mathcal{R}_P(D_{e^*\xi}, n_1)$. Thus, as $T_P$ is a sub-path of $T$ from $n_1$ which is a common node of $D_e$ and $D_{e^*\xi}$ then by induction hypothesis $P \in L(r(D_e))^{*\xi} \circ_\xi L(D_e, D_e)$. By definition of iterated substitution there exists $t \in \mathbb{N}$ such that $P \in \left( \bigcup_{t' \le t} L(r(D_e))^{t'\xi} \right) \circ_\xi L(D_e, D_e)$. We have $P \in L(r(D_e))^{(t+1)\xi}$.

2. otherwise, assume that $\gamma(r(D_e)) = \rho(x_1, \ldots, x_m)$. There are two cases according to $\gamma(r_j)$, for all $j \in [m]$ :

   **a.** If there exists $i \in [m]$ such that $\gamma(r_i) = \xi$ then, by construction, $k = m$, $out(r(D_e))$ and $out(r(D_{e^*\xi}))$ have the same destination nodes that are $r_1, \ldots, r_m$, and $out(n)$ has as destination (after $\xi$-normalization) $n_1, \ldots, n_k = r_1, \ldots, r_{i-1}, y, r_{i+1}, \ldots, r_m$. In addition, $\rho' \equiv \rho^{*x_i}$ and for all $j \in [k]$, $\gamma(r_j) = \gamma'(n_j)$. If there exists $i' \in [k]$ such that $(x'_1, \ldots, x'_k) = 1_{i'}$ then $T = (v, T_P)$ with $v$ labeled by $(n, (x'_1, \ldots, x'_k))$ and $T_P \in \mathcal{R}_P(D_{e^*\xi}, n_{i'})$. In this case we deduce that $L(D_{e^*\xi}, n_{i'}) \subseteq L(D_{e^*\xi}, n)$. If $i' = i$ then $P = \xi$ and $P \in L(r(D_e))^{0\xi}$.

   Otherwise, By Lemma 33, $(x'_1, \ldots, x'_k) \in L(\rho)$. That implies that $L(D_e, r_{i'}) \subseteq L(r(D_e))$. Furthermore, as $T_P$ is a sub-path of $T$ from $r_{i'}$ which is a common node of $D_e$ and $D_{e^{*\xi}}$ then by induction hypothesis $P \in L(r(D_e))^{*\xi} \circ_\xi L(D_e, r_{i'})$. By definition of iterated substitution there exists $t \in \mathbb{N}$ such that $P \in \left( \bigcup_{t' \leq t} L(r(D_e))^{t'\xi} \right) \circ_\xi L(D_e, r_{i'})$.
We have $P \in L(r(D_e))^{(t+1)\xi}$.

In case $\sum_{j \in [k]} x'_j > 1$, $P = \|_{j \in [k]} \|_{s \in [x'_j]} P_{j,s}$ where each $P_{j,s} \in L(D_{e^{*\xi}}, n_j)$ is the label of some direct sub-path of $T$ from $n_j$. Then, by induction hypothesis, for all $j \in [k]$, $j \neq i$, for all $s \in [x'_j]$, $P_{j,s} \in L(r(D_e))^{*\xi} \circ_\xi L(D_e, r_j)$. Then

$$P \in L(r(D_e))^{*\xi} \circ_\xi L(D_e, r_1)^{\|^{x'_1}} \| \cdots \| \xi^{\|^{x'_i}} \| \cdots \| L(r(D_e))^{*\xi} \circ_\xi L(D_e, r_k)^{\|^{x'_k}}$$

We have, by Lemma 60, $P \in L(r(D_e))^{*\xi}$;

b. otherwise, by construction, $k = m + 1$, $n_1, \ldots, n_k = r_1, \ldots, r_m, y$, $\rho'(x_1, \ldots, x_k) \equiv (\rho(x_1, \ldots, x_m) \wedge x_k = 0) \vee (\wedge_{j \in [m]} x_j = 0 \wedge x_k = 1)$, $\gamma'(n_k) = \xi$. Observe that $L(\rho') = \{(x_1, \ldots, x_m, 0) : (x_1, \ldots, x_m) \in L(\rho)\} \cup \{1_k\}$. If $(x'_1, \ldots, x'_k) = 1_k$ then $P = \xi$ and, by definition of iterated substitution, $P \in L(r(D_e))^{0\xi}$.

Otherwise, if there exists $i' \in [m]$ such that $(x'_1, \ldots, x'_k) = 1_{i'}$ then $P \in L(D_{e^{*\xi}}, n_{i'})$. That implies that $L(D_e, r_{i'}) \subseteq L(r_P(D_e))$. Furthermore, as $T_P$ is a sub-path of $T$ from $r_{i'}$ which is a common node of $D_e$ and $D_{e^{*\xi}}$ then by induction hypothesis $P \in L(r(D_e))^{*\xi} \circ_\xi L(D_e, r_{i'})$. By definition of substitution operation there exists $t \in \mathbb{N}$ such that $P \in \left( \bigcup_{t' \leq t} L(r(D_e))^{t'\xi} \right) \circ_\xi L(D_e, r_{i'})$. We have $P \in L(r(D_e))^{(t+1)\xi}$.

Otherwise, $P = \|_{j \in [m]} \|_{s \in [x'_j]} P_{j,s}$ where each $P_{j,s} \in L(D_{e^{*\xi}}, n_j)$ and the label of some direct sub-path of $T$ from $n_j$. Then by induction hypothesis, for all $j \in [m]$, for all $s \in [x'_j]$, $P_{j,s} \subseteq L(r(D_e))^{*\xi} \circ_\xi L(D_e, r_j)$. Furthermore, we know by definition of iterated substitution that for all $j \in [m]$, for all $s \in [x'_j]$ there exists $t_{j,s} \in \mathbb{N}$ such that $P_{j,s} \subseteq \left( \bigcup_{t'_{j,s} \leq t_{j,s}} L(r(D_e))^{t'_{j,s}\xi} \right) \circ_\xi L(D_e, r_j)$. Let $t = \max\{t_{j,s} : j \in [m], s \in [x'_j]\}$. Then

$$P \in \left( \bigcup_{t' \leq t} L(r(D_e))^{t'\xi} \right) \circ_\xi \left( L(D_e, r_1)^{\|^{x'_1}} \| \cdots \| L(D_e, r_m)^{\|^{x'_m}} \right)$$

As $\left( L(D_e, r_1)^{\|^{x'_1}} \| \cdots \| L(D_e, r_m)^{\|^{x'_m}} \right) \subseteq L(r(D_e))$, we have $P \in L(r(D_e))^{(t+1)\xi}$.

◄

▶ **Lemma 62.** *Let $D_e$ be the D-graph of some rational expression $e$. Then $L(r(D_{e^{*\xi}})) \circ_\xi L(r(D_e)) \subseteq L(r(D_{e^{*\xi}}))$.*

**Proof.** Assume that $out(r(D_e)) = r(D_e) \rightarrow r_1, \ldots, r(D_e) \rightarrow r_k$ and $out(r(D_{e^{*\xi}})) = r(D_{e^{*\xi}}) \rightarrow r'_1, \ldots, r(D_{e^{*\xi}}) \rightarrow r'_m$ for some $k, m \in \mathbb{N} \setminus \{0\}$. Let $\gamma$ and $\gamma'$ be the maps labeling respectively $D_e$ and $D_{e^{*\xi}}$. There are two cases according to $\gamma(r(D_e))$.

1. If $\gamma(r(D_e)) \notin \mathcal{P}$ then by construction, $m = 2$, $\gamma'(r(D_{e^{*\xi}})) = \rho'(x_1, x_2) \equiv x_1 + x_2 = 1$, $r'_1 = r(D_e)$, $\gamma'(r'_1) = \gamma(r(D_e))$ and $\gamma'(r'_2) = \xi$. By Lemma 59, $L(r(D_{e^{*\xi}})) \circ_\xi L(r(D_e)) \subseteq L(D_{e^{*\xi}}, r'_1)$. Thus, since we have $1_1 \in L(\rho')$, $L(r(D_{e^{*\xi}})) \circ_\xi L(r(D_e)) \subseteq L(r(D_{e^{*\xi}}))$;

2. otherwise, assume that $\gamma(r(D_e)) = \rho(x_1, \ldots, x_k)$. By construction, $r(D_e) = r(D_{e^{*\xi}})$. There are two cases according to $\gamma(r_j)$, for all $j \in [k]$.

**a.** If there exists $i \in [k]$ such that $\gamma(r_i) = \xi$, by construction $\mathrm{out}(r(D_e)) = \mathrm{out}(r(D_{e^{*\xi}}))$ and $\gamma'(r(D_{e^{*\xi}})) = \rho'(x_1, \ldots, x_k) \equiv \rho^{*x_i}$. Let $P \in L(r(D_{e^{*\xi}})) \circ_\xi L(r(D_e))$. By definition of substitution operation there exists $P' \in L(r(D_e))$ such that $P \in L(r(D_{e^{*\xi}})) \circ_\xi P'$. In addition there exists $(x_1, \ldots, x_k) \in L(\rho)$ such that

$$P' \in L(D_e, r_1)^{\|x_1} \parallel \cdots \parallel \xi^{\|x_i} \parallel \cdots \parallel L(D_e, r_k)^{\|x_k}$$

Assume wlog that the $x_i$ occurrences of $\xi$ in $P'$ are replaced by $P_1, \ldots, P_{x_i} \in L(r(D_{e^{*\xi}}))$ to obtain $P$. Then

$$P \in L(r(D_{e^{*\xi}})) \circ_\xi L(D_e, r_1)^{\|x_1} \parallel \cdots \parallel P_1 \parallel \cdots \parallel P_{x_i} \parallel \cdots \parallel L(r(D_{e^{*\xi}})) \circ_\xi L(D_e, r_k)^{\|x_k}$$

Note that for all $s \in [x_i]$, there exists $(x'_{s,1}, \ldots, x'_{s,k}) \in L(\rho')$ such that

$$P_s \in L(D_{e^{*\xi}}, r'_1)^{\|x'_{s,1}} \parallel \cdots \parallel \xi^{\|x'_{s,i}} \parallel \cdots \parallel L(D_{e^{*\xi}}, r'_k)^{\|x'_{s,k}}$$

By Lemma 59, for all $j \in [k]$, $j \neq i$, $L(r(D_{e^{*\xi}})) \circ_\xi L(D_e, r_j) \subseteq L(D_{e^{*\xi}}, r'_j)$. In addition, by construction we have $L(D_e, r_i) = L(D_{e^{*\xi}}, r'_i) = \{\xi\}$. Then

$$P \in L(D_{e^{*\xi}}, r'_1)^{\|y_1} \parallel \cdots \parallel L(D_{e^{*\xi}}, r'_i)^{\|y_i} \parallel \cdots \parallel L(D_{e^{*\xi}}, r'_k)^{\|y_k}$$

where $y_i = \sum_{s \in [x_i]} x'_{s,i}$ and for all $j \in [k] \setminus \{i\}$, $y_j = x_j + \sum_{s \in [x_i]} x'_{s,j}$. Observe that for all $s \in [x_i]$ there exists $t_s \in \mathbb{N}$ such that $(x'_{s,1}, \ldots, x'_{s,k}) \in L(\rho^{t_s x_i})$ and $t_s$ is the smallest possible. Let $t = \max\{t_s : s \in [x_i]\}$. We have $(y_1, \ldots, y_k) \in L(\rho^{(t+1)x_i})$. Thus $P \in L(r(D_{e^{*\xi}}))$;

**b.** otherwise, by construction, $m = k+1$, $\gamma'(r(D_{e^{*\xi}})) = \rho'(x_1, \ldots, x_m) \equiv (\rho(x_1, \ldots, x_k) \wedge x_m = 0) \vee (\wedge_{j \in [k]} x_j = 0 \wedge x_m = 1)$, $\gamma'(r'_m) = \xi$ and for all $j \in [k]$, $r_j = r'_j$ and $\gamma'(r'_j) = \gamma(r_j)$. Observe that for all $(x_1, \ldots, x_k) \in L(\rho)$, there exists $(x_1, \ldots, x_k, 0) \in L(\rho')$. By Lemma 59, for all $j \in [k]$, $L(r(D_{e^{*\xi}})) \circ_\xi L(D_e, r_j) \subseteq L(D_{e^{*\xi}}, r'_j)$. Thus $L(r(D_{e^{*\xi}})) \circ_\xi L(r(D_e)) \subseteq L(r(D_{e^{*\xi}}))$.

◄

▶ **Lemma 63.** *Let $D_e$ be the D-graph of some rational expression $e$. Then $L(r(D_e))^{*\xi} \subseteq L(r(D_{e^{*\xi}}))$.*

**Proof.** Let $P \in L(r(D_e))^{*\xi}$. By definition of iterated substitution, there exists $i \in \mathbb{N}$ such that $P \in L(r(D_e))^{i\xi}$. The lemma can be easily proved by induction on $i$ using Lemma 58 and Lemma 62. ◄

▶ **Lemma 64.** *Let $D_e$ be the D-graph of some rational expression $e$. Then $L(r(D_{e^{*\xi}})) \subseteq L(r(D_e))^{*\xi}$.*

**Proof.** Assume that $\mathrm{out}(r(D_e)) = r(D_e) \to r_1, \ldots, r(D_e) \to r_k$ and $\mathrm{out}(r(D_{e^{*\xi}})) = r(D_{e^{*\xi}}) \to r'_1, \ldots, r(D_{e^{*\xi}}) \to r'_m$ for some $k, m \in \mathbb{N} \setminus \{0\}$. Let $\gamma$ and $\gamma'$ be the maps labeling respectively $D_e$ and $D_{e^{*\xi}}$. By construction, $\gamma'(r(D_{e^{*\xi}})) \in \mathcal{P}_m$. Assume that $\gamma'(r(D_{e^{*\xi}})) = \rho'(x_1, \ldots, x_m)$. Let $P \in L(r(D_{e^{*\xi}}))$. Then for any $T \in \mathcal{R}_P(D_{e^{*\xi}}, r(D_{e^{*\xi}}))$, the root of $T$ is labeled by $(r(D_{e^{*\xi}}), (x'_1, \ldots, x'_m))$ for some $(x'_1, \ldots, x'_m) \in L(\rho')$. There are two cases according to $\gamma(r(D_e))$.

**1.** If $\gamma(r(D_e)) \notin \mathcal{P}$ then by construction, $m = 2$, $\gamma'(r(D_{e^{*\xi}})) = \rho'(x_1, x_2) \equiv x_1 + x_2 = 1$, $r'_1 = r(D_e)$, $\gamma'(r'_1) = \gamma(r(D_e))$ and $\gamma'(r'_2) = \xi$. Observe that $L(\rho') = \{1_1, 1_2\}$. If $(x'_1, x'_2) = 1_2$ then $P = \xi$ and by definition of iterated substitution $P \in L(r(D_e))^{0\xi}$.

Otherwise, $(x_1', x_2') = 1_1$ and $P \in L(D_{e*\xi}, r_1')$. As a consequence of Lemma 61, $P \in L(r(D_e))^{*\xi} \circ_\xi L(r(D_e))$. Then by definition of iterated substitution there exists $t \in \mathbb{N}$ such that $P \in \left( \bigcup_{t' \leq t} L(r(D_e))^{t'\xi} \right) \circ_\xi L(r(D_e))$. Thus $L(r(D_{e*\xi})) \subseteq L(r(D_e))^{(t+1)\xi}$;

2. otherwise, assume that $\gamma(r(D_e)) = \rho(x_1, \ldots, x_k)$. By construction, $r(D_e) = r(D_{e*\xi})$. There are two cases according to $\gamma(r_j)$, for all $j \in [k]$.

   a. If there exists $i \in [k]$ such that $\gamma(r_i) = \xi$, by construction $\text{out}(r(D_e)) = \text{out}(r(D_{e*\xi}))$ and $\rho' \equiv \rho^{*x_i}$. By Lemma 61, for all $j \in [k], j \neq i$, $L(D_{e*\xi}, r_j') \subseteq L(r(D_e))^{*\xi} \circ_\xi L(D_e, r_j)$. In addition, $L(D_{e*\xi}, r_i') = L(D_e, r_i) = \{\xi\}$. Then

   $$P \in L(r(D_e))^{*\xi} \circ_\xi L(D_e, r_1)^{\|x_1'} \| \cdots \| \xi^{\|x_i'} \| \cdots \| L(r(D_e))^{*\xi} \circ_\xi L(D_e, r_k)^{\|x_k'}$$

   and, by Lemma 60, $P \in L(r(D_e))^{*\xi}$;

   b. otherwise, by construction $m = k + 1$ and for all $j \in [k]$, $r_j' = r_j$. In addition, $\rho'(x_1, \ldots, x_m) \equiv (\rho(x_1, \ldots, x_k) \wedge x_m = 0) \vee (\wedge_{j \in [k]} x_j = 0 \wedge x_m = 1)$, $\gamma'(r_m') = \xi$ and for all $j \in [k]$, $\gamma'(r_j') = \gamma(r_j)$. Observe that $L(\rho') = \{1_m\} \cup \{(x_1, \ldots, x_k, 0) : (x_1, \ldots, x_k) \in L(\rho)\}$. If $(x_1', \ldots, x_m') = 1_m$ then $P = \xi$ and by definition of iterated substitution $P \in L(r(D_e))^{0\xi}$.

   Otherwise, $(x_1', \ldots, x_k') \in L(\rho)$. By Lemma 61, $L(D_{e*\xi}, r_j') \subseteq L(r(D_e))^{*\xi} \circ_\xi L(D_e, r_j)$ for all $j \in [k]$. Then, by definition of iterated substitution, for all $j \in [k]$ there exists $t_j$ such that $L(D_{e*\xi}, r_j') \subseteq \left( \bigcup_{t_j' \leq t_j} L(r(D_e))^{t_j'\xi} \right) \circ_\xi L(D_e, r_j)$. Let $t = \max\{t_j : j \in [k]\}$. Then

   $$P \in \left( \bigcup_{t' \leq t} L(r(D_e))^{t'\xi} \right) \circ_\xi \left( L(D_e, r_1)^{\|x_1'} \| \cdots \| L(D_e, r_k)^{\|x_k'} \right)$$

   We have $P \in L(r(D_e))^{(t+1)\xi}$.                                                     ◄

Thanks to Lemma 63 and Lemma 64 we have the following corollary.

▶ **Corollary 65.** *Let $D_e$ be the D-graph of some rational expression $e$. Then $L(r(D_{e*\xi})) = L(r(D_e))^{*\xi}$.*

### 5.4.3 All cases together

We are now ready to prove Proposition 55.

**Proof of Proposition 55.** When $f$ is a rational expression the corresponding $>1$-expression is denoted by $e$. We proceed by induction on $f$.

If $f = a \in A$ then $e = a$, $D_e$ is simply a node $n$ labeled by $a$ and by Definition 48, $\bigcup_{P \in SP^\diamond(A)} \mathcal{R}_P(D_e, r(D_e)) = \{(n, a)\}$. Then $L(D_e) = L(f) = \{a\}$.

If $f = \epsilon$ then $e = \epsilon$, $D_e$ is simply a node $n$ labeled by some closed Presburger tautology $\rho$ and by Definition 48, $\bigcup_{P \in SP^\diamond(A)} \mathcal{R}_P(D_e, r(D_e)) = \{(n, ())\}$. Then $L(D_e) = L(f) = \{\epsilon\}$.

In the case where $f = f_1 \| f_2$, $e$ has the form $e_1 \| e_2$. Assume first that none of $r(D_{e_1})$ and $r(D_{e_2})$ are labeled in $\mathcal{P}$, then by construction of $D_e$, $T \in \mathcal{R}_P(D_e, r(D_e))$ for some $P \in SP^{\diamond+}(A)$, if and only if $T$ has the form $(m, T_1, T_2)$ with $m$ labeled by $(r(D_e), (1, 1))$, $T_i \in \mathcal{R}_{P_i}(D_{e_i}, r(D_{e_i}))$ for some non-empty $P_i \in L(D_{e_i})$, $i \in [2]$ and $P =$ $P_1 \parallel P_2$. Then $L(D_e) = L(D_{e_1}) \parallel L(D_{e_2})$ and by induction hypothesis $L(D_e) = L(f)$. Otherwise, assume that $r(D_{e_1})$ is labeled by some $\rho_1(x_{1,1}, \ldots, x_{1,k})$ and $\mathrm{out}(r(D_{e_1})) = r(D_{e_1}) \to r_1, \ldots, r(D_{e_1}) \to r_k$ for some $k \in \mathbb{N}$ and $r(D_{e_2})$ is not labeled in $\mathcal{P}$. Then by construction of $D_e$, $r(D_e)$ is labeled by $\rho(x_{1,1}, \ldots, x_{1,k}, x_2) \equiv \rho_1 \circ_{x_1} (x_1 = x_2 = 1)$ and $\mathrm{out}(r(D_e)) = r(D_e) \to r_1, \ldots, r(D_e) \to r_k, r(D_e) \to r_{D_{e_2}}$. As a consequence of Lemma 31, $L(\rho) = \{(y_1, \ldots, y_k, 1) : (y_1, \ldots, y_k) \in L(\rho_1)\}$. In particular, when $e_1 = \epsilon$ then $\rho_1$ is a closed tautology, $L(\rho) = \{(1)\}$ and $\mathrm{out}(r(D_e)) = r(D_e) \to r(D_{e_2})$. Due to the construction of $D_e$, $T \in \mathcal{R}_P(D_e, r(D_e))$ for some $P \in SP^{\diamond+}(A)$, if and only if $T$ has the form $(m, (T_{P_{i,j}})_{i \in [k], j \in [y_i]}, T')$ with $m$ labeled by some $(r(D_e), (y_1, \ldots, y_k, 1))$ with $(y_1, \ldots, y_k) \in L(\rho_1)$, $T_{P_{i,j}} \in \mathcal{R}_{P_{i,j}}(D_{e_1}, r_i)$ for some non-empty $P_{i,j} \in L(D_{e_1}, r_i)$, $i \in [k]$, $j \in [y_i]$, and $T' \in \mathcal{R}_{P'}(D_{e_2}, r(D_{e_2}))$ for some non-empty $P' \in L(D_{e_2}, r(D_{e_2}))$ such that $P = P_1 \parallel P'$ where $P_1 = \parallel_{i \in [k]} \parallel_{j \in [y_i]} P_{i,j}$ ($P_1 \in L(D_{e_1})$). Then $L(D_e) = L(D_{e_1}) \parallel L(D_{e_2})$ and by induction hypothesis $L(D_e) = L(f)$. We prove the case when only $r(D_{e_2})$ is labeled in $\mathcal{P}$ using similar arguments. Assume now that $r(D_{e_1})$ is labeled by some $\rho_1(x_{1,1}, \ldots, x_{1,k})$, $\mathrm{out}(r(D_{e_1})) = r(D_{e_1}) \to r_1, \ldots, r(D_{e_1}) \to r_k$, $r(D_{e_2})$ is labeled by some $\rho_2(x_{2,1}, \ldots, x_{2,k'})$ and $\mathrm{out}(r(D_{e_2})) = r(D_{e_2}) \to r'_1, \ldots, r(D_{e_2}) \to r'_{k'}$ for some $k, k' \in \mathbb{N}$. Then by construction of $D_e$, $r(D_e)$ is labeled by $\rho(x_{1,1}, \ldots, x_{1,k}, x_{2,1}, \ldots, x_{2,k'}) \equiv \rho_2 \circ_{x_2} \rho_1 \circ_{x_1} (x_1 = x_2 = 1)$ and $\mathrm{out}(r(D_e)) = r(D_e) \to r_1, \ldots, r(D_e) \to r_k, r(D_e) \to r'_1, \ldots, r(D_e) \to r'_{k'}$. As a consequence of Lemma 31 $L(\rho) = \{(y_1, \ldots, y_k, y'_1, \ldots, y'_{k'}) : (y_1, \ldots, y_k) \in L(\rho_1), (y'_1, \ldots, y'_{k'}) \in L(\rho_2)\}$. In particular, $\rho$ is a closed tautology if and only if $\rho_1$ and $\rho_2$ are both closed tautologies too ($e_1 = e_2 = \epsilon$). Then due to the construction of $D_e$, $T \in \mathcal{R}_P(D_e, r(D_e))$ for some $P \in SP^{\diamond}(A)$, if and only if $T$ has the form $(m, (T_{P_{i,j}})_{i \in [k], j \in [y_i]}, (T_{P'_{i,j}})_{i \in [k'], j \in [y'_i]})$ with $m$ labeled by some $(r(D_e), (y_1, \ldots, y_k, y'_1, \ldots, y'_{k'}))$, $(y_1, \ldots, y_k) \in L(\rho_1)$ and $(y'_1, \ldots, y'_{k'}) \in L(\rho_2)$, $T_{P_{i,j}} \in \mathcal{R}_{P_{i,j}}(D_{e_1}, r_i)$ for some non-empty $P_{i,j} \in L(D_{e_1}, r_i)$, $i \in [k]$, $j \in [y_i]$, $T_{P'_{i,j}} \in \mathcal{R}_{P'_{i,j}}(D_{e_2}, r'_i)$ for some non-empty $P'_{i,j} \in L(D_{e_2}, r'_i)$, $i \in [k']$, $j \in [y'_i]$, such that $P = P_1 \parallel P_2$ where $P_1 = \parallel_{i \in [k]} \parallel_{j \in [y_i]} P_{i,j}$ ($P_1 \in L(D_{e_1})$), and $P_2 = \parallel_{i \in [k']} \parallel_{j \in [y'_i]} P'_{i,j}$ ($P_2 \in L(D_{e_2})$). Then $L(D_e) = L(D_{e_1}) \parallel L(D_{e_2})$ and by induction hypothesis $L(D_e) = L(f)$.

The proof when $f = f_1 + f_2$ is similar to the previous case.

When $f = f_1 \cdot f_2$ then $e$ has the form $e_1 \cdot^{>1} e_2 +_{\epsilon \in L(e_1)} e_2 +_{\epsilon \in L(e_2)} e_1$. Assume first that $\epsilon \notin L(e_1) \cup L(e_2)$. Then $e = e_1 \cdot^{>1} e_2$. By construction of $D_e$, $T \in \mathcal{R}_P(D_e, r(D_e))$ for some $P \in SP^{\diamond+}(A)$, if and only if $T$ has the form $(m, T_1, T_2)$ with $m$ labeled by $(r(D_e), 2)$ and $T_i \in \mathcal{R}_{P_i}(D_{e_i}, r(D_{e_i}))$ for some non-empty $P_i \in L(r(D_{e_i}))$, $i \in [2]$, and $P = P_1 P_2$. Then $L(D_e) = L(D_{e_1}) \cdot^{>1} L(D_{e_2})$ and by induction hypothesis $L(D_e) = L(f)$. In the case where $\epsilon \in L(e_1) \cap L(e_2)$, $e$ has the form $e_1 \cdot^{>1} e_2 + e_1 + e_2$. Using similar arguments that when $e$ had the form $e_1 + e_2$ we have $L(D_e) = L(D_{e_1 \cdot^{>1} e_2}) \cup L(D_{e_1}) \cup L(D_{e_2})$. Then, as $L(D_{e_1 \cdot^{>1} e_2}) = L(D_{e_1}) \cdot^{>1} L(D_{e_2})$, by induction hypothesis, $L(D_e) = L(f)$. We prove similarly the cases where $\epsilon \in L(e_2) \setminus L(e_1)$ and $\epsilon \in L(e_1) \setminus L(e_2)$.

In the case where $f = f_1 \diamond f_2$, $e$ has the form $e_1 \diamond^{>1} e_2 + e_1 +_{\epsilon \in L(e_1)} e_2$. Assume first that $\epsilon \notin L(e_1)$. In this case $e$ has the form $e_1 \diamond^{>1} e_2 + e_1$. Using similar arguments that when $e$ had the form $e_1 + e_2$ we have $L(D_e) = L(D_{e_1 \diamond^{>1} e_2}) \cup L(D_{e_1})$. Let $e' = e_1 \diamond^{>1} e_2$. Then by construction of $D_{e'}$, $T \in \mathcal{R}_P(D_{e'}, r(D_{e'}))$ for some $P \in SP^{\diamond+}(A)$, if and only if $T$ has the form $(m, (T_j)_{j \in J \cup \hat{J}^*})$ with $m$ labeled by $(r(D_{e'}), J \cup \hat{J}^*)$, $J \in \mathcal{S} \setminus \{0, 1\}$, $T_j \in \mathcal{R}_{P_j}(D_{e_1}, r(D_{e_1}))$ when $j \in J$ for some $P_j \in L(r(D_{e_1}))$, $T_j \in \mathcal{R}_{P_j}(D_{e_2}, r(D_{e_2}))$ when $j \in \hat{J}^*$ for some $P_j \in L(r(D_{e_2}))$, $P = \prod_{j \in J \cup \hat{J}^*} P_j$ and there must exist $j, j' \in J \cup \hat{J}^*$ such that $j \neq j'$ and $P_j, P_{j'} \neq \epsilon$. Then $L(D_{e'}) = L(D_{e_1}) \diamond^{>1} L(D_{e_2})$ and $L(D_e) = L(D_{e_1}) \diamond^{>1} L(D_{e_2}) \cup L(D_{e_1})$. Then by induction hypothesis $L(D_e) = L(f)$. Assume now that $\epsilon \in L(e_1)$. Then $e$ has the form $e_1 \diamond^{>1} e_2 + e_1 + e_2$. Using similar arguments that when $e$ had the form $e_1 + e_2$ we have $L(D_e) = L(D_{e_1 \diamond^{>1} e_2}) \cup L(D_{e_1}) \cup L(D_{e_2})$. Then, as $L(D_{e_1 \diamond^{>1} e_2}) = L(D_{e_1}) \diamond^{>1} L(D_{e_2})$ we have, by induction hypothesis, $L(D_e) = L(f)$.

When $f = f_1^*$, $e$ has the form $e_1^{*^{>1}} + e_1 + \epsilon$. Let $e' = e_1^{*^{>1}}$. By construction of $D_{e'}$, $T \in \mathcal{R}_P(D_{e'}, r(D_{e'}))$ for some $P \in SP^{\diamond+}(A)$, if and only if $T$ has the form $(m, T_{P_1}, \ldots, T_{P_n})$ with $m$ labeled by $(r(D_{e'}), n)$, $n \in \mathbb{N} \setminus \{0, 1\}$, for all $j \in [n]$, $T_{P_j} \in \mathcal{R}_{P_j}(D_{e_1}, r(D_{e_1}))$ for some $P_j \in L(r(D_{e_1}))$, $P = \prod_{j \in [n]} P_j$ and there must exist $j, j' \in [n]$ such that $j \neq j'$ and $P_j, P_{j'} \neq \epsilon$. Then $L(D_{e'}) = L(D_{e_1})^{*^{>1}}$. Using similar arguments that when $e$ had the form $e_1 + e_2$ we have $L(D_e) = L(D_{e_1^{*^{>1}}}) \cup L(D_{e_1}) \cup L(D_\epsilon)$. Then, as $L(D_{e_1^{*^{>1}}}) = L(D_{e_1})^{*^{>1}}$, we have, by induction hypothesis, $L(D_e) = L(f)$.

We prove similarly to the previous case, the cases when $f = f_1^{\natural^{>1}}$ and $f = f_1^{-\natural^{>1}}$.

When $f = f_1^{\omega^{>1}}$, $e$ has the form $e_1^{\omega^{>1}} + _{\epsilon \in L(e_1)} (e_1^{*^{>1}} + e_1 + \epsilon)$. Assume first that $\epsilon \notin L(e_1)$. Then $e$ has the form $e_1^{\omega^{>1}}$. By construction of $D_e$, $T \in \mathcal{R}_P(D_e, r(D_e))$ for some $P \in SP^{\diamond+}(A)$, if and only if $T$ has the form $(m, (T_j)_{j \in \omega})$ with $m$ labeled by $(r(D_{e_1}), \omega)$, $T_j \in \mathcal{R}_{P_j}(D_{e_1}, r(D_{e_1}))$ for some $P_j \in L(r(D_{e_1}))$, $j \in \omega$, $P = \prod_{j \in \omega} P_j$ and there must exist $j, j' \in \omega$ such that $j \neq j'$ and $P_j, P_{j'} \neq \epsilon$. Then $L(D_e) = L(D_{e_1})^{\omega^{>1}}$ and by induction hypothesis $L(D_e) = L(f)$. In the case where $\epsilon \in L(e_1)$, $e$ has the form $e_1^{\omega^{>1}} + e_1^{*^{>1}} + e_1 + \epsilon$. Using similar arguments that when $e$ had the form $e_1 + e_2$ we have $L(D_e) = L(D_{e_1^{\omega^{>1}}}) \cup L(D_{e_1^{*^{>1}} + e_1 + \epsilon})$. Then, as $L(D_{e_1^{\omega^{>1}}}) = L(D_{e_1})^{\omega^{>1}}$ and $L(D_{e_1^{*^{>1}} + e_1 + \epsilon}) = L(D_{e_1})^{*^{>1}} + L(D_{e_1}) + L(\epsilon)$, we have, by induction hypothesis, $L(D_e) = L(f)$.

We prove the case when $f = f_1^{-\omega^{>1}}$ similarly to the previous case.

The cases when $f$ has the form $f_1^{*\xi}$ or $f_1 \circ_\xi f_2$ are treated respectively in Corollary 57 and Corollary 65. ◄

## 6 Coloring

Let $C$ be a non-empty finite set whose elements are named *colors* and $P \in SP^\diamond$. We let $F_s(P)$ denote the class of all sequential factors of $P$.

▶ **Definition 66.** *Let $P \in SP^\diamond$ and $C$ be a non-empty finite set of colors. A s-coloring $\mathfrak{c} \colon F_s(P) \to C$ of $P$ with $C$ is a partial map from $F_s(P)$ to $C$. It is compatible if for any different $F, F' \in F_s(P)$ such that $\mathfrak{c}(F)$ and $\mathfrak{c}(F')$ are defined, either:*

- *$F \cap F' = \emptyset$ and, if $FF' \in F_s(P)$, then $\mathfrak{c}(F) \neq \mathfrak{c}(F')$;*
- *one is strictly included into the other, say wlog $F \subsetneq F'$, and if $\mathfrak{c}(F) = \mathfrak{c}(F')$ there is some $x \in F' \setminus F$ such that $x$ is incomparable to all the elements of $F$.*

In Sub-section 6.2 we show how compatible s-colorings can be encoded by means of MSO. The technique relies on another one, called *ms-coloring*, specific to some particular case of sequential factors (Sub-section 6.1). In Sub-section 6.3 we link compatible s-colorings and paths in D-graphs of rational expressions.

Before starting we need some additional definitions on posets.

A factor $F$ of $P$ is *sequentially maximal* if there is no $R, S$ with at least one of them non-empty such that $R + F + S$ is a factor of $P$. We let $F_{ms}(P)$ denote the class of all elements of $F_s(P)$ that are sequentially maximal. A factor $F \in F_{ms}(P)$ is *direct* if there is no $P' \in F_{ms}(P) \setminus \{P, F\}$ such that $F \in F_{ms}(P')$. We let $DF_{ms}(P)$ denote the subclass of $F_{ms}(P)$ consisting of all strict direct factors of $P$.

▶ **Example 67.** Let $P = (a \cup b) + c + (d \cup (e + (f \cup (g + h))) \cup i)$. Then:

- *$a$, $a \cup b$ and $(a \cup b) + c$ are strict factors of $P$;*
- *$a \cup b$ and $d \cup (e + (f \cup (g + h))) \cup i$ are parallel factors of $P$;*

- $(a \cup b) + c$ and $c + (d \cup (e + (f \cup (g + h))) \cup i)$ are sequential factors of $P$;
- $g + h$ and $d \cup i$ are sequentially maximal factors of $P$;
- $a$, $e + (f \cup (g + h))$ and $P$ are direct sequentially maximal sequential factors of $P$.

## 6.1 ms-coloring

A *ms-coloring* of $P \in SP^\diamond$ with $C$ is a total map $\mathsf{c} \colon F_{ms}(P) \to C$. In general functions are not MSO-expressible, but a ms-coloring can be encoded using three sets $X_c^w, X_c^s, X_c^p$ of elements of $P$, for each $c \in C$. For short let $ms_C^X$ denote the sets $(X_c^w, X_c^s, X_c^p)_{c \in C}$. Also, given $c \in C$, $x \in P$ and $\alpha \in \{w, s, p\}$, we denote by $ms_C^X +_c^\alpha x$ the subsets $(X_{c'}^{\prime w}, X_{c'}^{\prime s}, X_{c'}^{\prime p})_{c' \in C}$ of $P$ defined by

$$X_{c'}^{\prime \beta} = \begin{cases} X_{c'}^\beta \cup \{x\} & \text{when } \beta = \alpha \text{ and } c' = c; \\ X_{c'}^\beta & \text{otherwise.} \end{cases}$$

When $(y, x, x') \in P^3$ we denote by $ms_C^X +_c (y, x, x') = ms_C^X +_c^s y +_c^p x +_c^p x'$. Also, we denote by $ms_C^X +_c x = ms_C^X +_c^w x$ when $x \in P$.

The three following definitions lay the foundations on how $ms_C^X$ is used to encode a ms-coloring of a poset.

▶ **Definition 68.** *Let $P \in SP^\diamond$, $ms_C^X$ some subsets of $P$, and let $c \in C$. Then $(y, x, x') \in X_c^s \times X_c^p \times X_c^p$ is directly bound by $c$ for $F \in F_{ms}(P)$ in $ms_C^X$ if the following conditions are true:*

1. $F = F_1 + F_2 + F_3$, $F_2 = F_x \cup F_{x'}$, $x \in F_x$, $x' \in F_{x'}$ and $y \in F \setminus F_2$, for some $F_1, F_2, F_3, F_x, F_{x'}$;
2. *there exists $z \in F \setminus F_2$ incomparable to (and distinct from) $y$;*
3. *there is no $F' \in F_{ms}(P) \setminus \{F\}$ such that $F' = F_1' + F_2' + F_3'$, $F_2' = F_{2,1}' \cup F_{2,2}'$ for some $F_1', F_2', F_3', F_{2,1}', F_{2,2}'$, such that one of the following conditions is true:*
   a. *there exist $x_1, x_2 \in X_c^p$ such that $x_1 \in F_{2,1}'$, $x_2 \in F_{2,2}'$ and $y \in F' \setminus F_2'$;*
   b. *there exist $y_1 \in X_c^s$ and $x_1 \in X_c^p$ such that $x \in F_{2,1}'$, $x_1 \in F_{2,2}'$ and $y_1 \in F' \setminus F_2'$;*
   c. *there exist $y_1 \in X_c^s$ and $x_1 \in X_c^p$ such that $x' \in F_{2,1}'$, $x_1 \in F_{2,2}'$ and $y_1 \in F' \setminus F_2'$.*

*In this case $y$, $x$, and $x'$ are respectively directly s-bound, p-bound and p-bound by $c$ for $F$ in $ms_C^X$. In addition, we say that $(y, x, x')$ (resp. $y$, $x$ and $x'$) is indirectly bound (resp. s-bound, p-bound and p-bound) by $c$ in $ms_C^X$ for all $F' \in F_{ms}(P) \setminus \{F\}$ such that $F \in F_{ms}(F')$.*

Observe that if $y, x$ and $(y, x, x')$ are directly respectively s-bound, p-bound and bound by $c \in C$ for some $F \in F_{ms}(P)$ in $ms_C^X$ then they are indirectly respectively s-bound, p-bound and bound in $(X_c^w \cap F', X_c^s \cap F', X_c^p \cap F')_{c \in C}$ for all $F' \in F_{ms}(P)$ strictly containing $F$. Also, it follows from Definition 68 that if $(y, x, x')$ is directly bound by some $c \in C$ for $P$ in $ms_C^X$ then there is no $(y, s, s'), (t, x, s'), (t, s, x') \in X_c^s \times X_c^p \times X_c^p$ that is bound by $c$ for some $F' \in F_{ms}(P) \setminus \{P\}$ in $ms_C^X$. An element of $P^3$ (resp. $P$) is *bound* (resp. *s-bound, p-bound*) by some $c \in C$ for some $F \in F_{ms}(P)$ in some $ms_C^X$ if it is indifferently directly or indirectly bound (resp. s-bound, p-bound) by $c \in C$ for $F \in F_{ms}(P)$ in $ms_C^X$.

▶ **Definition 69.** *Let $P \in Seq$ of irreducible sequential factorization $\sum_{j \in J} P_j$ and $ms_C^X$ some subsets of $P$. The set $\mathcal{C}_{ms_C^X}(P)$ of candidates for the ms-coloring of $P$ is the smallest subset of $P \cup P^3$ such that:*

- *if there exist $j \in J$ such that $|P_j| = 1$ and $c \in C$ such that $P_j \subseteq X_c^w$ then $P_j \subseteq \mathcal{C}_{ms_C^X}(P)$;*
- *f there is no $j \in J$ such that $|P_j| = 1$ and there exist $c \in C$, $y, x, x' \in P$ such that $(y, x, x')$ is directly bound by $c$ for $P$ in $ms_C^X$ then $(y, x, x') \in \mathcal{C}_{ms_C^X}(P)$.*

▶ **Definition 70.** *Let* $P \in Seq$ *of irreducible sequential factorization* $\sum_{j \in J} P_j$ *and* $ms_C^X$ *some subsets of* $P$. *Then* $ms_C^X$ *ms-colors* $P$ *in* $c \in C$, *denoted* $ms_C^X(P) = c$, *if* $\mathcal{C}_{ms_C^X}(P)$ *is some singleton* $\mathcal{C}_{ms_C^X}(P) = \{x\}$ *and either* $x \in X_c^w$ *or* $x \in X_c^s \times X_c^p \times X_c^p$. *In this case we say that* $P$ *is ms-colored in* $c$ *by* $x$. *In addition* $ms_C^X$ *completely ms-colors a poset* $P \in SP^\diamond$ *if* $ms_C^X(F)$ *is defined for all* $F \in F_{ms}(P)$. *Finally, when* $P \in SP^\diamond$ *and* $\mathfrak{c} \colon F_{ms}(P) \to C$ *is a ms-coloring of* $P$ *with* $C$ *then the subsets* $ms_C^X$ *of* $P$ *encode* $\mathfrak{c}$ *if* $ms_C^X(F) = \mathfrak{c}(F)$ *for all* $F \in F_{ms}(P)$.

At the end of this sub-section we show how to encode the ms-coloring of some $P \in SP^\diamond$ by some $ms_C^X$, by induction on $r_X(P)$. The ms-coloring in $c \in C$ of a factor of $F_{ms}(P)$ that has no candidate for its ms-coloring and has an element $x$ comparable to all the others is easy, since it suffices to add $x$ to $X_c^w$. Note that this does not influence the ms-coloring of other factors of $F_{ms}(P)$. However, adding an element of $P$ into $X_c^s$ or $X_c^p$ in order to ms-color some factor of $F_{ms}(P)$ may influence the ms-coloring of other factors. This leads us to introduce the following definition.

▶ **Definition 71.** *Let* $P \in SP^\diamond$, $ms_C^X$ *some subsets of* $P$, $F \in F_{ms}(P)$ *and* $c \in C$. *Then* $x \in P$ *is* s-free *(resp.* p-free*) for* $(F, c)$ *in* $ms_C^X$ *if it is not s-bound (resp. p-bound) by* $c$ *for* $F$ *in* $ms_C^X +_c^s x$ *(resp.* $ms_C^X +_c^p x$*).*

Observe that if $x \in P$ is $\alpha$-free for some $(F, c)$ with $F \in F_{ms}(P)$, $c \in C$ and $\alpha \in \{s, p\}$ then $x$ is $\alpha$-free for $(F', c)$ in $ms_C^X$ and in $(X_c^w \cap F', X_c^s \cap F', X_c^p \cap F')_{c \in C}$ for all $F' \in F_{ms}(F)$ such that $x \in F'$.

▶ **Example 72.** Let $C = \{\texttt{red}, \texttt{blue}\}$ be a set of colors. Consider the poset $P \in SP^\diamond$ of Figure 6. Let $P_1 = \{x_i : i \in [2; 16]\}$, $P_2 = \{x_i : i \in [17; 25]\}$ and $P_3 = \{x_{19}, x_{20}, x_{22}, x_{23}\}$. Then $P_1, P_2, P_3 \in F_{ms}(P)$. Let $\mathfrak{c} \colon F_{ms}(P) \to C$ be a ms-coloring of $P$ such that $\mathfrak{c}(P_1) = \texttt{red}$ and $\mathfrak{c}(P_2) = \mathfrak{c}(P_3) = \texttt{blue}$. We define $ms_C^X$ that encodes $\mathfrak{c}(P_i)$, $i \in [3]$. Note that in this

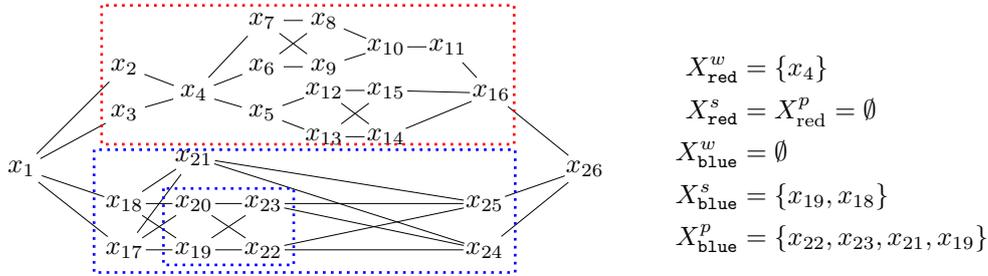

$$X_{\texttt{red}}^w = \{x_4\}$$
$$X_{\texttt{red}}^s = X_{\texttt{red}}^p = \emptyset$$
$$X_{\texttt{blue}}^w = \emptyset$$
$$X_{\texttt{blue}}^s = \{x_{19}, x_{18}\}$$
$$X_{\texttt{blue}}^p = \{x_{22}, x_{23}, x_{21}, x_{19}\}$$

■ **Figure 6** The Hasse diagram of a poset $P \in SP^\diamond$ and a ms-coloring of $P_1, P_2, P_3 \in F_{ms}(P)$

example we encode only the restriction of $\mathfrak{c}$ to $P_1, P_2, P_3$: the generalization of the method in order to encode the total map $\mathfrak{c} \colon F_{ms}(P) \to C$ is easy. We start with all sets of $ms_C^X$ empty. Observe that $P_1$ contains two elements comparable to all the other elements of $P_1$: $x_4$ and $x_{16}$. Choose one of them, indifferently $x_4$. Then add $x_4$ to $X_{\texttt{red}}^w$. This suffices to encode that $P_1$ is ms-colored in $\texttt{red}$, since there is no other candidate in $ms_C^X$ for the ms-coloring of $P_1$. Let us now encode that $\mathfrak{c}(P_3) = \texttt{blue}$. Note that since $P_3$ has no element comparable to all the others, we have to use a 3-tuple in $P_3$. Choose for example $(x_{19}, x_{22}, x_{23})$, and add $x_{19}$ in $X_{\texttt{blue}}^s$, $x_{22}, x_{23}$ in $X_{\texttt{blue}}^p$. Then $(x_{19}, x_{22}, x_{23})$ is directly bound by $\texttt{blue}$ for $P_3$ in $ms_C^X$. This encodes that $P_3$ is ms-colored in $\texttt{blue}$, since there is no other $c \in C$, $(z, z', z'') \in X_c^s \times X_c^p \times X_c^p$ that can also encode a color for $P_3$, and that this does not interfere with encoding of ms-coloring of other factors of $F_{ms}(P)$. Let us turn now to $P_2$. It has no element comparable to all the others, so we have to choose a 3-tuple $(y, y', y'')$ as we did for $P_3$. Note that $x_{19}$ is not a possible choice for $y$ since it is already s-bound by $\mathtt{blue}$ for $P_3$ in $ms_C^X$. Also, $x_{20}$ is not a possible choice for $y$ since it is not s-free for $(P_3, \mathtt{blue})$ and thus not s-free for $(P_2, \mathtt{blue})$. Indeed, adding it in $X_{\mathtt{blue}}^s$ would make $(x_{20}, x_{22}, x_{23})$ another candidate for ms-coloring $P_3$ in blue and as a consequence, regardless of $y'$ and $y''$, $(x_{20}, y', y'')$ could not be bound by $\mathtt{blue}$ for $P_2$ in $ms_C^X$. Choose, for example, $y = x_{18}$ and $y' = x_{21}$. As we can not take $y'' \in \{x_{22}, x_{23}\}$ since they are already p-bound by $\mathtt{blue}$ for $P_3$ in $ms_C^X$, then necessarily $y''$ is indifferently one of $x_{20}, x_{19}$, say $x_{19}$. Add $y$ in $X_{\mathtt{blue}}^s$ and $y', y''$ in $X_{\mathtt{blue}}^p$ and this is done. Finally, $(X_{\mathtt{red}}^w, X_{\mathtt{red}}^s, X_{\mathtt{red}}^p) = (\{x_4\}, \emptyset, \emptyset)$ and $(X_{\mathtt{blue}}^w, X_{\mathtt{blue}}^s, X_{\mathtt{blue}}^p) = (\emptyset, \{x_{19}, x_{18}\}, \{x_{22}, x_{23}, x_{21}, x_{19}\})$.

▶ **Proposition 73.** *Let $C$ be a non-empty finite set of colors and $P \in SP^\diamond$. For any ms-coloring $\mathfrak{c}$ of $P$ with $C$ there exist some subsets $ms_C^X$ of $P$ that encode $\mathfrak{c}$.*

The remainder of this section is devoted to the proof of Proposition 73. The following lemma shows in particular that any poset $P$ of $Seq$ that is completely ms-colored by some $ms_C^X$ contains, for all $c \in C$, at least a s-free and a p-free element for $(P, c)$ in $ms_C^X$.

▶ **Lemma 74.** *Let $P \in Seq$ completely ms-colored by some $ms_C^X$. If for all $c \in C$, all the elements of $X_c^p$ (resp. $X_c^s$) are p-bound (resp. s-bound) by $c$ for $P$ in $ms_C^X$ then there exist $x, y \in P$ such that for all $c \in C$, $x$ is p-free and $y$ is s-free for $(P, c)$ in $ms_C^X$.*

**Proof.** We prove the lemma by induction on $r_X(P)$. If $r_X(P) = (0, i)$ for $i \in \{0, 1\}$ then $P$ is either the singleton or a finite linear ordering. In both cases and for all $c \in C$, there is no $(y, x, x') \in P^3$ bound by $c$ for $P$ in $ms_C^X$. Thus, $\bigcup_{c \in C} X_c^s = \bigcup_{c \in C} X_c^p = \emptyset$ and for all $x \in P$, for all $c \in C$, $x$ is naturally s-free and p-free for $(P, c)$ in $ms_C^X$. Otherwise, assume that the irreducible sequential factorization of $P$ is $\sum_{j \in J} P_j$, for some $J \in \mathcal{S} \setminus \{0, 1\}$. Assume that $P$ is ms-colored in $e$ by $z \in X_e^w$. Then by Definitions 69–70, $z$ is comparable to each element of $P \setminus \{z\}$. In this case, by Definition 71, $z$ is s-free and p-free for $(P, c)$ in $ms_C^X$, for all $c \in C$.

Otherwise, assume that $P$ is ms-colored in $e$ by $(y, x, x') \in X_e^s \times X_e^p \times X_e^p$. Then by Definition 70, $\mathcal{C}_{ms_C^X}(P) = \{(y, x, x')\}$ and by Definition 69 there is no $j \in J$ such that $|P_j| = 1$, and $(y, x, x')$ is directly bound by $e$ for $P$ in $ms_C^X$. By Definitions 68, $P = P_1 + P_2 + P_3$, $P_2 = P_x \cup P_{x'} \cup P_{x''}$ for some $P_x, P_{x'} \in DF_{ms}(P)$ and $P_1, P_2, P_3, P_{x''}$ some factors of $P$, such that $x \in P_x, x' \in P_{x'}$ and $y \in P \setminus P_2$. Observe that there exists $P_y \in DF_{ms}(P)$ such that $y \in P_y$ and all the elements of $P_y$ are comparable to those of $P_2$. We have $r_X(P_y), r_X(P_x), r_X(P_{x'}) < r_X(P)$. Furthermore, $x$ is not p-bound, $x'$ is not p-bound and $y$ is not s-bound by $e$ for respectively $P_x$, $P_{x'}$ and $P_y$ in $ms_C^X$, otherwise $(y, x, x')$ cannot be directly bound by $e$ for $P$ in $ms_C^X$ according to Definition 68. So, let $\overline{X_e^p} = X_e^p \cap P_y$, $\overline{X_e^s} = (X_e^s \cap P_y) \setminus \{y\}$ and $\overline{X_c^p} = X_c^p \cap P_y$, $\overline{X_c^s} = X_c^s \cap P_y$ for all $c \in C \setminus \{e\}$ and let $ms_C^{\overline{X}} = (X_c^w, \overline{X_c^s}, \overline{X_c^p})_{c \in C}$. Observe that $P_y$ and $ms_C^{\overline{X}}$ satisfy the hypothesis of the lemma. Then by induction hypothesis, there exists $z \in P_y$ such that $z$ is p-free for $(P_y, c)$ in $ms_C^{\overline{X}}$, for all $c \in C$. In addition, for all $c \in C$, $z$ is still p-free for $(P_y, c)$ in $ms_C^X$. In fact, assume by contradiction that there exists $c' \in C$ such that $z$ is not p-free for $(P_y, c')$ in $ms_C^X$. Then by Definition 71, there exist $y_1 \in X_{c'}^s$ and $x_2 \in X_{c'}^p$ such that $(y_1, x_2, z)$ is bound by $c'$ for $P_y$ in $ms_C^{X'} = ms_C^X +_{c'}^p z$. Note that as $y$ is used in $(y, x, x')$ to ms-color $P$ in $e$ then $y$ is s-free for $(P_y, e)$ in $ms_C^X$. Thus $z$ may not be p-free only when $y_1 = y$ and $c' = e$. Indeed, despite $y_1 \in \overline{X_{c'}^s}$ and $x_2 \in \overline{X_{c'}^p}$ when $y_1 \neq y$ or $c' \neq e$, $(y_1, x_2, z)$ did not form a bound tuple since $z$ was proved p-free for $(P_y, c')$ in $ms_C^X$.

By the lemma hypothesis, $x_2$ is p-bound by $e$ for $P$ in $ms_C^X$. Then there exist $y_2 \in X_e^s$ and $x_2' \in X_e^p$ such that $(y_2, x_2, x_2')$ is bound by $e$ for $P$ in $ms_C^X$. In addition, $(y_2, x_2, x_2')$ is bound by $e$ for $P_y$ in $ms_C^X$ since if $y_2, x_2' \in P \setminus P_y$ then $(y_2, x_2, x_2')$ will be another candidate for ms-coloring $P$ in $e$ by $ms_C^X$ which is not allowed by definition of the complete ms-coloring of $P$. Thus, $(y, x_2, z)$ and $(y_2, x_2, x_2')$ are bound by $e$ for $P_y$ in respectively $ms_C^{X'}$ and $ms_C^{\overline{X}}$. Assume that $(y, x_2, z)$ and $(y_2, x_2, x_2')$ are directly bound by $e$ for respectively $P_y'$ in $ms_C^{X'}$ and $P_y''$ in $ms_C^{\overline{X}}$ for some $P_y', P_y'' \in F_{ms}(P_y)$. Observe that $(y_2, x_2, x_2')$ is bound by $e$ for $P_y''$ in $ms_C^{\overline{X}}$ too since $y_2 \neq y$ (the contrary would contradict the fact that $(y, x, x')$ is bound by $e$ for $P$ in $ms_C^X$). Observe also that $ms_C^{X'}$ contains $ms_C^X$. There are three cases:

1. if $P_y' \in F_{ms}(P_y'') \setminus \{P_y''\}$ then necessarily each of $y, z$ and $x_2$ are incomparable to $x_2'$. Then, naturally $(y_2, z, x_2')$ is also directly bound by $e$ for $P_y''$ in $ms_C^{\overline{X}} +_c^p z$, which contradicts the fact that $z$ is p-free for $(P_y, e)$ in $ms_C^{\overline{X}}$;

2. in the case where $P_y' = P_y''$ then by Definition 68, $P_y' = P_{y,1}' + P_{y,2}' + P_{y,3}'$ such that $x_2, x_2', z \in P_{y,2}'$, $z$ and $x_2'$ are incomparable to $x_2$ and $y, y_2 \in P_y' \setminus P_{y,2}'$. In this case $(y_2, x_2, z)$ is also directly bound by $e$ for $P_y'$ in $ms_C^{\overline{X}} +_{c'}^p z$, which contradicts the fact that $z$ is p-free for $(P_y, e)$ in $ms_C^{\overline{X}}$;

3. if $P_y'' \in F_{ms}(P_y') \setminus \{P_y'\}$ then $(y, x_2, z)$ cannot be bound by $e$ in $ms_C^{X'}$ by Definition 68.

Thus for all $c \in C$, $z$ is still p-free for $(P_y, c)$ in $ms_C^X$.

Let us prove that for all $c \in C$, $z$ is still also p-free for $(P, c)$ in $ms_C^X$. By contradiction suppose that there exists $c' \in C$ such that $z$ is not p-free for $(P, c')$ in $ms_C^X$. Then, there exist $y_1 \in X_{c'}^s$ and $x_2 \in X_{c'}^p$ such that $(y_1, x_2, z)$ is bound by $c'$ for $P$ in $ms_C^{X'}$. As $z$ is p-free for $(P_y, c')$ in $ms_C^X$ then $y_1, x_2 \notin P_y$. That means that $(y_1, x_2, z)$ is directly bound for $P$ in $ms_C^{X'}$. Then by Definition 68 there exist $P_{y_1}, P_{x_2} \in DF_{ms}(P)$ such that $y_1 \in P_{y_1}$, $x_2 \in P_{x_2}$ and all the elements of $P_y$ are comparable to those of $P_{y_1}$ and incomparable to those of $P_{x_2}$. By hypothesis, $y_1$ and $x_2$ are respectively s-bound and p-bound by $c'$ for $P$ in $ms_C^X$. Then there exist $y_2 \in X_{c'}^s$ and $x_1, x_1', x_2' \in X_{c'}^p$ such that $(y_1, x_1, x_1')$ and $(y_2, x_2, x_2')$ are bound by $c'$ for $P$ in $ms_C^X$. By hypothesis $P$ is ms-colored by $(y, x, x')$ in $ms_C^X$. Then the complete ms-coloring of $P$ by $ms_C^X$ implies that $(y_1, x_1, x_1')$ and $(y_2, x_2, x_2')$ are bound by $c'$ for respectively $P_{y_1}$ and $P_{x_2}$ in $ms_C^X$. Observe that $ms_C^{X'}$ contains $ms_C^X$ and $P_{y_1}, P_{x_2} \in F_{ms}(P) \setminus \{P\}$. Thus by Definition 68, $(y_1, x_2, z)$ cannot be bound by $c'$ in $ms_C^{X'}$ which is a contradiction.

Now, let $\overline{X_e^s} = X_e^s \cap P_x$, $\overline{X_e^p} = (X_e^p \cap P_x) \setminus \{x\}$ and $\overline{X_c^p} = X_c^p \cap P_x$, $\overline{X_c^s} = X_c^s \cap P_x$ for all $c \in C \setminus \{e\}$ and let $ms_C^{\overline{X}} = (X_c^w, \overline{X_c^s}, \overline{X_c^p})_{c \in C}$. Observe that $P_x$ and $ms_C^{\overline{X}}$ satisfy the hypothesis of the lemma. Then by induction hypothesis, for all $c \in C$, there exists $z \in P_x$ such that $z$ is s-free for $(P_x, c)$ in $ms_C^{\overline{X}}$. The proof that $z$ is still s-free for $(P, c)$, for all $c \in C$, in $ms_C^X$ uses similar arguments.                                                                                                  ◄

Using the previous lemma, we show that the ms-coloring of a poset $P \in Seq$ can be encoded without changing the ms-coloring of its sequentially maximal strict sequential factors.

▶ **Lemma 75.** *Let $P \in Seq$ and $ms_C^X$ some subsets of $P$. If the following conditions are true:*

- $\mathcal{C}_{ms_C^X}(P) = \emptyset$;
- *for all $P' \in F_{ms}(P) \setminus \{P\}$, $P'$ is completely ms-colored by $ms_C^X$;*
- *for all $c \in C$, all the elements of $X_c^p$ (resp. $X_c^s$) are p-bound (resp. s-bound) by $c$ for $P$ in $ms_C^X$,*

*then for all $e \in C$, there exist $z \in P \cup P^3$, $ms_C^{X'} = ms_C^X +_e z$, such that:*

- for all $P' \in F_{ms}(P) \setminus \{P\}$, $x \in P'^3$ and $c \in C$, $x$ is bound by $c$ in $ms_C^X$ for $P'$ if and only if $x$ is bound by $c$ in $ms_C^{X'}$ for $P'$;
- $z$ ms-colors $P$ in $e$.

**Proof.** Assume we want to ms-color $P$ in $e \in C$. If $P = \{x\}$ then each set of $ms_C^X$ is empty. Define $ms_C^{X'}$ by all sets empty except $X_e'^w = \{x\}$. In this case, $\mathcal{C}_{ms_{C}X'}(P) = \{x\}$. If $|P| > 1$, assume that its irreducible sequential factorization is $P = \sum_{j \in J} P_j$, for some $J \in \mathcal{S} \setminus \{0,1\}$. If there exists $j \in J$ such that $P_j = \{x\}$ then it suffices to set $ms_C^{X'} = ms_C^X +_e x$ to reach the lemma. Assume now that the irreducible sequential factorization of $P$ is $P = \sum_{j \in J} P_j$, for some $J \in \mathcal{S} \setminus \{0,1\}$ and each $P_j$ is a parallel poset. Let $j, j' \in J$ such that $j \neq j'$. Then $P_j = P_{j,1} \cup P_{j,2} \cup P_{j,3}$ and $P_{j'} = P_{j',1} \cup P_{j',2} \cup P_{j',3}$ such that $P_{j,1}, P_{j,2}, P_{j',1}, P_{j',2} \in DF_{ms}(P)$. By hypothesis, each of $P_{j,1}, P_{j,2}, P_{j',1}, P_{j',2}$ are completely ms-colored by $ms_C^X$. In addition, all the elements of $P_{k,t} \cap X_c^p$ and $P_{k,t} \cap X_c^s$ are respectively p-bound and s-bound by $c$ for $P_{k,t}$ in $ms_C^X$, $k \in \{j, j'\}$, $t \in [2]$, for all $c \in C$. By Lemma 74 there exist $y_{j,1} \in P_{j,1}$ s-free, $x_{j',1} \in P_{j',1}$ p-free and $x_{j',2} \in P_{j',2}$ p-free for respectively $(P_{j,1}, e)$, $(P_{j',1}, e)$ and $(P_{j',2}, e)$ in $ms_C^X$. In addition, one can prove, by contradiction and using the same arguments of Lemma 74's proof, that $y_{j,1}, x_{j',1}$ and $x_{j',2}$ are still respectively s-free, p-free and p-free for $(P, e)$ in $ms_C^X$. So, by setting $ms_C^{X'} = ms_C^X +_e (y_{j,1}, x_{j',1}, x_{j',2})$, then $(y_{j,1}, x_{j',1}, x_{j',2})$ is directly bound by $e$ for $P$ in $ms_C^{X'}$ and $(y_{j,1}, x_{j',1}, x_{j',2})$ ms-colors $P$ in $e$. Furthermore, since $y_{j,1}, x_{j',1}$ and $x_{j',2}$ are respectively s-free, p-free, p-free for $(P, e)$ in $ms_C^X$ then for all $P' \in F_{ms}(P) \setminus \{P\}$, $x \in P'^3$ and $c \in C$, $x$ is bound by $c$ in $ms_C^X$ for $P'$ if and only if $x$ is bound by $c$ in $ms_C^{X'}$ for $P'$. ◄

We are now ready to prove Proposition 73.

**Proof of Proposition 73.** The case of the empty poset is trivial, so assume $P \in SP^{\diamond+}$. We build, by induction on $r_X(P)$, a ms-coloring $ms_C^X = (X_c^w, X_c^s, X_c^p)_{c \in C}$ encoding $\mathfrak{c}$, making sure that for all $c \in C$, $X_c^p$ and $X_c^s$ contain only respectively p-bound and s-bound elements for $P$ in $ms_C^X$. We start with all sets of $ms_C^X$ empty.

The first case is when $r_X(P) = (0,0)$. In this case, $P$ is the singleton $\{x\}$. Then, set $X_e^w = \{x\}$, where $e = \mathfrak{c}(P)$, and all the other sets composing $ms_C^X$ to $\emptyset$ and this is done.

Otherwise, assume that $P \in Seq$ and its irreducible sequential factorization is $\sum_{j \in J} P_j$, for some $J \in \mathcal{S} \setminus \{0,1\}$. For all $j \in J$, let $P_j = \bigcup_{i \in [k_j]} P_{j,i}$ be the irreducible parallel factorisation of $P_j$. Note that each $r_X(P_{j,i}) < r_X(P)$ for all $P_{j,i}$. Then by induction hypothesis, there exists $ms_C^{X_{j,i}}$ satisfying the hypothesis of the proposition for each $P_{j,i} \in DF_{ms}(P)$. Let $ms_C^{X'}$ be the union, set by set, of the different $ms_C^{X_{j,i}}$, for all $j \in J$, $i \in [k_j]$, $k_j > 1$. By construction of $ms_C^{X'}$, for all $c \in C$, each element $x$ of $X_c^p$ (resp. $X_c^s$) is p-bound (resp. s-bound) by $c$ for some element $F_x \in F_{ms}(P) \setminus \{P\}$ in $ms_C^{X'}$. This ensures that $\mathcal{C}_{ms_{C}^{X'}}(P) = \emptyset$ and $ms_C^{X'}(F) = \mathfrak{c}(F)$ for all $F \in F_{ms}(P) \setminus \{P\}$. It suffices to apply Lemma 75 to conclude.

Finally, assume that $P$ is of the form $\bigcup_{j \in [k]} P_j$ for some $k > 1$ where each $P_j \in DF_{ms}(P)$. Since each $P_j \in Seq$, building a ms-coloring $ms_C^{X_j}$ satisfying the hypothesis of the proposition for each $P_j$ can be achieved using the same arguments that those used in the previous paragraph. We define $ms_C^X$ as the union, set by set, of the different $ms_C^{X_j}$s, for all $j \in [k]$. ◄

We proved that any ms-coloring $\mathfrak{c}$ of $P$ with $C$ can be encoded by some $ms_C^X$ as above. Furthermore, we claim that there exist MSO formulæ:

- $\mathtt{ms}_C^X(F) = c$ which is satisfied if and only if $ms_C^X(F) = c$ (cf. Definition 70), assuming $F \in F_{ms}(P)$. It expresses that

- if $F$ has at least one element comparable to all the others, then there is a unique $(e, x) \in C \times F$ such that $x$ is comparable to all the other elements of $F$, and $x \in X_e^w$. If such $(e, x)$ exists then $e = c$;

- or $F$ has no element comparable to all the others, and there is a unique $(e, x) \in C \times F^3$ such that $x$ is directly bound by $e$ for $F$ in $ms_C^X$ (cf. Definition 68). If such $(e, x)$ exists then $e = c$.

■ `ms-Coloring`$(P, ms_C^X)$ which is satisfied if and only if $ms_C^X$ is a complete ms-coloring of $P$. It expresses that for each $F \in F_{ms}(P)$ there exists $c \in C$ such that $ms_C^X(F) = c$.

These formulæ are obtained by a direct translation into MSO of Definitions 68–70.

## 6.2 Encoding a compatible s-coloring with MSO

We encode a compatible s-coloring by means of MSO with a set $X_c^v$ for each $c \in C$, and a complete ms-coloring of $P$ with $2^C$ encoded by some $ms_{2^C}^X$. For short we let $s_C^X$ denote the sets $((X_c^v)_{c \in C}, ms_{2^C}^X)$.

▶ **Definition 76.** *Let $P \in SP^{\diamond+}$, $F \in F_s(P)$ of sequentially irreducible factorisation $F = \sum_{j \in J} F_j$, $X_c^v$ a subset of $P$ for each $c \in C$, and $ms_{2^C}^X$ a complete ms-coloring of $P$. Then $s_C^X$ s-colors $F$ in $c \in C$, denoted $s_C^X(F) = c$, if for all $j \in J$,*

1. *$F_j \in X_c^v$ when $|F_j| = 1$;*
2. *when $|F_j| > 1$, $c \in ms_{2^C}^X(F_{j,i})$ for all $F_{j,i}$ such that $F_j = \bigcup_{i \in [n_j]} F_{j,i}$ is the irreducible parallel factorisation of $F_j$;*
3. *there is no $F'$, $F''$ not both empty such that $F' + F + F'' \in F_s(P)$ and $s_C^X(F' + F + F'') = c$.*

*Finally, when $P \in SP^{\diamond}$ and $\mathfrak{c}: F_s(P) \to C$ is a s-coloring of $P$ with $C$ then the subsets $s_C^X$ of $P$ encode $\mathfrak{c}$ if $s_C^X(F) = \mathfrak{c}(F)$ for all $F \in F_s(P)$.*

As a consequence of Condition 3 of Definition 76, observe that when $F$ and $F'$ are two sequential factors of $P$ such that $FF'$ is also a sequential factor of $P$, there is no $s_C^X$ encoding a s-coloring such that $s_C^X(F) = s_C^X(F')$.

▶ **Example 77.** Let $C = \{\mathtt{red}, \mathtt{green}\}$ be a set of colors and $P = a + (b \cup c) + d + (e \cup ((g \cup h) + (i \cup j)) \cup f)$. Let $F_1 = a + (b \cup c)$, $F_2 = P \setminus F_1$, $F_3 = (g \cup h) + (i \cup j)$ and $F_4 = e$. Then $F_1, F_2, F_3, F_4 \in F_s(P)$. Note that $F_3 \subsetneq F_2$ and that there exists $x \in F_2 \setminus F_3$ such that $x$ is incomparable to any $y \in F_3$. Let also $\mathfrak{c}: F_s(P) \to C$ be the s-coloring defined by $\mathfrak{c}(F_1) = \mathfrak{c}(F_4) = \mathtt{green}$ and $\mathfrak{c}(F_2) = \mathfrak{c}(F_3) = \mathtt{red}$. Note that $\mathfrak{c}$ is compatible. Figure 7 represents $P$ with an encoding $s_C^X$ of $\mathfrak{c}$.

The set $F_{ms}(P)$ consists of the singleton posets $b, c, e, g, h, i, j, f$, of $F_3$ and of $P$. We have $ms_{2^C}^X(b) = ms_{2^C}^X(c) = \{\mathtt{green}\}$ and $ms_{2^C}^X(e) = ms_{2^C}^X(g) = ms_{2^C}^X(h) = ms_{2^C}^X(i) = ms_{2^C}^X(j) = ms_{2^C}^X(f) = \{\mathtt{red}\}$. We also have $ms_{2^C}^X(F_3) = \{\mathtt{red}\}$ since $\mathcal{C}_{ms_{2^C}^X}(F_3) = \{(h, i, j)\}$, $(h, i, j) \in X_{\{\mathtt{red}\}}^s \times X_{\{\mathtt{red}\}}^p \times X_{\{\mathtt{red}\}}^p$ and there is no decomposition of $F_3$ into $F_3 = X + Y + Z$ for some $X, Y, Z$ such that $Y$ is a singleton. As $\mathcal{C}_{ms_{2^C}^X}(P) = \{a\}$ and $a \in X_\emptyset^w$ we have $ms_{2^C}^X(P) = \emptyset$, and thus $P$ is completely ms-colored by $ms_{2^C}^X$.

We have $s_C^X(F_1) = \mathtt{green}$ since $a \in X_{\mathtt{green}}^v$, $d \notin X_{\mathtt{green}}^v$, $DF_{ms}(F_1) = \{\{b\}, \{c\}\}$, and $\mathtt{green} \in ms_{2^C}^X(F)$ for all $F \in DF_{ms}(F_1)$. We have $s_C^X(F_2) = \mathtt{red}$ since $d \in X_{\mathtt{red}}^v$, $\mathtt{red} \notin ms_{2^C}^X(b)$, $DF_{ms}(F_2) = \{\{f\}, F_3, F_4\}$ and $\mathtt{red} \in ms_{2^C}^X(F)$ for all $F \in DF_{ms}(F_2)$. We have $s_C^X(F_3) = \mathtt{red}$ since $DF_{ms}(F_3) = \{\{g\}, \{h\}, \{i\}, \{j\}\}$, $\mathtt{red} \in ms_{2^C}^X(F)$ for all $F \in DF_{ms}(F_3)$, and there is no $X, Y$ non both empty such that $X + F_3 + Y$ is a factor of $P$. Finally $s_C^X(F_4) = \mathtt{green}$ since $e \in X_{\mathtt{green}}^v$ and there is no $X, Y$ non both empty such that $X + e + Y$ is a factor of $P$.

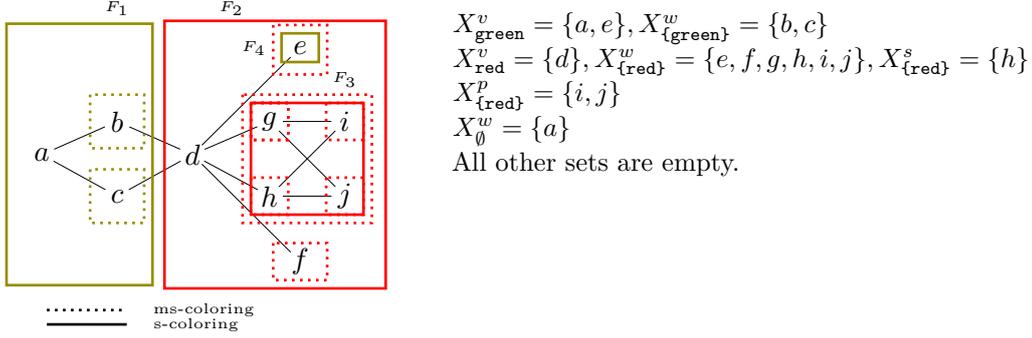

$X_{\texttt{green}}^v = \{a, e\}, X_{\{\texttt{green}\}}^w = \{b, c\}$

$X_{\texttt{red}}^v = \{d\}, X_{\{\texttt{red}\}}^w = \{e, f, g, h, i, j\}, X_{\{\texttt{red}\}}^s = \{h\}$

$X_{\{\texttt{red}\}}^p = \{i, j\}$

$X_{\emptyset}^w = \{a\}$

All other sets are empty.

**Figure 7** The Hasse diagram of a poset $P \in SP^{\diamond}$ and a s-coloring $s_C^X$ of $P$.

▶ **Proposition 78.** *Let $P \in SP^{\diamond}$, $C$ be a non-empty finite set of colors and $\mathfrak{c}: F_s(P) \to C$ a compatible s-coloring of $P$. There exist some subsets $s_C^X$ of $P$ that encode $\mathfrak{c}$.*

**Proof.** From $\mathfrak{c}$ define a ms-coloring $\mathfrak{c}': F_{ms}(P) \to 2^C$ as follows: for all $F \in F_{ms}(P)$,

$$\mathfrak{c}'(F) = \{c : F \in DF_{ms}(F') \text{ for some } F' \in F_s(P) \text{ such that } \mathfrak{c}(F') = c\}$$

According to Proposition 73, $\mathfrak{c}'$ is encoded by some $ms_{2^C}^X$. In order to define an encoding $s_C^X = ((X_c^v)_{c \in C}, ms_{2^C}^X)$ of $\mathfrak{c}$ it suffices now to define $X_c^v$ as follows, for each $c \in C$: $x \in X_c^v$ if and only if there exists $F \in F_s(P)$ such that $\mathfrak{c}(F) = c$ and $x$ is comparable to all the elements of $F \setminus \{x\}$. We verify that $s_C^X(F) = \mathfrak{c}(F)$ for all $F \in F_s(P)$. Assume first that $\mathfrak{c}(F) = c$ for some $F \in F_s(P)$ and $c \in C$, and let $F = \sum_{j \in J} F_j$ be its irreducible sequential factorisation. Let $j \in J$ such that $F_j$ consists of a unique element $x$, which is thus comparable to all the elements of $F \setminus \{x\}$. By definition $x \in X_c^v$. Thus Item 1 of Definition 76 is verified. Similar arguments apply when $F_j$ is not a singleton: Item 2 of Definition 76 is also verified. Since there is no $F' \in F_s(P)$ such that $\mathfrak{c}(F') = c$ and $FF'$ or $F'F \in F_s(P)$, and there is no $X, Y$ not both empty such that $\mathfrak{c}(XFY) = c$, then Item 3 of Definition 76 is verified, and thus $s_C^X(F) = c$. Similar arguments are used to show that if $s_C^X(F) = c$ then $\mathfrak{c}(F) = c$. ◀

We proved that any compatible s-coloring $\mathfrak{c}$ of $P$ with $C$ can be encoded by some $s_C^X$ as above. Furthermore, we claim that there exist MSO formulæ:

- $\texttt{s}_C^X(F) = c$ which is satisfied if and only if $s_C^X$ s-colors $F$ in $c \in C$, assuming $F \in F_s(P)$. It is obtained by a direct translation into MSO of Definition 76;
- $\texttt{s-Coloring}(P, s_C^X)$ which is satisfied if and only if $\texttt{ms-Coloring}(P, ms_{2^C}^X)$ is true and for each non-empty parallel factor $F$ of $P$ there exists $c \in 2^C$ such that $ms_{2^C}^X(F') = c$, for all $F' \in DF_{ms}(F)$.

## 6.3 Compatible s-colorings and D-graphs

s-coloring is a key argument in the transformation of the D-graph of a rational expression into a P-MSO formula (Section 7). We use s-coloring in order to associate a special edge to some $F \in F_s(P)$ as follows.

▶ **Proposition 79.** *Let $D_e$ be the D-graph of a rational expression $e$, $C = \mathbb{B} \times E_S(D_e)$, $n$ a node of $D_e$. Assume that there is a path $T_P$ from $n$ labeled by some $P \in SP^{\diamond}(A)$. There exist a compatible s-coloring of $P$ with $C$ and its encoding $s_C^X$ with MSO such that $\pi_2(s_C^X(F)) = f \in E_S(D_e)$ if and only if $F$ is marked by $f$ in $T_P$, for any $F \in F_s(P)$.*

The encoding for s-colorings does not allow $s_C^X(F) = s_C^X(F')$ when $F, F', FF' \in F_s(P)$. Alternation of the booleans in $C$ is used when we need a s-coloring to associate the same special edge to $F$ and $F'$.

**Proof.** Let $\mathfrak{c} : F_s(P) \to C$ be a s-coloring of $P$ such that, for all $F \in F_s(P)$, $\mathfrak{c}(F) = (b, f)$ if and only if the following conditions are verified:

- $F$ is marked by $f$ in $T_P$;
- for all $F' \in F_s(P)$, if $FF' \in F_s(P)$ and $\mathfrak{c}(F') = (b', f)$ then $b \neq b'$.

Obviously $\mathfrak{c}$ verifies the specification of the proposition (see also Remark 52), and the existence of $s_C^X$ comes from Proposition 78. ◀

▶ **Example 80.** Let us continue Example 54. Denote by $b_l, c_l, d_l$ (resp. $b_r, c_r, d_r$) the leftmost (resp. rightmost, cf. Figure 5) elements of $P$ labeled by $b, c$ and $d$. Let $C = \mathbb{B} \times E_S(D_e)$. By Proposition 79 there exists a compatible s-coloring and its encoding $s_C^X = ((X_c^v)_{c \in C}, ms_{2^C}^X)$ such that $s_C^X(F_1) = (b_1, n_4 \to n_7)$, $s_C^X(F_2) = (b_2, n_4 \to n_7)$, $s_C^X(F_3) = (b_3, n_4 \to n_3)$, $s_C^X(F_4) = (b_4, n_4 \to n_2)$ for some $b_1, b_2, b_3, b_4 \in \mathbb{B}$, and $s_C^X(F)$ is undefined when $F \neq F_1, F_2, F_3, F_4$. Note that $b_1$ must be different from $b_2$ since $F_1 F_2 \in F_s(P)$ and $\pi_2(s_C^X(F_1)) = \pi_2(s_C^X(F_2)) = n_4 \to n_7$. Following the techniques of Section 6 we have

- $X_{(b_1, n_4 \to n_7)}^v = \{b_l\}$;
- $X_{(b_2, n_4 \to n_7)}^v = \{b_r\}$;
- $X_{(b_3, n_4 \to n_3)}^v = \{b_l, b_r\}$;
- $X_{(b_4, n_4 \to n_2)}^v = F_4$;
- $X_c^v = \emptyset$ for all $c \in C \setminus \{(b_1, n_4 \to n_7), (b_2, n_4 \to n_7), (b_3, n_4 \to n_3), (b_4, n_4 \to n_2)\}$;
- $ms_{2^C}^X(c_l) = ms_{2^C}^X(d_l) = \{(b_3, n_4 \to n_3), (b_1, n_4 \to n_7)\}$;
- $ms_{2^C}^X(c_r) = ms_{2^C}^X(d_r) = \{(b_3, n_4 \to n_3), (b_2, n_4 \to n_7)\}$;
- $ms_{2^C}^X(F) = \emptyset$ when $F \in F_{ms}(P) \setminus \{\{c_l\}, \{d_l\}, \{c_r\}, \{d_r\}\}$.

Proposition 79 states that the s-coloring $\mathfrak{c}$ of $P$ issued from marks in a path can be encoded with MSO using some $s_C^X$.

## 7 From D-graphs to P-MSO

Let $e$ be a rational expression, $D_e$ the D-graph of its >1-expression and $C = \mathbb{B} \times E_S(D_e)$. We are going to parse $D_e$ in order to compute a P-MSO sentence $\phi_{D_e}$ such that $P$ is a model for $\phi_{D_e}$ if and only if $P \in L(D_e)$, for all $P \in SP^\diamond(A)$. For each node $n$ of $D_e$, we define a P-MSO formula $\phi_n$ accordingly to the definition of a path from $n$. The formula $\phi_n$ depends on a second-order parameter $X$, and we want $\phi_n$ to be satisfied if and only if $X$ is a factor of $P$ labeling a path $T_X$ from $n$. The formula $\phi_n(X)$ depends on the label of $n$ and its edging $out(n) = n \to n_1, \ldots, n \to n_k$. If $n$ is labeled by some letter $a \in A$, then $\phi_n(X) \equiv |X| = 1 \wedge \forall x \; (x \in X \to a(x))$. In the case where $n$ is labeled by $\cdot^{>1}$ then $n$ has two direct descendants $n_1$ and $n_2$, and $\phi_n(X)$ expresses that there exists a partition of $X$ into non-empty $X_1, X_2$ such that $X_1 < X_2$ and $X_i$, $i \in [2]$, satisfies $\phi_{n_i}(X_i)$. This construction for $\cdot^{>1}$ is P-MSO definable. The cases of other labels are mere adaptations of the case of linear orderings [5], except for Presburger formulæ. Indeed, recall that by construction, $D_e$ without its special edges has a structure of acyclic graph (Property DAG), and that only nodes labeled in $\mathcal{P}$ may be sources of special edges, which may cause circular dependencies between the $\phi_n$s. Recall also that a sub-path $T_R$ is started by a special edge $e$ if and only if its label $R$ is marked by $e$ in $T_R$ (Definition 48), and that $R$ is necessarily a sequential poset because of Property SS. Circular dependencies between the $\phi_n$s are avoided as follows. We encode by means of MSO the marking of labels of sub-paths using s-coloring: we assume $R$ to be the label of a sub-path marked by some special edge $e$ if and only if $R$ is s-colored by $(b, e)$ for some boolean $b$. Proposition 79 states that this can always be done. When $n$ is the source of a special edge $n \to m$, instead of making $\phi_n$ dependant of $\phi_m$ for testing that a sequential factor $F \in L(m)$, we make $\phi_n$ dependant of a formula that checks if $F$ is s-colored by $(b, n \to m)$ for some boolean $b$. This supposes that it is known that every sequential factor $F$ s-colored by $(b, n \to m)$ satisfies $\phi_m$.

Formally, set $C = \mathbb{B} \times E_S(D_e)$. When the label of a node $n$ is some $\rho(x_1, \ldots, x_k)$ and $\mathrm{out}(n) = n \to n_1, \ldots, n \to n_k$, set

$$\phi_n(X) \equiv \mathcal{Q}(X, \chi_1, \ldots, \chi_k, \rho(x_1, \ldots, x_k))$$

where

$$\chi_i \equiv \begin{cases} \forall Y (\forall y \; y \in Y) \to \phi_{n_i}(Y) & \text{when } n \to n_i \in E_N(D_e); \\ \forall Y (\forall y \; y \in Y) \to \vee_{b \in \mathbb{B}} s_C^X(Y) = (b, n \to n_i) & \text{when } n \to n_i \in E_S(D_e). \end{cases}$$

Informally speaking, the sentence $\phi_{D_e}$ encodes that $P \in L(D_e)$ if and only if there is a path $T_P$ in $D_e$ from $r(D_e)$ and labeled by $P$; in order to avoid circular dependencies it uses an encoding $s_C^X$ of a compatible s-coloring of $P$ such that $\pi_2(s_C^X(F)) = f$ if and only if $F$ is marked by $f$ in $T_P$, for any sequential factor $F$ of $P$. It guarantees that when $F$ is marked by $f = n \to m$ then $F$ satisfies $\phi_m(F)$. Formally

$$\begin{aligned} \phi_{D_e} \equiv & \exists R \exists s_C^X \; (\forall x \; x \in R) \wedge \mathtt{s\text{-}Coloring}(R, s_C^X) \wedge \phi_{r(D_e)}(R) \\ & \wedge \big( \bigwedge_{n \to m \in E_S(D_e)} \forall F \; (\mathtt{F}_s(F, R) \wedge \bigvee_{b \in \mathbb{B}} \mathtt{s}_C^X(F) = (b, n \to m)) \to \phi_m(F) \big) \\ & \bigvee_{\epsilon \in L(e)} \forall X \; (\forall x \; x \in X) \to |X| = 0 \end{aligned}$$

▶ **Example 81.** Let $e = a \circ_\xi (a(\xi \parallel \xi))^{*\xi}$ be the rational expression $e_1$ of Example 53. In this example we detail the construction of the P-MSO formula $\phi_{D_e}$ from $D_e$, using the techniques of Section 7. The D-graph $D_e$ and the poset $P = a(a \parallel a(a(a \parallel a(a \parallel a)) \parallel a(a \parallel a)))$ of $L(D_e)$ are represented on the left side of Figure 8. Let $n_1, \ldots, n_6$ be the pre-order traversal of $D_e$ without its special edges. We have $E_S(D_e) = \{n_4 \to n_2\}$, $C = \mathbb{B} \times E_S(D_e)$ and

- $\phi_{n_1}(X) \equiv \mathcal{Q}(X, \overline{\phi_{n_2}(Y)}, \overline{\phi_{n_6}(Y)}, x_1 + x_2 = 1)$;
- $\phi_{n_2}(X) \equiv \exists X_1, X_2, \; X = X_1 + X_2 \wedge \phi_{n_3}(X_1) \wedge \phi_{n_4}(X_2)$;
- $\phi_{n_3}(X) \equiv \phi_{n_5}(X) \equiv \phi_{n_6}(X) \equiv |X| = 1 \wedge \forall x \; (x \in X \to a(x))$;
- $\phi_{n_4}(X) \equiv \mathcal{Q}(X, \vee_{b \in \mathbb{B}} \mathtt{s}_C^X(Y) = (b, n_4 \to n_2), \overline{\phi_{n_5}(Y)}, x_1 + x_2 = 2)$

where $\overline{\psi(Y)} \equiv \forall Y \; (\forall y \; y \in Y) \to \psi(Y)$. Then

$$\begin{aligned} \phi_{D_e} \equiv & \exists R \exists s_C^X \; (\forall x \; x \in R) \wedge \mathtt{s\text{-}Coloring}(R, s_C^X) \wedge \phi_{n_1}(R) \\ & \wedge (\forall F \; (F_s(F, R) \wedge \bigvee_{b \in \mathbb{B}} \mathtt{s}_C^X(F) = (b, n_4 \to n_2)) \to \phi_{n_2}(F)) \end{aligned}$$

On the right side of Figure 8 is pictured a path $T_P$ in $D_e$ from $n_1$ labeled by $P$. The root of $T_P$ is labeled by $(n_1, (1, 0))$ with $(1, 0) \in L(x_1 + x_2 = 1)$. Then $T_P$ has a unique direct sub-path $T_P'$ from $n_2$ also labeled by $P$. Thus $P \in L(n_2)$. Observe that $T_P' = (m, T_{P_1}, T_{P_2})$ where $m$ is labeled by $(n_2, 2)$, and $T_{P_1}$ and $T_{P_2}$ are some paths in $D_e$ from respectively $n_3$ and $n_4$ labeled respectively by $P_1 = a$ and $P_2 = a \parallel a(a(a \parallel a(a \parallel a)) \parallel a(a \parallel a))$. Hence, $P = P_1 P_2$. Let $F_i$, $i \in [4]$, be the labels of the sub-paths $T_{F_i}$, $i \in [4]$, of $T_P$ started by the special edge $n_4 \to n_2$, as in Figure 8. A factor of $P$ is marked by $n_4 \to n_2$ in $T_P$ if and only if it is one of the $F_i$s. Therefore, observe that there is some $x \in F_i \setminus F_j$ which is incomparable to all the elements of $F_j$ when $F_j \subsetneq F_i$, as mentioned in Item ($C_2$) of Remark 52. By Proposition 79, this marking can be expressed by means of MSO by some encoding $s_C^X$ of a compatible s-coloring. A factor of $P$ marked in $T_P$ satisfies $\vee_{b \in \mathbb{B}} \mathbf{s}_C^X(Y) = (b, n_4 \to n_2)$, for some $b \in \mathbb{B}$, if and only if it is one of the $F_i$s. The maximum elements of $P_2$ satisfy $\phi_{n_5}$, $P_1$ satisfies $\phi_{n_3}$ and $P_2$ satisfies $\phi_{n_4}$. Thus $P$ satisfies $\phi_{n_2}$ as well as $\phi_{n_1}$ and is a model for $\phi_{D_e}$.

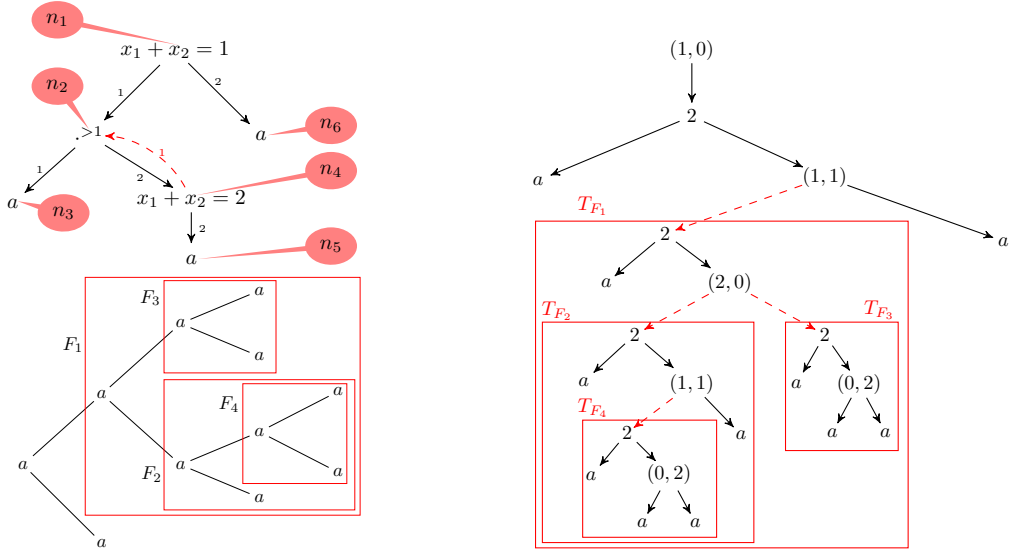

■ **Figure 8** The D-graph $D_e$ of the $>$1-expression of $e = a \circ_\xi (a(\xi \parallel \xi))^{*\xi}$, a poset $P = a(a \parallel a(a(a \parallel a(a \parallel a)) \parallel a(a \parallel a)))$ of $L(e)$ and a path in $D_e$ from $n_1$ labeled by $P$

## 8 From P-MSO to rational expressions

The transformation of a P-MSO formula $\varphi$ into a rational expression $e$ such that $L(\varphi) = L(e)$ is based on several known results and uses only well-known techniques. Thus we give here only the main arguments, and refer the reader to the bibliography for the details. Automata over posets of $SP^\diamond(A)$ effectively equivalent to rational expressions were introduced in [6]. Melting the techniques of the translation of a MSO formula to an automaton over countable and scattered linear orderings [5], and those of the translation of a P-MSO formula to an automaton over finite N-free posets [3], we build by induction on $\varphi$ a branching automaton $\mathcal{A}_\varphi$ such that $L(\mathcal{A}_\varphi) = L(\varphi)$. The only inductive step which can not be directly deduced from [5] or [3] is the construction of $\mathcal{A}_{\neg\varphi'}$ from $\mathcal{A}_{\varphi'}$. This case is a direct consequence of Theorem 18.

## 9 Conclusion

Links between automata theory and formal logic have been initiated by Büchi in his earlier works, in particular in order to use automata as a basis for decision procedures for MSO. The results presented in this paper have the same use. The construction of a branching automaton $\mathcal{A}_\varphi$ from a formula $\varphi$ of P-MSO such that $L(\mathcal{A}_\varphi) = L(\varphi)$ is effective. Furthermore, it is decidable whether $L(\mathcal{A}_\varphi) = \emptyset$ or not. As a consequence:

▶ **Theorem 82.** *Let A be an alphabet. The P-MSO theory of $SP^\diamond(A)$ is decidable.*

Since the pioneer works of Büchi, links between automata and logic have been used in many ways, such as the characterization of fragments of MSO and classification of languages of words, trees and other structures. Our work is a step in that direction for languages of transfinite posets.